\documentclass[12pt]{iopart}
\usepackage{iopams}
\usepackage{amsthm,dsfont,bm,amssymb}
\expandafter\let\csname equation*\endcsname\relax
\expandafter\let\csname endequation*\endcsname\relax
\usepackage{amsmath}
\usepackage{graphicx}
\usepackage{enumerate}
\usepackage[mathscr]{eucal}
\usepackage{tensor}
\usepackage{cite}

\begin{document}

\review{Stochastic Tools Hidden Behind the Empirical Dielectric Relaxation Laws} %Anomalous Time Evolution of Complex Systems}

\author{Aleksander Stanislavsky$^{1,2}$ and Karina Weron$^{3}$}
\address{$^1$Institute of Radio Astronomy, NASU, 4 Mystetstv St., 61002 Kharkiv, Ukraine.\\
$^2$V.N. Karazin Kharkiv National University, Svobody Sq. 4, 61022 Kharkiv, Ukraine.\\
$^3$Faculty of Fundamental Problems of Technology, Wroc{\rm\l}aw University of Science and Technology, Wybrze${\rm\dot z}$e Wyspia${\rm\acute n}$skiego 27, 50-370 Wroc{\l}aw, Poland.}
\begin{abstract}
The paper is devoted to recent advances in stochastic modeling of anomalous kinetic processes observed in dielectric materials which are prominent examples of disordered (complex) systems. Theoretical studies of dynamical properties of those ``structures with variations'' [Goldenfield and Kadanoff 1999 {\em Science} \textbf{284} 87--9] require application of such mathematical tools by means of which their random nature can be analyzed and, independently of the details differing various systems (dipolar materials, glasses, semiconductors, liquid crystals, polymers, etc.), the empirical universal kinetic patterns can be derived. We begin with a brief survey of the historical background of the dielectric relaxation study.  After a short outline of the theoretical ideas providing the random tools applicable to modeling of relaxation phenomena, we present probabilistic implications for the study of the relaxation-rate distribution models. In the framework of the probability distribution of relaxation rates we consider description of complex systems, in which relaxing entities form random clusters interacting with each other and single entities.  Then we focus on stochastic mechanisms of the relaxation phenomenon. We discuss the diffusion approach and its usefulness for understanding of anomalous dynamics of the relaxing systems.  We also discuss extensions of the diffusive approach to the systems under tempered random processes. Useful relationships among different stochastic approaches to the anomalous dynamics of complex systems allow us to get a fresh look on this subject. The paper closes with a final discussion on achievements of stochastic tools describing the anomalous time evolution of the complex systems.
\end{abstract}

\maketitle

\newtheorem{theorem}{Theorem}[section]
\newtheorem{prop}{Proposition}[section]
\newtheorem{cor}{Corollary}[section]
\theoremstyle{definition}
\newtheorem{defi}{Definition}[section]
\newtheorem{ex}{Example}[section]

\tableofcontents
\pagestyle{plain}

\newpage
\hfill{\it Dedicated in memory of Leo Kadanoff}

%%%%%%%%%%%%%%
%%%%%%%%%%%%%%
\section{Introduction}
%%%%%%%%%%%%%%
%%%%%%%%%%%%%%
Motion of charges, their accumulation and discharge are the basis of many physical, chemical and biological processes in nature. Undoubtedly, many-body interactions \cite{gold,dh83,ngai11} play an appreciable role in the time evolution of such systems. Besides, the systems themselves are weakly or strongly disordered. This aspect is very versatile. Defects, vacancies and dislocations are frequently present in real materials \cite{cusack}. Amorphous materials possess a marked departure from crystalline order \cite{elliott}, and a perfect (ideally ordered) crystal is difficult to find in nature. There exists a great variety of materials that have a local order in few atoms or molecules, but their structure becomes disordered on larger length scales. In a consequence, these effects induce relaxation processes inseparably linked with disorder in the systems. The relaxing entities -- dipoles, traps, ions and so on, interact not only among themselves, but also with the surrounding medium to modify disorder of this medium and to affect other entities. The transformations include time fluctuations in potentials seen by each entity and essentially act as a noise source. On the other hand, they form a complex potential landscape with many local minima separated by barriers of all scales, trapping and untrapping the entity orbits in a self-similar hierarchy of cantori. Consequently, the motion of entities can be very similar to a random walk. It is not surprising that parallels, suggested in literature \cite{shlesinger84,ks86,gy95,fy95}, to be drawn between relaxation and diffusion.

The relaxation properties of various complex systems (amorphous semiconductors and insulators, polymers, molecular solid solutions, glasses, etc) have attracted an immediate interest of scientists and technologists for a long time \cite{jonscher83} (and the references therein). This gave a huge wealth of experimental data. The data analysis discovered the ``universality'' of relaxation patterns \cite{jonscher96} {\it per se} that is enclosed into fractional power laws of relaxation responses (in frequency and time) for a very wide range of materials. The fascinating behavior covers 17 decades ($\sim$10$^{-5}$-10$^{12}$ Hz in frequency or $\sim$10$^{-12}$-10$^5$ s in time), and a theoretical explanation of the universal relaxation response is one of the most difficult problems of Physics today, as any realistic physical treatment of relaxation has to take into account the stochastic or probabilistic representation of the system's behavior. Most of the interpretations in  literature (see their comprehensive discussion, for example, in \cite{jonscher83,jonscher96}) explain only a limited number of characteristics of the relaxation processes in complex systems. Without any doubt the physical processes going on in disordered  media are complex and independent on the details of the systems under investigation. Therefore, any simple interpretation based on one or two observational facts will not explain all the features of relaxation patterns self consistently. This review is just devoted to recent advances in the theory of relaxing complex media. Our approach is based on limit theorems of probability theory. The usefulness of the theorems is that they allow us to connect microscopic stochastic dynamics of relaxing entities with the macroscopic deterministic behavior of the systems as a whole. The universal macroscopic relaxation response appears not to be attributed to any particular entity taken from those forming such a system. Any excited complex system tending to equilibrium passes from less disordered states to more disordered ones. Macroscopic evolution of the system is a result of averaging over local random properties of system's entities. The complex system is a marvelous ``supercomputer'' capable for the procedure, and the limit theorems of probability theory play the role of its ``software''.

The complex systems and the investigation of their structural and dynamical properties have established on the physics agenda almost three decades ago. These ``structure with variations'' \cite{gold} are characterized through (i) a large diversity of elementary units, (ii) strong interactions between the units, (iii) a non-predictable or anomalous evolution in course in time \cite{peliti}. Their study play a dominant role in exact and life sciences, including a richness of systems such as glasses, liquid crystals, polymers, proteins, biopolymers, organisms or even ecosystems.

\subsection{Origins of the theory of relaxation}

Experimental and theoretical studies of relaxation phenomena have a long history. The first measurements of electrical relaxation were carried out for alkali ions in the Leyden jar (a glass) in 1847 and 1854 by R. Kohlrausch \cite{kohl47,kohl54}, and the observations of mechanical relaxation in the natural polymer, silk, in 1863 and 1866 were continued by his son, F. Kohlrausch \cite{kohl63}. The concept of ``relaxation time'' into physics and engineering was introduced by J. C. Maxwell in 1867 \cite{maxw67}. As was shown by M. J. Curie \cite{curie89} and E. von Schweidler \cite{schweid07} the dielectric relaxation response in the time domain can be described by a short-time power-law dependence. Perhaps, P. Debye in 1913 was the first who derived soundly the relaxation relationship based on principles of statistical mechanics \cite{d12,debye13}. For this purpose he used Einstein's theory of Brownian motion \cite{einst05,einst06} to consider the collisions between a rotating dipolar molecule and its neighboring other molecules in the liquid under the assumption that the only electric field acting on the molecule is an external field. Consequently, the Debye relaxation law is expressed in terms of rotational Brownian motion, and it has an exponentially decaying form in time domain. The physical mechanism underlying the Debye law is obviously simpler than the one underlying the stretched exponential relaxation found by Kohlrausch \cite{kohl54}. Rapid developments of science and technology in the twentieth and twenty-first centuries created a wide variety of new materials with non-exponential behavior in relaxation properties. Over the past 100 years, many empirical relaxation laws, which are regarded as generalizations of the Debye (D) relaxation law, have been discovered. Among the most known, and frequently utilized to analyze the frequency domain measurements, are the Cole-Cole (CC) law (1941-1942) \cite{cole41,cole42}, the Cole-Davidson (CD) law (1950-1951) \cite{david50,david51} and the Havriliak-Negami (HN) law (1966-1967) \cite{hn66,hn67}. In fact, all they were established empirically. Their form is convenient to write as
\begin{equation}
\phi^*_{\rm HN}(\omega)=\frac{1}{[1+(i\omega/\omega_p)^\alpha]^\gamma}\,,\quad 0<\alpha,\gamma\leq1\,,\label{eqi0a}
\end{equation}
$\alpha=\gamma=1$ being the D case; $\alpha=1,0<\gamma<1$ being CD; $0<\alpha<1,\gamma=1$ being CC. The stretched exponential relaxation pattern, known also as the Kohlrausch-Williams-Watts (KWW) function, has a simple form in time domain (1970) \cite{ww70}, namely it reads
\begin{equation}
\phi_{\rm KWW}(t)=e^{-(t/\tau_p)^\alpha} \label{eqi0b}
\end{equation}
with $0<\alpha<1$. Here $\tau_p=\omega_p^{-1}$ is the time constant characteristic for a given material. The KWW function takes the simple exponential form if $\alpha=1$.

\subsection{Experimental peculiarity of relaxation}\label{subsection_ib}

\begin{figure}
\centerline{\includegraphics[clip,width=1.15\columnwidth]{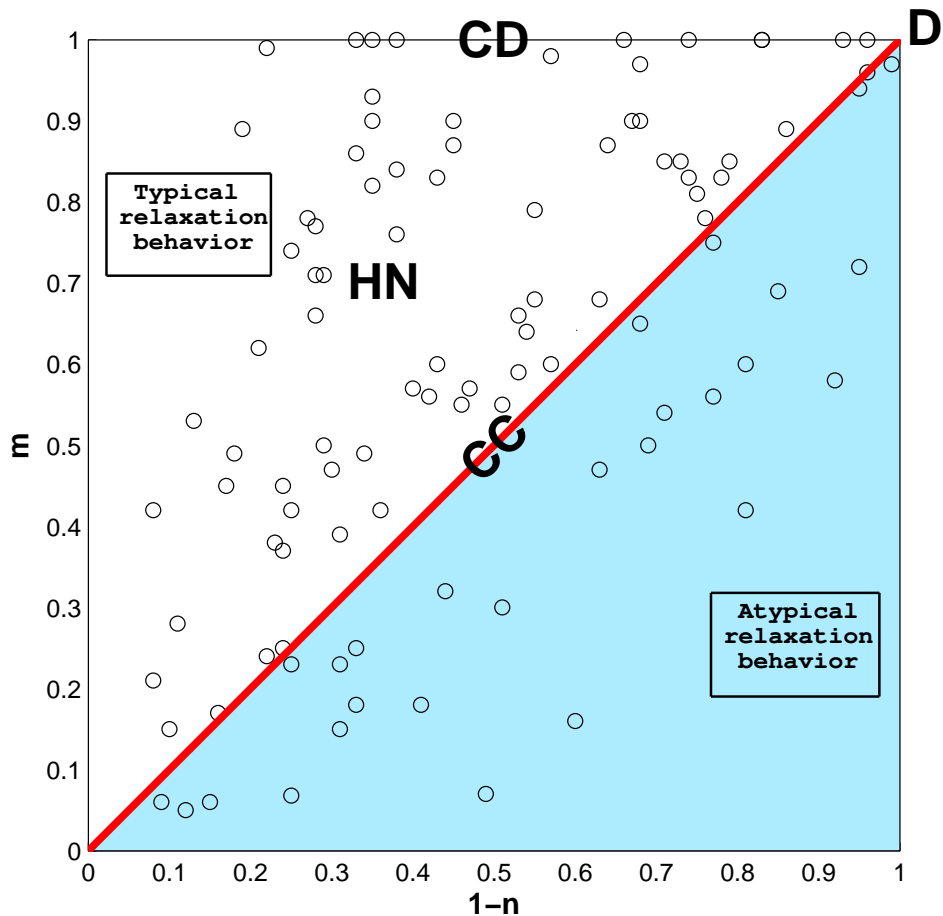}}
\caption{(Color on-line) Relaxation diagram positioning different
two-power laws of relaxation. The power-law exponents $m$ and $1-n$ focus on
declination of the imaginary susceptibility $\chi''(\omega)$ for
low and high frequencies. The circles are experimental points (for
various materials) taken from the book \cite{jonscher83}. The area with $m<1-n$ contains fewer data points compared to the other part with $m>1-n$. The abbreviations have the following meanings: D -- Debye; CC --
Cole-Cole; CD -- Cole-Davidson; HN -- Havriliak-Negami.}
\label{diag-1}
\end{figure}

In his two monographs A. K. Jonscher \cite{jonscher83,jonscher96} has shown that a common property of the empirical relaxation laws is that they exhibit the fractional-power dependence in the complex dielectric susceptibility $\chi(\omega) = \chi'(\omega)-i\chi''(\omega)$, i.\,e.
\begin{equation}
\chi(\omega)\propto\left(\frac{i\omega}{\omega_p}\right)^{n-1}\qquad{\rm for\ }\omega\gg\omega_p\,,\label{eqi1}
\end{equation}
and
\begin{equation}
\Delta\chi(\omega)\propto\left(\frac{i\omega}{\omega_p}\right)^m\qquad{\rm for\ }\omega\ll\omega_p\,,\label{eqi2}
\end{equation}
where positive constant $\omega_p$ is the loss peak frequency characteristic for the investigated material, and $\Delta\chi(\omega)=\chi_0-\chi(\omega)$. It is worth noticing that this unique property is independent on any particular details of the examined systems and a  great majority of dielectric materials may be characterized by two parameters (power-law exponents) $m$ and $1-n$, both falling strictly within the range (0,1]. This evidently shows Figure~\ref{diag-1}, where the exponents were defined for one hundred different materials (see Table 5.1 in \cite{jonscher83}). In this connection it should be noticed that any satisfactory theory of relaxation must be capable of explaining this very general feature being so largely independent of the detailed physical nature of the materials involved.

Experimental data of relaxation can be characterized in both time and frequency domains. The inverse Stieltjes-Fourier transform
\begin{equation}
\phi^*(\omega)=\int_0^\infty e^{-it\omega}\,d(1-\phi(t))\label{eqii8}
\end{equation}
relates the time-domain relaxation function $\phi(t)$  to the complex susceptibility $\chi(\omega)$ by the formula $\chi(\omega)=\phi^*(\omega)(\chi_0-\chi_\infty)+\chi_\infty$, where $\phi^*(\omega)$ is the frequency domain relaxation function (shape function). The constant $\chi_\infty$ represents the asymptotic value of the susceptibility $\chi(\omega)$ at high frequencies, and $\chi_0$ is the value of the opposite limit.
Note that the derivative $f(t)=-d\phi(t)/dt$, called the response function, is connected with the shape function $\phi^*(\omega)$ by the following relation
\begin{equation}
\phi^*(\omega)=\int_0^\infty e^{-it\omega}\,f(t)\,dt\label{eqii9}
\end{equation}
that will be very useful for finding the latter function hereinafter. The existence of the following asymptotic responses, corresponding to Eqs.(\ref{eqi1}) and (\ref{eqi2}),
\begin{equation}
f(t)\propto
\begin{cases} (t/\tau_p)^{-n} &
{\rm for}\quad t \ll \tau_p\\ V(t) & {\rm
for}\quad t \gg \tau_p\end{cases}\label{eqii10}
\end{equation}
in relaxation dynamics of complex systems has been established \cite{jonscher83,jonscher96,hav94}, the function $V(t)$ being regarded as
\begin{displaymath}
V(t)=
\begin{cases} (t/\tau_p)^{-m-1} &
{\rm for\ HN\ and\ CC\ relaxation}\,,\\ \exp\left[-(t/\tau_p)^{\alpha}\right] & {\rm for\ KWW\ relaxation}\,,\\
\exp\left(-t/\tau_p\right) & {\rm for\ CD\ relaxation}\,.\end{cases}
\end{displaymath}
The typical, fractional two-power-law behavior (\ref{eqi1}) and (\ref{eqi2}) is usually fitted with the HN function. In this case one has $m=\alpha$, $n=1-\alpha\gamma$ and $m\geq1-n$. The exponents $m=1$ and $n=1-\gamma$ correspond to the CD relaxation, characterized by the short-time fractional power law only. The same property, however of different origins, concerns the KWW response for which the short-time power-law exponent $n=1-\alpha$.

Observe that fitting with the HN function of the atypical relaxation data (see Figure~\ref{diag-1}), for which the power-law exponents fulfill the opposite relation $m<1-n$, requires for the parameter $\gamma$ values greater than 1 \cite{hav94}. As it will be shown below, the stochastic scenarios of relaxation do not allow us to derive this function with $\gamma>1$.

%%%%%%%%%%%%%%
%%%%%%%%%%%%%%
\section{Definitions and terminology}
%%%%%%%%%%%%%%
%%%%%%%%%%%%%%

\subsection{Limit theorems}

The probability theory considers a chance of the occurrence of an event in multiple repeating random experiments so as, for example, in a series of throws of a coin, where we can observe either its head or tail many times \cite{feller66}. In stochastic modeling of kinetic processes the basic notation involves random variables. They are characterized by the distribution function $F(x)=\Pr(X\leq x)$ providing information about the probability $\Pr(x\leq X<x+dx)$, that the random variable $X$ takes a value between $x$ and $x+dx$, is equal to the difference $F(x+dx)-F(x)$. If the distribution function $F(x)$, $F(-\infty)=0$ and $F(+\infty)=1$, of the random variable $X$ fulfills the condition
\begin{displaymath}
F(x) =\int_{-\infty}^xf(y)\,dy
\end{displaymath}
for every real $x$, then the function $f(x)$ is called the probability density function (pdf). The $n$th moment of a pdf $f(x)$ is the expected value of $X^n$, namely
\begin{displaymath}
\left\langle X^n\right\rangle =\int_{-\infty}^\infty x^n\,f(x)\,dx\,.
\end{displaymath}
More generally, for any integrable function $g(\cdot)$ the expected value of $g(X)$ reads
\begin{displaymath}
\left\langle g(X)\right\rangle=\int_{-\infty}^\infty g(x)\,f(x)\,dx\,.
\end{displaymath}
Note, the last definition will be often used in the present paper.

The study of sequences of random independent and identically distributed (iid) variables $X_1,\,X_2,\dots,\,X_n$ is one of the central topics in the probability theory. This is explained by some causes. At first, the statistical properties of the above sequence can be analyzed only asymptotically, i.\,e. when  the number of variables $n\to\infty$. The distribution characteristics such as moments are calculated in this way. On the other hand, often the set $X_1$, $X_2,\ \dots$ is a sequence of observations where the variable $X$ is observed repeatedly in time. Each individual observation is unpredictable, but the frequency of different outcomes over a large number of such observations becomes predictable. In particular, following the Bernoulli law of large numbers, in the experiments with only two results (``success'' and ``failure'') the frequency of the success will oscillate around the probability $p$ of the success \cite{feller66}. The number of the success in $n$ trials is defined by the sum $\zeta_n=X_1+\dots+X_n$ having the binomial distribution. The strong law of large numbers states  that the random variable $\frac{\zeta_n}{n}$ loses its randomness as the number $n$ of trials tends to infinity. Further studies of the deviation estimate $\left|\frac{\zeta_n}{n}-p\right|$ led to the first central limit theorem, i.\,e. the sum $\zeta_n$ for sufficiently large $n$, independently of distribution  of a single component $X_i$ (but with the finite second moment), has the law close to the normal one. Namely, the central limit theorem answers why in so many uses (like the theory of errors, for example) one can find the probability distributions closely connected with the Gaussian one. Moreover, a wide circle of practical applications extends an essence of this theorem so that it has been generalized in many different ways. One of such generalizations concerns those distributions of $X_i$ that have no finite variance nor even mean value. Other direction to new limit theorems considers the operations on the sequence $X_1$, $X_2, \dots$ different from the summation, as for example, in the extreme value theory \cite{llr86} where the minimum and maximum operations are taken into account. Any case of the limit theorems indicates an asymptotic tendency of the sequence of random variables, as a result of an operation, to some non-degenerate random variable belonging to the class of limiting distributions (domain of attraction) different for every operation.

\subsection{L\'evy $\alpha$-stable distributions}

In this subsection we present some basic facts on the L\'evy $\alpha$-stable ($L\alpha S$ in short notation) distributions useful for the purpose of this article. The principal feature of these distributions is that they are completely described as limits of the normalized sums of iid summands \cite{feller66}. Consequently, $L\alpha S$ distributions represent some kind of a universal law.

The distribution function $F(x)=\Pr(X\leq x)$ is called stable if for every $a_1>0$, $b_1$, $a_2>0$, $b_2$ there are constants $a>0$ and $b$ such that the equation
\begin{equation}
F(a_1x+b_1)\ast F(a_2x+b_2)=F(ax+b) \label{eqii1}
\end{equation}
holds. The symbol $F_1\ast F_2$ indicates the convolution of two distributions in the sense
\begin{equation}
F_1\ast F_2=\int F_1(x-y)\,dF_2(y)\,.\label{eqii2}
\end{equation}
It turns out that always
\begin{equation}
a = (a_1^\alpha+a_2^\alpha)^{1/\alpha}\qquad{\rm with\ }0<\alpha\leq2 \label{eqii3}
\end{equation}
and the constant $\alpha$ is called the characteristic exponent of $L\alpha S$ distribution. Equation (\ref{eqii1}) can be solved in terms of characteristic functions, i.\,e., via Fourier transform
\begin{displaymath}
h(s) =\int_{-\infty}^\infty\exp(isx)\,dF(x)\,.
\end{displaymath}
For the distribution function $F(x)$ to be $L\alpha S$ it is necessary and sufficient that its characteristic function $h(s)$ is represented by the formula
%\begin{displaymath}
%\log h(s)= \begin{cases}
%i\gamma s-\sigma|s|^\alpha\{1-i\beta\,{\rm sgn}(s)\tan(\pi\alpha/2)\} \\ \hfill \mbox{if } \alpha\neq 1\,,\\ \\ i\gamma s-\sigma|s|-i\beta(2/\pi)\sigma s\log|s| \\ \hfill \mbox{if } \alpha=1\,,\end{cases}
%\end{displaymath}
\begin{displaymath}
\log h(s)= \begin{cases}
i\gamma s-\sigma|s|^\alpha\{1-i\beta\,{\rm sgn}(s)\tan(\pi\alpha/2)\} \qquad \mbox{if } \alpha\neq 1\,,\\ i\gamma s-\sigma|s|-i\beta(2/\pi)\sigma s\log|s| \qquad\qquad\quad \mbox{  if } \alpha=1\,,\end{cases}
\end{displaymath}
where $\alpha$, $\beta$, $\gamma$ and $\sigma$ are real constants with $\sigma\geq0$, $0<\alpha\leq2$ and $|\beta|\leq 1$. Here, $\alpha$ is the characteristic exponent, $\gamma$ and $\sigma$ determine location and scale. The coefficient $\beta$ indicates whether the $L\alpha S$ distribution is symmetric ($0<\alpha\leq 2$, $\beta=0$) or completely asymmetric ($0<\alpha<1$, $|\beta|=1$). The values $\alpha=2$ and $\beta=0$ yield the Gaussian distribution. As $h(s)$ is absolutely integrable, the corresponding $L\alpha S$ distribution has a density $f(x)$. Beautiful animations of the stable pdf's with different values of the parameters are available on J. P. Nolan's website (http://fs2.american.edu/jpnolan/www/stable/stable.html).

The most convenient formulation of the limit theorem, which gives description of the distribution law $F(x)$ governing the sum of a large number of mutually iid random quantities $X_i$, $i=1,2,\dots,n$, can be given in the following form : {\it only L$\alpha$S distributions have a domain of attraction}, i.\,e., there exist normalizing constants $a_n>0$, $b_n$ such that the distribution of $a_n^{-1}(X_1+X_2+\dots+X_n)-b_n$ tends to $F(x)$ as $n\to\infty$. The normalizing constants can be chosen in such a way that $a_n= n^{1/\alpha}$.

Note, the random variable $X$ can describe an arbitrary physical magnitude (e.\,g. time, space, temperature, energy, etc). In particular, when $X$ is a waiting or residence time, the tail $\Pr(X>x)$ of the distribution $F(x)$ determines the survival probability. Let us add that the distribution of a non-negative random variable, say $X$, has a power asymptotic form if the tail $1-F(x)=\Pr(X >x)$ satisfies the condition
\begin{equation}
\lim_{x\to\infty}\frac{\Pr(X>x)}{\sigma x^{-a}}=1\label{eqii4}
\end{equation}
for some $a>0$ and $\sigma>0$; that is, if for large values of $x$ the tail decays as a fractional power law $\sigma x^{-a}$. There are many different continuous and discrete distributions satisfying condition (\ref{eqii4}). Classical examples of continuous ones are the $L\alpha S$ laws, also the Pareto and Burr distributions with an appropriate choice of their parameters \cite{feller66,john70}.

If the distribution of random variable $X$ has a heavy tail with the parameter $0<a<1$, then the expected value $\langle X\rangle=\int x\,f(x)\,dx=\int x\,dF(x)$ is infinite. Note that in general if $\Pr(X>x)\sim 1/x^a$ for $x\to\infty$, then the moments $\langle X^n\rangle$ are finite for $a>n$. Therefore, the two considered attributes, the finiteness of the expected value and the heavy-tail property (\ref{eqii4}), clearly exclude each other. Besides, both provide only limited information on the corresponding distributions. Hence, the conditions put on the distributions of the microscopic quantities in the proposed scheme are rather general. On the other hand, by utilizing the limit theorems of probability theory the macroscopic result is determined in any detail.

\subsection{Mixtures of distributions}\label{subsiicn}

Mixtures of distributions occur frequently in applications of the probability theory \cite{feller66}. They also are directly relevant to problems of non-exponential relaxation. In this instance we deal with random variables the distribution of which depends on various factors, and all relaxing systems consist of many subsystems interacting among each other in a random way. Therefore, we call the sort of systems as a complex one. If $X$ is the random variable with pdf $f_X(x)$, then the random variable $aX$ ($a$ is constant) has the pdf $f_X(x/a)\frac{1}{a}$, and the random variable $X+a$ obeys the pdf $f_X(x-a)$.  Let $Y$ be another random variable with pdf $f_Y(y)$. Now the product of random variables $XY$ takes the pdf in the integral form
\begin{displaymath}
\int f_X(x/y)\,f_Y(y)\,\frac{dy}{y}\,.
\end{displaymath}
On the other hand, the pdf of the random variable $X/Y$ is written as
\begin{displaymath}
\int f_X(xy)\,f_Y(y)\,y\,dy\,.
\end{displaymath}
The sum $X+Y$ is described by the convolution of pdfs, namely
\begin{displaymath}
\int f_X(x-y)\,f_Y(y)\,dy\,.
\end{displaymath}
The relaxation rates of complex systems can depend on many parameters: temperature, defects, pressure and so on. Each of them has a very different distribution during a specific experimental scenario. However, macroscopic behavior of such systems is only a result of averaging such random effects. Thus, the mixtures of distributions become very helpful for the study of relaxation mechanisms.

\subsection{Stochastic processes. Subordination}\label{pariid}
As J. L. Doob has defined \cite{doob53}, a stochastic process is ``the mathematical abstraction of an empirical process whose development is governed by probabilistic laws''. There are two equivalent points of view about what is the stochastic process: (i) an infinite collection of random variables indexed by an integer or a real number often interpreted as time, and (ii) a random function of two or several deterministic arguments, one of which is the time $t$. It is convenient to consider separately the cases of discrete and continuous time. A discrete stochastic process $X=\{X_n,\ n =0,1,2,\dots\}$ is a countable collection of random variables indexed by the non-negative integers, and a continuous stochastic process $X=\{X_t,\ 0\leq t<\infty\}$ is an uncountable collection of random variables indexed by the non-negative real numbers. The Bernoulli process is perhaps the simplest non-trivial stochastic process. It is a sequence, $X_0, X_1, \dots$, of iid binary random variables that take only two values, 0 and 1. The common interpretations of the values $X_i$ are true or false, success or failure, arrival or no arrival, yes or no, etc. Note that the simple model of Bernoulli process initiated a great development of the studies on the limit theorems and served as the building block for other more complicated stochastic processes (Poisson process, renewal processes and others). The most known continuous stochastic process is the Brownian motion. Starting with 1827, when the botanist R. Brown observed zigzag, irregular patterns in the movement of microscopic pollen grains suspended in water, the phenomenon has found a satisfactory explanation only in 1905-1906 due to the physicists A. Einstein and M. Smoluchowski. Their probabilistic models have been based on assumption that the Brownian motion is a result of continual collisions of the pollen grains by the molecules of the surrounding water \cite{sp12}. In 1923, the mathematician N. Wiener has proved the mathematical existence of Brownian motion as a stochastic process with the given properties. Any Brownian motion is a continuous time series of random variables whose increments are iid normally distributed with zero mean. It is plausible by the central limit theorem. Notice that this stochastic process is also a continuous-time analog to the simple symmetric random walk \cite{sp12}. If one considers a massive Brownian particle under the influence of friction, the Ornstein-Uhlenbeck process has a bounded variance and admits a stationary probability distribution \cite{uo30}. Eventually, this list of continuous stochastic processes unbarred doors to their study in different ways and under various conditions \cite{hf13}.

On the other hand, the diversity of stochastic processes may be extended notedly, if the parameter (index) $t$ varies stochastically. This approach, introduced by S. Bochner in 1949 \cite{bochner49} is called subordination \cite{feller66,wm09}. Then the process $Y(G(t))$ is obtained by randomizing the time variable of the stochastic process $Y(\tau)$ using a new ``timer'', which is a stochastic process $G(t)$ with nonnegative independent increments. The resulting process $Y(G(t))$ is said to be subordinated to $Y(\tau)$ and is directed by the process $G(t)$, which is called a directional process. The directional process is often referred to as randomized time or operational time \cite{feller66}. In general, the subordinated process $Y(G(t))$ can be non-Markovian, even if its parent process $Y(\tau)$ is Markovian.

%%%%%%%%%%%%%%
%%%%%%%%%%%%%%
\section{Relaxation function as the initial-state survival probability.\\ Probabilistic models}
%%%%%%%%%%%%%%
%%%%%%%%%%%%%%
The classical Debye relaxation law
\begin{equation}
\phi_{\rm D}(t) = \exp(-t/\tau_p)\label{eqiii1}
\end{equation}
is characterized by a single relevant relaxation time $\tau_p$. To account for the non-exponential relaxation phenomenon, in the historically oldest attempt E. von Schweidler \cite{schweid07} assumed different parts of the orientational polarization to decline exponentially with different relaxation times $\tau_i$, yielding
\begin{displaymath}
\phi(t)=\sum_{i=1}^{n}p_i\exp\left(-t/\tau_i\right)\,,
\end{displaymath}
where the weights $p_i$ of the exponential decays fulfill
\begin{displaymath}
\sum_{i=1}^np_i=1\,.
\end{displaymath}
Few years later, K.W. Wagner \cite{wagner13} proposed to use of a continuous distribution $w(\tau)$ of relaxation times
\begin{equation}
\phi(t) = \int_0^\infty\exp(-t/\tau)\,w(\tau)\,d\tau\,,\label{eqiii2}
\end{equation}
where $\int_0^\infty w(\tau)\,d\tau=1$.

This approach is microscopically arbitrary since it does not yield any constraints on the random microscopic scenario of relaxation. The probability density function $w(\tau)$, as well as, the weights $p_i$ are determined only by the empirical patterns of $\phi(t)$. This simple way to derive the non-exponential decay is associated with a picture of parallel relaxations, in which each degree of freedom (each relaxation channel) relaxes independently with random relaxation time \cite{fjst88,lp80,bb78,h83,mc85,sm84,mb84,shlesinger84,ks85,kb85,ks86,weron86}. From the probabilistic point of view, both above formulas reveal the weighted average $\left\langle\dots\right\rangle$ of an exponential relaxation
\begin{equation}
\phi(t) = \left\langle e^{-t/\widetilde T}\right\rangle\label{eqiii2a}
\end{equation}
with respect to the distribution of the random effective relaxation time $\widetilde T$ with support of $\tau\in\left[0,\infty\right)$.

Contrary to models that were based on a parallel addition of relaxation contributions, the model presented in \cite{psaa84} proposes a serial summation of a hierarchy of relaxations extending over the same spatial range. The authors pointed out that a group of dipoles must adopt to a specific configuration before a subset can relax, which then releases the constraints preventing a further subset from relaxing, and so on. Although it has been realized in many approaches that the individual dipoles and their environment do not remain independent during the regression of fluctuation, as yet no microscopic model has been based directly on this conclusion. The exception is the cluster model \cite{dh79,dh83,d84,dnh85,dh87,dh89}, which derived entirely new expressions from a consideration of the way in which the energy contained in fluctuation is distributed over a system of interacting clusters. This is also the only theory in which the results obtained are in agreement with empirical functions input to fit the experimental data for $\phi(t)$ in the short- $t\ll\tau_p$ and the long-time $t\gg\tau_p$ limits.

\subsection{Transition probability}

Relaxation properties of dielectric materials have been the subject of experimental and theoretical investigations for many years. This is not only due to the need for understanding of the electrical properties of various technological materials, but it has also been realized that the basic physics of the dielectric relaxation response leads to interesting questions about the theoretical description of physical relaxation mechanisms in disordered (complex) systems.

Relaxation phenomena are experimentally observed when a physical macroscopic magnitude (concentration, current, etc.), characteristic for the investigated system, monotonically decays or grows in time. In case of the dielectric relaxation this process is commonly defined as an approach to equilibrium of a dipolar system driven out of equilibrium by a step or alternating external electric field. The time-dependent response of relaxing systems to a steady electric field is described by the relaxation function $\phi(t)$ (satisfying $\phi(0)=1$ and $\phi(\infty)=0$) which is a solution of the two-state master equation
\begin{equation}
\frac{d\phi(t)}{dt}=-r(t)\phi(t)\,,\qquad\phi(0)=1\,.\label{eqii7}
\end{equation}
The non-negative quantity $r(t)$ is the transition rate of the system (i.\,e. the probability of transition per unit time), see e.\,g. \cite{fjst88}. Consequently, the function $\phi(t)$ has a meaning of the survival probability $\Pr(\tilde\theta\geq t)$ of a non-equilibrium initial state of the relaxing system \cite{ks86}. In other words, $\phi(t)$ is determined by the probability that the system as a whole will not make a transition out of its original state for at least time $t$ after entering it at $t=0$.

Note, if instead of the decay of polarization one observes its increase under the influence of the steady external electric field, the proper function to describe this relaxation process would be just $F(t) = 1-\phi(t)$ satisfying $F(0)=0$ and $F(\infty)=1$ (see, e.g. \cite{jonscher83}). From the probabilistic point of view, the response function $f(t)$ is just the probability density function of the waiting-time distribution $F(t) = 1-\phi(t)=\Pr(\tilde\theta<t)$.

Let us assume that a relaxing physical system undergoes an irreversible transition from initial state ${\bf 1}$, imposed at time $t=0$, to state ${\bf 2}$ that differs from ${\bf 1}$ in some physical parameter. The transition ${\bf 1}\to {\bf 2}$, defined as a change of this parameter, takes place at random instant of time. To be precise, we consider the conditional probability $p(t,dt)$ that the system as a whole will undergo the transition during a time interval $(t,t+dt)$ provided that the transition did not occur before time $t$, i.\,e.
\begin{equation}
p(t,dt)=\Pr\bigl(t\leq\tilde\theta\leq t+dt|\tilde\theta\geq t\bigr)\,,\label{eqiii3}
\end{equation}
where $\tilde\theta$ is the system's random waiting time for transition ${\bf 1}\to {\bf 2}$. The conditional probability
$p(t, dt)$ defined in (\ref{eqiii3}) can be expressed as
\begin{equation}
p(t,dt)=-\frac{\Pr(\tilde\theta\geq t+dt)-\Pr(\tilde\theta\geq t)}{\Pr(\tilde\theta\geq t)}\,,\label{eqiii3a}
\end{equation}
where $\Pr(\tilde\theta\geq t+dt)$ and $\Pr(\tilde\theta\geq t)$ are the survival probabilities, i.\,e. the probabilities that the system will remain in state ${\bf 1}$ until time $t + dt$ and $t$, respectively.
As $dt\to 0$, probability (\ref{eqiii3a}) can be rewritten in a form more useful for further considerations, namely
\begin{displaymath}
p(t,dt)=-d\ln\Pr\bigl(\tilde\theta\geq t\bigr)\,.
\end{displaymath}
In this case the function
\begin{displaymath}
r(t)=-\frac{d}{dt}\ln\Pr(\tilde\theta\geq t)
\end{displaymath}
is nothing else than the intensity of  time-dependent transition probability (transition rate) \cite{kampen84,jw99}.
Then the survival probability $\Pr(\tilde\theta\geq t)=\phi(t)$ can be found at once if one knows the explicit form of the intensity, namely
\begin{equation}
\phi(t)=\exp\Bigl(-\int_0^tr(s)\,ds\Bigr)\,.\label{eqiii3b}
\end{equation}
As the intensity of transition is essentially time dependent, the evolutionary law for the entity is of a non-exponential form. For the time-independent intensity $r(s) = \tau_p^{-1} = {\rm const}$ one obtains
\begin{displaymath}
\phi(t)=\exp(-t/\tau_p)\,,
\end{displaymath}
what recovers the classical exponential decay law (\ref{eqiii1}) with the value $\tau_p^{-1}$ determining the relaxation rate of the transition process.

Unfortunately, the time-dependent relaxation rate $r(t)$ in Eq. (\ref{eqii7}) can take the very cumbersome form for empirical laws of relaxation (especially, for the HN one). Moreover, the determinism appears from empirical laws, observed on the macroscopic level, while randomness is induced by variations in the local environment. In order to understand the phenomenon of the universal relaxation response one needs to consider relaxation in complex systems in a way that separates it from a particular physical context, and observation that the dynamics of such systems is characterized by seemingly contrary states, i.\,e., local randomness and global determinism, is crucially relevant to this issue. These two states, in a natural way, can coexist in the framework of the limit theorems of probability theory when the relaxation function $\phi(t)$, being the survival probability of the system in the initially imposed non-equilibrium state until time $t$, is described by the first-passage of the system.

\subsection{Microscopic stochastic scenario}

In any complex system, capable of responding to an external electric field, it is possible that only a part of the total number $N$ of dipoles in the system is able to follow significantly changes of the field \cite{jonscher83}. In general, the distribution of the random waiting time $\tilde\theta$ of the entire system is determined by the first passage of the system from its initial state \cite{wj93,jw99}.%,wk97}.

Consider (as in the preceding subsection) a system of $N$ entities, each waiting for transition ${\bf 1}\to {\bf 2}$ for a random time $\theta_{i}$, where $1\leq i\leq N$. Generally speaking, the waiting times $\theta_{1},\theta_{2},\ldots,\theta_{N}$ form an arbitrary sequence of iid non-negative random variables. The entities undergo transition in a certain order that can be reflected in the notion of order  statistics, that provide a non-decreasing rearrangement $\theta_{(1)}\leq\theta_{(2)}\leq\ldots\leq\theta_{(N)}$ of times $\theta_{1},\theta_{2},\ldots,\theta_{N}$ \cite{jw99,feller66}.  From this rearrangement follow two obvious statistics, the first and the $N$th one, $\theta_{(1)}$ and $\theta_{(N)}$, respectively. Now, denoting an unknown (random) number of individual transitions occurred until time $t>0$ by $\eta_N(t)$, we can connect the event number $\{\eta_N(t)=n\}$ with the order statistics via the relations
%\begin{eqnarray}
%\{\eta_N(t)=0\}&=&\{\theta_{(1)}>t\}=\{\min(\theta_{1},\theta_{2},\ldots,\theta_{N})>t\},\nonumber\\
%\{\eta_N(t)=n\}&=&\{\theta_{(n)}<t\,,\theta_{(n+1)}>t\}\nonumber\\
%&&\qquad\qquad\qquad{\rm for}\ \ n=1,2,\dots,N-1,\nonumber\\
%\{\eta_N(t)=N\}&=&\{\theta_{(N)}\leq t\}\nonumber\\
%&=&\{\max(\theta_{1},\theta_{2},\ldots,\theta_{N})\leq t\}\,.\label{eqiii6}
%\end{eqnarray}
\begin{eqnarray}
\{\eta_N(t)=0\}&=&\{\theta_{(1)}>t\}=\{\min(\theta_{1},\theta_{2},\ldots,\theta_{N})>t\},\nonumber\\
\{\eta_N(t)=n\}&=&\{\theta_{(n)}<t\,,\theta_{(n+1)}>t\}\quad{\rm for}\ \ n=1,2,\dots,N-1,\nonumber\\
\{\eta_N(t)=N\}&=&\{\theta_{(N)}\leq t\}=\{\max(\theta_{1},\theta_{2},\ldots,\theta_{N})\leq t\}\,.\label{eqiii6}
\end{eqnarray}
The first of them indicates that no transition has occurred in this system until
time $t$. The second one shows a tendency to ${\bf 1}\to{\bf 2}$. The population of entities in the state ${\bf 1}$ are decreased step-by-step in favor of the relaxation output ${\bf 2}$. The last expression of Eq.(\ref{eqiii6}) means that all
transitions have been finished up to time $t$.

Let us now introduce a notion of the initial-state survival probability of the entire $N$-dimensional system as ${\rm
Pr}\bigl(\widetilde\theta_N\geq t\bigr)$, i.\,e. the probability
that transition of the system as a whole has not occurred prior to a time
instant $t$, where $\widetilde\theta_N$ denotes the system's waiting time for transition from its initial, imposed state. The probability ${\rm
Pr}\bigl(\widetilde\theta_N\geq t\bigr)$ means also that there is no any individual
transition until time $t$. Therefore, from Eq.(\ref{eqiii6}) we have
%\begin{eqnarray}
%{\rm Pr}\bigl(\widetilde\theta_N\geq t\bigr)&=&{\rm Pr}\bigl(\eta_N(t)=0\bigr)\nonumber\\
%&=&{\rm Pr}\bigl(\min(
%\theta_{1},\theta_{2},\ldots,\theta_{N})\geq t\bigr)\,.\label{eqiii9}
%\end{eqnarray}
\begin{equation}
{\rm Pr}\bigl(\widetilde\theta_N\geq t\bigr)={\rm Pr}\bigl(\eta_N(t)=0\bigr)={\rm Pr}\bigl(\min(
\theta_{1},\theta_{2},\ldots,\theta_{N})\geq t\bigr)\,.\label{eqiii9}
\end{equation}
As a rule, the macroscopic systems consist of a large number of relaxing entities so that the relaxation function can be approximated by the weak limit in distribution
%\begin{eqnarray}
%\phi(t)&=& \Pr(\widetilde\theta\geq t)\nonumber\\
%&\stackrel{d}{=}&\lim_{N\to\infty}{\rm Pr}\bigl(A_N\min(\theta_{1},\theta_{2},\ldots,\theta_{N})\geq
%t\bigr)\,,\label{eqiii11b}
%\end{eqnarray}
\begin{equation}
\phi(t)= \Pr(\widetilde\theta\geq t)\stackrel{d}{=}\lim_{N\to\infty}{\rm Pr}\bigl(A_N\min(\theta_{1},\theta_{2},\ldots,\theta_{N})\geq
t\bigr)\,,\label{eqiii11b}
\end{equation}
where $A_N$ denotes a sequence of normalizing constants, and $\stackrel{d}{=}$ means ``equal in law''. In relation to the above definition of the relaxation function, the frequency-domain shape function (\ref{eqii9}) can be written as
\begin{equation}
\phi^*(\omega)=\langle e^{-i\omega\widetilde\theta}\rangle\,,\label{eqiii11bd}
\end{equation}
where $\langle\dots\rangle$ denotes an average with respect to the distribution of the system's effective waiting time $\widetilde\theta$.
It follows from  limit theorems  of the extremal value theory \cite{llr86}, that since the sequence of waiting times
$\theta_{1},\theta_{2},\ldots,\theta_{N}$ consists of iid non-negative random variables, the above definition of the relaxation function leads to the result
\begin{displaymath}
\phi(t)=\exp[-(At)^\alpha]\,,
\end{displaymath}
where $A$ and $\alpha$ are positive constants. Observe that this form of the relaxation function, being just the tail of the well-known Weibull distribution \cite{feller66}, contains three possible cases: the stretched exponential behavior if $0<\alpha<1$, the exponential if $\alpha=1$, and the compressed exponential one if
$\alpha>1$. At this point it is natural to ask how, within the proposed stochastic scenario, one can derive the stretched exponential KWW function, as well as, the other empirical relaxation patterns (see Eq.(\ref{eqi0a})).  To solve this problem one has to realize that, in general, in empirical relaxation evidence  we observe two classes of relaxation responses. Namely, a class exhibiting the short-time power law only (fitted with the KWW or CD function), and a class exhibiting both short- and long-time power laws (fitted with the HN function). Hence, the first step to solve this problem is to find a rigorous mathematical condition yielding the KWW short-time power law in the framework of the above stochastic scenario.

\subsection{Stretched exponential relaxation}

Traditional interpretation of the non-exponential relaxation phenomena is based on the concept of a system of independent, exponentially relaxing species (dipoles) with different relaxation rates \cite{bb78}. The exponential relaxation of an individual dipole in this case is conditioned only by the value taken by its relaxation rate. So, taking into account influence of the local random environment on the entity, one can conclude \cite{wj93,jw99}: if the relaxation rate $\beta_i$ of {\it i}th dipole has taken the value $b$, then the probability that this dipole has not changed its initial aligned position up to the moment $t$, is
\begin{equation}
\Pr(\theta_i\geq t|\beta_i=b)=\exp(-bt)\ \ {\rm for}\ \ t\geq0\ \ \ b>0\,.\label{eqiii21}
\end{equation}
The random variable $\beta_i$ denotes the relaxation rate of {\it i}th dipole and the variable $\theta_i$, the time needed for changing its initial orientation; $\beta_1,\beta_2,\dots$ and $\theta_1,\theta_2,\dots$ form sequences of non-negative, iid random variables. The randomness of the individual relaxation rate is motivated by the fact that in a complex system its entity can be into many states or even pass through a whole hierarchy of substates within states, and the distribution of individual relaxation rates effectively accounts for the transition intensity between the states and substates.

From the law of total probability \cite{feller66}, we have
\begin{equation}
\Pr(\theta_i\geq t)=\int_0^\infty\exp(-bt)\,dF_{\beta_i}(b)\,,\label{eqiii22}
\end{equation}
where $F_{\beta_i}(b)$ is the distribution function of each relaxation rate $\beta_i$. In other words, $F_{\beta_i}(b)$ denotes the probability that the relaxation rate of {\it i}th dipole has taken a value less than or equal to $b$. Formula (\ref{eqiii22}) shows that if one takes into account influence of the local random environment on relaxation behavior of a dipole, its initial-state survival probability decays non-exponentially. Only if the influence is deterministic, i.e., the individual relaxation rate $\beta_i$ takes the value $b_0$ with probability 1, given by the pdf of the Dirac $\delta$-function form $dF_{\beta_i}(b)=\delta(b-b_0)\,db$, the individual survival probability decays exponentially $\Pr(\theta_i\geq t)=e^{-b_0t}$.

In order to obtain an explicit form of the relaxation function $\phi(t)$ defined in (\ref{eqiii11b}), let us observe that the right-hand expression in (\ref{eqiii22}) is just the Laplace transform of the distribution function $F_{\beta_i}(b)$ at point $t$,
\begin{displaymath}
\Pr(\theta_i\geq t)={\cal L}(F_{\beta_i} (b); t)\,.
\end{displaymath}
Because $\theta_i$ are independent random variables, we get
%\begin{eqnarray*}
%\Pr\left(A_N\min(\theta_1,\dots,\theta_N)\geq t\right)&=&\left[\Pr\left(\theta_i\geq\frac{t}{A_N}\right)\right]^N\\
%&=&\left[{\cal L}\left(F_{\beta_i} (b); \frac{t}{A_N}\right)\right]^N\,.
%\end{eqnarray*}
\begin{displaymath}
\Pr\left(A_N\min(\theta_1,\dots,\theta_N)\geq t\right)=\left[\Pr\left(\theta_i\geq\frac{t}{A_N}\right)\right]^N=\left[{\cal L}\left(F_{\beta_i} (b); \frac{t}{A_N}\right)\right]^N\,.
\end{displaymath}
The {\it N}th power of the Laplace transform of the non-degenerate distribution function converges to the non-degenerate limiting transform, as $N$ tends to infinity, if and only if $F_{\beta_i}(b)$ belongs to the domain of attraction of the completely asymmetric $L\alpha S$ law $F_{\widetilde\beta}(b)$. Then, for some $\alpha$, $0<\alpha<1$, we have
%\begin{eqnarray}
%\lim_{N\to\infty}\left[{\cal L}\left(F_{\beta_i}(b); \frac{t}{N^{1/\alpha}}\right)\right]^N={\cal L}\left(F_{\widetilde\beta}(b),t\right)\nonumber\\
%=\exp[-(At)^\alpha]\,,\label{eqiii24}
%\end{eqnarray}
\begin{equation}
\lim_{N\to\infty}\left[{\cal L}\left(F_{\beta_i}(b); \frac{t}{N^{1/\alpha}}\right)\right]^N={\cal L}\left(F_{\widetilde\beta}(b),t\right)=\exp[-(At)^\alpha]\,,\label{eqiii24}
\end{equation}
where $A$ is a positive constant. Hence, the limiting transform in (\ref{eqiii24}) is the Laplace transform of the  $L\alpha S$ distribution with the non-negative support and the stable parameter $\alpha$ belonging to the range (0,1).

It is not necessary to know the detailed nature of $F_{\beta_i}(b)$ to obtain the above stretched exponential (KWW) limiting form. In fact, this is determined only by tail behavior of $F_{\beta_i}(b)$ for large $b$, see Eq.(\ref{eqii4}), and so a good deal may be said about the asymptotic properties based on rather limited knowledge of the properties of $F_{\beta_i}(b)$. In other words, the necessary and sufficient condition for the relaxation rate $\beta_i$ to have the limiting transform in (\ref{eqiii24}) is the self-similar property in taking the value greater than $b$ and the value greater than $xb$, where $x$ is a positive constant, and $b$ takes a large value. It has been suggested \cite{dh87,ks86} that self-similarity (fractal behavior) is a fundamental feature of relaxation in real materials. This result, obtained here by means of pure probabilistic techniques, independently of the physical details of dipolar systems, is in agreement with models \cite{ks86,weron86,psaa84} identifying this region of fractal behavior.

Let us observe that the right-hand side of formula (\ref{eqiii22}) can be also interpreted as the weighted average of an exponential decay
\begin{displaymath}
\Pr(\theta_i\geq t)=\langle\exp(-\beta_i t)\rangle\,,
\end{displaymath}
where the mean value $\langle\dots\rangle$ is taken with respect to the relaxation-rate probability distribution $F_{\beta_i}(b)$. This leads to
%\begin{eqnarray}
%{\rm Pr}\bigl(A_N\min(\theta_{1},\theta_{2},\ldots,\theta_{N})\geq
%t\bigr)\nonumber\\
%= \exp\left(-t\sum^N_{i=1}\beta_i/A_N\right)\,,\label{eqiii10}
%\end{eqnarray}
\begin{equation}
{\rm Pr}\bigl(A_N\min(\theta_{1},\theta_{2},\ldots,\theta_{N})\geq
t\bigr)= \exp\left(-t\sum^N_{i=1}\beta_i/A_N\right)=e^{-\widetilde\beta_Nt}\,,\label{eqiii10}
\end{equation}
where $(\beta_1,\beta_2,...\beta_N)$ are the non-negative iid random relaxation rates of individual
transitions. If $\beta_i$ has a finite  mean, i.\,e. $\langle\beta_i\rangle<\infty$, then the macroscopic
development gives nothing new because the relaxation evolves exponentially with a constant rate $b_0$, and
$\langle\widetilde\beta_N=\sum^N_{i=1}\beta_i/N\rangle=b_0$ as $N\to\infty$. But the
stochastic picture changes drastically, if the sum
\begin{equation}
\widetilde\beta_N=\sum^N_{i=1}\beta_i/A_N\label{eqiii11}
\end{equation}
consists of rates no having any mean. Summation of iid random variables is well known in literature \cite{feller66,zolot86} and the resulting completely asymmetric $L\alpha S$ distribution $F_{\widetilde\beta}(b)$ of the effective relaxation rate $\tilde\beta$ can be approximated by the weak limit
\begin{equation}
\widetilde\beta\stackrel{d}{=}\lim_{N\to\infty}\widetilde\beta_N\,.\label{eqiii11a}
\end{equation}
In practice, even $N\approx 10^5-10^6$ can suffice to replace adequately $\widetilde\beta_N$ in (\ref{eqiii10}) by the limit $\widetilde\beta$. Taking into account Eqs. (\ref{eqiii9})-(\ref{eqiii11a}), we get
\begin{equation}
\phi(t)=\Pr\bigl(\widetilde\theta\geq t\bigr)
=\langle e^{-\widetilde\beta t}\rangle=\int_0^\infty e^{-bt}\,dF_{\tilde\beta}(b)\,,\label{eqiii4}
\end{equation}
what again yields the KWW stretched exponential decay (\ref{eqiii24}).

Therefore, the relaxation function (\ref{eqiii11b}) with $A_N=N^{1/\alpha}$, for some $0<\alpha<1$, is well defined and equals
\begin{equation}
\phi(t)=\exp[-(At)^\alpha]\,,\label{eqiii25}
\end{equation}
where $A=\tau_p^{-1}$ (see Eq.(\ref{eqi0b})). When $\alpha\to 1$, the theoretically derived KWW function (\ref{eqiii25}) obtains the D form (\ref{eqiii1}). From the mathematical point of view \cite{feller66,zolot86,uchzol99} this corresponds to the case of degenerate distribution function $F_{\widetilde\beta}(b)$, i.\,e. to the case when the effective random relaxation rate $\widetilde\beta$ can take only one value. The corresponding pdf is then of the Dirac $\delta$-function form. At this point we have to stress that the degenerate distributions (of different, studied below physical magnitudes) yield the limiting value 1 of the HN and KWW exponents (see Eqs. (\ref{eqi0a}) and (\ref{eqi0b})). So, to avoid confusion between the theoretical (0,1) and the experimental (0,1] ranges of possible values, taken by the characteristic exponents, we will always in our theoretical studies include the degenerate distributions.

Following the historically oldest approach to non-exponential relaxation \cite{bb78}, the relaxation function can be expressed as $\phi(t)=\langle e^{-t/\widetilde T}\rangle$, since it has been assumed that non-exponential relaxation function takes the form of a weighted average of an exponential decay $e^{-t/\tau}$ with respect to the distribution $w(\tau)\,d\tau$ of the random effective relaxation time $\widetilde T$, see Eq.(\ref{eqiii2}). As the effective relaxation rate $\widetilde\beta = 1/\widetilde T$, the formula (\ref{eqiii2}) can be rewritten as follows
\begin{equation}
\phi(t)=\langle e^{-\widetilde\beta t}\rangle=\int_0^\infty e^{-bt}\,g(b)\,db\,,\label{eqiii5}
\end{equation}
where $\left(\widetilde\beta: b\in [0,\infty)\right)$. This representation assigns any non-exponential relaxation function $\phi(t)$ to the Laplace transform of the effective relaxation-rate distribution $g(b)$. The probability density functions $w(\tau)$ (see Eq.(\ref{eqiii2})) and $g(b)$ (see Eq.(\ref{eqiii5})) are related to each other, namely $g(b)=b^{-2}w(b^{-1})$. The relationship between $g(b)$ and $w(\tau)$, corresponding to the KWW relaxation, allows us to show \cite{weron86} that in contrast to the momentless distribution $dF_{\widetilde\beta}(b)=g(b)\,db$ of the effective relaxation rate $\widetilde\beta$, the distribution $dF_{\widetilde T}(\tau)=w(\tau)\,d\tau$ possesses finite average and higher moments of effective relaxation time $\widetilde T$. Notice, the relaxation rates are additive, but the relaxation times are not. Therefore, the relaxation rates as random variables are more convenient for the probabilistic formalism based on the limit theorems of probability theory. Hence, in further study only formula (\ref{eqiii5}) will be utilized.

Let us observe that independently on a statistical distribution of relaxation rates $\beta_i$ we find in expression (\ref{eqiii21}) a hidden assumption. Namely, each relaxing dipole after a sufficiently long time (after removing the electric field) changes its initial position with probability 1, i.\,e.
\begin{equation}
\Pr(\theta_i\geq t|\beta_i=b)=\exp(-bt)=\begin{cases} 1 &
\mbox{for}\ \ t=0\\ 0 & \mbox{for}\ \ t\to\infty\,.\end{cases}\label{eqiii26}
\end{equation}
 Such an assumption is the main reason why the relaxation function (\ref{eqiii11b}) cannot have any other form than the KWW one (\ref{eqiii25}). The above analysis gives also an insight into the physical origins of the short-time power law observed in all non-exponential relaxation responses. For the simplest non-exponential case (\ref{eqiii25}), the response function reads
\begin{displaymath}
f(t)=\alpha A(At)^{\alpha-1}e^{-(At)^\alpha}\propto\begin{cases} (At)^{-n} &
\mbox{for}\ \ t\ll 1/A\\ e^{-(At)^\alpha} & \mbox{for}\ \ t\gg 1/A\,,\end{cases}
\end{displaymath}
where $n=1-\alpha$ results from the $L\alpha S$ distribution of the effective relaxation rate $\widetilde\beta$. The power-law exponent $n$ is determined by the long-tailed properties (see Eq.(\ref{eqii4})) of this distribution $1-F_{\widetilde\beta}(b)\propto b^{-\alpha}$ for $b\to\infty$.

In order to obtain a class of dielectric responses exhibiting both the short- and the long-time power law, one should modify either the assumption (\ref{eqiii21}) to define the random waiting time which can be infinite with some non-zero probability or modify the definition of the relaxation function (\ref{eqiii11b}) to account for the random number of individual relaxation contributions (relaxation channels). As we will see below, such a modification, being in agreement with physical intuition on relaxation mechanisms, leads us directly to the non-exponential responses (\ref{eqii10}). In proposed schemes of relaxation the KWW and D functions are included as special cases.

\subsection{Conditionally exponential decay model}

Let us assume independent exponential relaxations constrained by the maximal time of a structural reorganization in all surrounding clusters (each consisting of a dipole and its non-polar environment). In a system composed of $N$ of relaxing dipoles, the probability \cite{weron91,weron92}
that the {\it i}th dipole has not changed its initial position up to the moment $t$ equals $\exp[-b\min(t,s)]$, if its relaxation rate has taken the value $b$ and the maximal time of the structural reorganization $\eta_{i,\rm max}= a_N^{-1}\max(\eta_1,\dots,\eta_{i-1},\eta_{i+1},\dots,\eta_N)$ in all surrounding clusters (under the suitable normalization) has been equal to $s$, i.\,e.
\begin{equation}
\Pr\Bigl(\theta_{i}\geq t|\beta_i = b, \eta_{i,\rm max}=s\Bigr)=\exp[-b\min(t,s)]\label{eqiii27}
\end{equation}
for $b>0, s>0, t\geq0$. The random variable $\beta_i$ denotes the relaxation rate of the {\it i}th dipole and the variable $\eta_i$, the time needed for the structural reorganization of {\it i}th cluster.
The variable $\theta_{iN}$ denotes the time needed for changing the orientation by the {\it i}th dipole in the system consisting of $N$ relaxing dipoles. $\beta_1,\beta_2,\dots$ and $\eta_1,\eta_2,\dots$ form independent sequences of non-negative, iid random variables. The variables $\theta_{1},\dots,\theta_{N}$ are also non-negative, iid for each $N$. It follows from (\ref{eqiii27}) that the random variable $\theta_{i}$ depends on the random variable $\beta_i$ and on the sequence of random variables $\eta_1,\dots,\eta_{i-1},\eta_{i+1},\dots,\eta_N$.

In contrast to (\ref{eqiii26}) we have
%\begin{eqnarray*}
%&&\Pr\Bigl(\theta_{i}\geq t | \beta_i=b, \eta_{i,\rm max}=s\Bigr)\\
%&&=\begin{cases} 1\,\ \ \
%\mbox{for}\ \ t=0\\ \exp(-bt)\ \ \ \mbox{for}\ \ t<s\\ \exp(-bs)={\rm const}>0\ \ \mbox{for}\ \ t\to\infty\,.\end{cases}
%\end{eqnarray*}
\begin{displaymath}
\Pr\Bigl(\theta_{i}\geq t | \beta_i=b, \eta_{i,\rm max}=s\Bigr)=\begin{cases} 1\,\ \ \
\mbox{for}\ \ t=0\\ \exp(-bt)\ \ \ \mbox{for}\ \ t<s\\ \exp(-bs)={\rm const}>0\ \ \mbox{for}\ \ t\to\infty\,.\end{cases}
\end{displaymath}
It means that dipoles altered by the external electric field do not have to change their initial positions with probability 1 after removing the field as $t$ tends to infinity (with some probability their initial states are ``frozen''). In this case, because of the improper form of the distribution (\ref{eqiii27}), the relaxation function (\ref{eqiii11b}) cannot be expressed as the waiting average with respect to the effective relaxation rate distribution (see Eq.(\ref{eqiii5})). Instead, a general relaxation equation, fulfilled by function (\ref{eqiii11b}), can be derived \cite{weron92}.

Since sequences $\beta_1,\beta_2,\dots$ and $\eta_1,\eta_2,\dots$ are independent, we have from the law of total probability
\begin{displaymath}
\Pr(\theta_{i}\geq t|\beta_i=b)=\int_0^\infty\exp[-b\min(t,s)]\,dF_{\eta,N}(s)\,,
\end{displaymath}
where $F_{\eta,N}(s)$ denotes the distribution function of the random variable which has the form $a_N^{-1}\max(\eta_1,\dots,\eta_{i-1},\eta_{i+1},\dots,\eta_N)$, i.\,e. the probability that this random variable has taken a value less than or equal to $s$. Since $\eta_j$ are iid random variables, we have $F_{\eta,N}(s)=[F_\eta(a_Ns)]^{N-1}$, where $F_\eta(s)$ denotes the distribution function of each $\eta_j$. Assuming $F_\eta(s)$ differentiable, we have $F_{\eta,N}(s)$ differentiable, too, and
%\begin{eqnarray*}
%&&\frac{d}{dt}\Pr\left(\theta_{i}\geq \frac{t}{A_N}\ |\ \beta_i=b\right)\\
%&&=\left[1-F_{\eta,N}
%\left(\frac{t}{A_N}\right)\right]\frac{d}{dt}\left(-b\frac{t}{A_N}\right)\,.
%\end{eqnarray*}
\begin{displaymath}
\frac{d}{dt}\Pr\left(\theta_{i}\geq \frac{t}{A_N}\ |\ \beta_i=b\right)
=\left[1-F_{\eta,N}
\left(\frac{t}{A_N}\right)\right]\frac{d}{dt}\left(-b\frac{t}{A_N}\right)\,.
\end{displaymath}
From the law of total probability once again, and from the Lebesgue theorem \cite{feller66}, we have
%\begin{eqnarray}
%&&\frac{d}{dt}\Pr\left(\theta_{i}\geq \frac{t}{A_N}\right)\nonumber\\
%&=&\left[1-F_{\eta,N}
%\left(\frac{t}{A_N}\right)\right]\frac{d}{dt}{\cal L}\left(F_{\beta_i}(b); \frac{t}{A_N}\right)\,,\label{eqiii29}
%\end{eqnarray}
\begin{equation}
\frac{d}{dt}\Pr\left(\theta_{i}\geq \frac{t}{A_N}\right)=\left[1-F_{\eta,N}
\left(\frac{t}{A_N}\right)\right]\frac{d}{dt}{\cal L}\left(F_{\beta_i}(b); \frac{t}{A_N}\right)\,,\label{eqiii29}
\end{equation}
where ${\cal L}(F_{\beta_i}(b); t)$ is the Laplace transform of the distribution function $F_{\beta_i}(b)$ of each $\beta_i$ at the point $t$.

Because $\theta_{iN}$ are iid random variables for each $N$, we have
\begin{equation}
\phi(t)=\lim_{N\to\infty}\left[\Pr\left(\theta_{i}\geq\frac{t}{A_N}\right)\right]^N\,.\label{eqiii30}
\end{equation}
Using the mathematical trick
\begin{displaymath}
\frac{d}{dt}\left[{\cal L}\left(F_{\beta_i}(b); \frac{t}{A_N}\right)\right]^N=N\left[{\cal L}\left(F_{\beta_i}(b); \frac{t}{A_N}\right)\right]^{N-1}\frac{d}{dt}{\cal L}\left(F_{\beta_i}(b); \frac{t}{A_N}\right)\,,
\end{displaymath}
it follows from (\ref{eqiii29}) that
%\begin{eqnarray}
%&&\frac{d}{dt}\Pr\left[\left(\theta_{i}\geq \frac{t}{A_N}\right)\right]^N\nonumber\\
%&=&\Pr\left[\left(\theta_{i}\geq \frac{t}{A_N}\right)\right]^{N-1}\left[1-F_{\eta,N}\left(\frac{t}{A_N}\right)\right]\nonumber\\
%&\times&\left[{\cal L}\left(F_\beta; \frac{t}{A_N}\right)\right]^{-N+1}\frac{d}{dt}\left[{\cal L}\left(F_\beta; \frac{t}{A_N}\right)\right]^N\,.\label{eqiii31}
%\end{eqnarray}
\begin{eqnarray}
\frac{d}{dt}\left[\Pr\left(\theta_{i}\geq \frac{t}{A_N}\right)\right]^N
&=&\left[\Pr\left(\theta_{i}\geq \frac{t}{A_N}\right)\right]^{N-1}\left[1-F_{\eta,N}\left(\frac{t}{A_N}\right)\right]\nonumber\\
&\times&\left[{\cal L}\left(F_\beta; \frac{t}{A_N}\right)\right]^{-N+1}\frac{d}{dt}\left[{\cal L}\left(F_\beta; \frac{t}{A_N}\right)\right]^N\,.\label{eqiii31}
\end{eqnarray}
As we know from the preceding section, the $N$th power of the Laplace transform of a non-degenerate distribution function $F_{\beta_i}(b)$ converges to the non-degenerate limiting transform, as $N$ tends to infinity, if and only if $F_{\beta_i}(b)$ belongs to the domain of attraction of the $L\alpha S$ law, and, for some $0<\alpha<1$, we have
\begin{equation}
\lim_{N\to\infty}\left[{\cal L}\left(F_{\beta_i}(b); \frac{t}{N^{1/\alpha}}\right)\right]^N=\exp(-(At)^\alpha)\,,\label{eqiii32}
\end{equation}
where $A$ is a positive constant. At the same time, the value
\begin{displaymath}
F_{\eta_i}\left(\frac{t}{N^{1/\alpha}}\right)=\Pr\bigl(N^{1/\alpha}\eta_{i,\rm max}\leq t\bigr)
\end{displaymath}
tends to a non-degenerate distribution function of non-negative random variable, as $N$ tends to infinity, if and only if $F_{\eta_i}(s)$, the distribution function of each $\eta_i$, belongs to the domain of attraction of the max-stable law of type II \cite{llr86}. Then, for the normalizing constant $a_N$ proportional to $N^{1/\alpha}\inf\{t : F_\eta(t)\geq1-(1/N - 1)\}$ we have
\begin{equation}
\lim_{N\to\infty}F_{\eta_i,N}\left(\frac{t}{N^{1/\alpha}}\right)=\exp\left(-\frac{(At)^{-\alpha}}{\kappa}\right)
\label{eqiii33}
\end{equation}
for some positive constants $\alpha$ and $\kappa$, and $A$ taken from Eq.(\ref{eqiii32}). To obtain the limiting forms (\ref{eqiii32}) and (\ref{eqiii33}) we need not know the detailed nature of $F_{\beta_i}(b)$ and $F_{\eta_i}(s)$. In fact, this is determined only by the behavior of the tail of $F_{\beta_i}(b)$ for large $b$ and of the tail of $F_{\eta_i}(s)$ for large $s$, i.\,e. the necessary and sufficient conditions for the relaxation rate $\beta_i$ and for the structural reorganization time $\eta_i$ to have the limits in Eqs. (\ref{eqiii32}) and (\ref{eqiii33}) are the self-similar properties, firstly of $\beta_i$, in taking the value greater than $b$ and the value greater than $xb$, and secondly of $\eta_i$ in taking the value greater than $s$ and the value greater than $xs$.

The relaxation function in Eq.(\ref{eqiii11b}) with $A_N=N^{1/\alpha}$ is well defined and, by Eqs. (\ref{eqiii30})--(\ref{eqiii33}), fulfills the general relaxation equation (a kinetic equation with a time-dependent transition rate $r(t)$, see Eq.(\ref{eqii7}))
\begin{equation}
\frac{d\phi(t)}{dt}=-\alpha A(At)^{\alpha-1}\left[1-\exp\left(-\frac{(At)^{-\alpha}}{\kappa}\right)\right]\phi(t)\,.
\label{eqiii34}
\end{equation}
Recall that the parameter $A$ has the sense of $\tau_p^{-1}$.
The coefficient $\kappa$ is a consequence of normalization in the limiting procedure in Eq.(\ref{eqiii33}). It decides how fast the structural reorganization of clusters is spread out in a system; $\kappa\to 0$ means the case in which cluster structure is neglected. If $\kappa\to 0$, Eq.(\ref{eqiii34}) takes the well-known form \cite{dh87,ks86,fjst88}%,p86,nrt88,p91}
\begin{equation}
\frac{d\phi(t)}{dt}=-\alpha A(At)^{\alpha-1}\phi(t)\label{eqiii35}
\end{equation}
with the solution (\ref{eqiii25}). In the general case we get the solution in an integral form
\begin{displaymath}
\phi(t)=\exp[-cS(t)]\,,
\end{displaymath}
where $c=\kappa^{-1}$ and
\begin{displaymath}
S(t)=\int_0^{\kappa(At)^\alpha}[1-\exp(-s^{-1})]\,ds\,.
\end{displaymath}
A similar form has been obtained as a result of the studies of different approaches (the F\"{o}rster direct-transfer model, the hierarchically constrained dynamics model, and the defect-diffusion model) analyzing non-exponential relaxations, with emphasis on the stretched exponential KWW form \cite{ks86,p86,p91}. Although each model describes a different mechanism, they have the same underlying reason for the stretched exponential pattern: the existence of scale invariant relaxation rates. Presenting one more approach, we have obtained the KWW relaxation function (\ref{eqiii25}) as a special case of Eq.(\ref{eqiii34}) when $\kappa\to 0$. We have also shown that the underlying reason for this is the existence of a type of self-similarity in the behavior of relaxation rates.

For practical  purposes, according to \cite{wk97}, the  solution of Eq.(\ref{eqiii34}) can be presented in the following form
%\begin{eqnarray*}
%\ln\phi(t)=&-&(At)^\alpha\Biggl[1-\exp\Biggl(-\frac{1}{\kappa(At)^\alpha}\Biggr)\Biggr]\\
%&-&\frac{1}{\kappa}\,
%\Gamma\Biggl(0,\frac{1}{\kappa(At)^\alpha}\Biggr)\,,
%\end{eqnarray*}
\begin{displaymath}
\ln\phi(t)=-(At)^\alpha\Biggl[1-\exp\Biggl(-\frac{1}{\kappa(At)^\alpha}\Biggr)\Biggr]
-\frac{1}{\kappa}\,\Gamma\Biggl(0,\frac{1}{\kappa(At)^\alpha}\Biggr)\,,
\end{displaymath}
where $\Gamma(a,z)$ is the incomplete gamma function defined as
\begin{displaymath}
\Gamma(a,z)=\int_z^\infty x^{a-1}\,e^{-x}\,dx\,.
\end{displaymath}
It follows from Eq.(\ref{eqiii34}) that the relaxation response may be written as
\begin{equation}
f(t)=\alpha A(At)^{\alpha-1}\left[1-\exp\left(-\frac{(At)^{-\alpha}}{\kappa}\right)\right]\phi(t)\,.
\label{eqiii36}
\end{equation}
Then, for the short-time regime its asymptotic behavior is
\begin{equation}
\lim_{t\to 0}\frac{f(t)}{(At)^{\alpha-1}}=\alpha A\lim_{t\to 0}\phi(t)=\alpha A\,,
\label{eqiii37}
\end{equation}
since $\phi(0)=1$. On the other hand, the long-time trend follows
\begin{equation}
\lim_{t\to\infty}\frac{f(t)}{(At)^{-\alpha/\kappa-1}}=A\kappa^{-1-1/\kappa}e^{-(1-\gamma_E)/\kappa}\,,\label{eqiii38}
\end{equation}
where $\gamma_E\approx0.577216$ is the Euler constant \cite{wk97}.
Thus, the response function $f(t)$ can exhibit the power-law properties in both short- and long-time limits, namely
\begin{equation}
f(t)\propto
\begin{cases} (At)^{-n} &
\mbox{as}\ \ At\ll 1\\ (At)^{-m-1} & \mbox{as}\ \ At\gg1\,,\end{cases}\label{eqiii39}
\end{equation}
where $n=1-\alpha$ and $m=\alpha/\kappa$.
The relaxation function $\phi(t)$ is determined by three parameters: $0<\alpha<1$, $A>0$ and $\kappa>0$. The parameter $\kappa$ distinguishes the fractional two-power-law behavior from the one-power-law KWW response, i.\,e. if $\kappa$ is small, the general relaxation solution of Eq.(\ref{eqiii34}) takes the form which is just the KWW relaxation function. Moreover, if $\alpha\to1$ we obtain the rarely observed D case. For $\kappa<1$ the relaxation function of Eq.(\ref{eqiii34}) describes the typical case of Figure~\ref{diag-1}, and for $\kappa>1$ we obtain the less typical relaxation behavior. Note also that the formalism of coupled cluster interactions finds a good support in experimental studies \cite{bfj01,jswbkw13,rv14}.

\subsection{Relaxation of hierarchically clustered systems}

The cluster model concept \cite{dh83,d84} presents a radical departure from the traditional interpretation of relaxation based on independent exponentially relaxing entities. The realistic idea originates from imperfectly ordered states of complex systems and their evolution. In this case the systems, which exhibit position or orientation relaxation, are composed of spatially limited regions (clusters). Because the structural order within any cluster is incomplete, there are internal and external dynamics of clusters. When an external field acts on such a system, entities of this system take positions along the field direction, but the positions will be very dependent on the local structure of the system, i.\,e. from  defects of different types. With regard to the imperfect structure the arrangement of entities after removing the external field starts to lose spatial uniformity. During this process of relaxation to an equilibrium geometry, the strongly coupled local motions are expected to arise firstly, thereby breaking down the arrangements into clusters that leads to weakly coupled inter-cluster motions forming a constraint hierarchy of interacting clusters and their long-range compositions. Each of these processes have their own characteristic contribution to the macroscopic evolution of the system as a whole. This is the reasonably natural way to modify a traditional approach of independent relaxing entities with random relaxation rates into a multilevel summation of a hierarchy of cluster relaxations with their random relaxation rates.

\subsubsection{Havriliak-Negami function}

Before going into details of the random-cluster relaxation model \cite{wjj01,jw02,jjw03} let us first discuss the stochastic representation (\ref{eqiii11bd}) of the HN function. Take into account the random effective waiting time $\widetilde\theta$ for transition of the relaxing system \cite{jw00b} as a mixture of random variables
\begin{equation}
\widetilde\theta_{\rm HN}=\tau_p\,S_\alpha\,(\varGamma_\gamma)^{1/\alpha}\,,\qquad 0<\alpha,\gamma<1\,.\label{eqiii12}
\end{equation}
Here $S_\alpha$ is such a positive random variable that its Laplace transform is the stretched exponential function
\begin{equation}
\langle e^{-sS_\alpha}\rangle=\int_0^\infty e^{-st}\,h_\alpha(t)\,dt=e^{-s^\alpha}\label{eqiii13}
\end{equation}
with $0<\alpha<1$. It is a well-known fact \cite{feller66} that in the above relation the random variable $S_\alpha$ has to be distributed according to the completely asymmetric $L\alpha S$ law with the pdf $h_\alpha(t)$ (for details see \cite{zolot86,uchzol99}). The pdf of the random variable $S_\alpha$ tends to the degenerate form $h_1(t)=\delta(t-1)$ (given by the Dirac $delta$-function) as $\alpha\to 1$.

The positive random variable $\varGamma_\gamma$ in Eq.(\ref{eqiii12}) is independent of $S_\alpha$ and distributed according to the gamma law $G_\gamma(t)$ defined by the pdf of the form
\begin{displaymath}
g_\gamma(t)=\frac{1}{\Gamma(\gamma)}\,t^{\gamma-1}\,e^{-t}\,,\qquad t>0\,,
\end{displaymath}
with $\Gamma(\gamma)$ being the Euler's gamma function \cite{john70}. It is worth noting that the Laplace transform of $\varGamma_\gamma$ reads
\begin{equation}
\langle e^{-s\varGamma_\gamma}\rangle=\int_0^\infty e^{-st}\,g_\gamma(t)\,dt=\frac{1}{(1+s)^\gamma}\,.\label{eqiii14}
\end{equation}
Using the properties (\ref{eqiii13}) and (\ref{eqiii14}), we derive the explicit form of Eq.(\ref{eqiii11bd})
\begin{eqnarray}
\phi^*(\omega)&=&\langle e^{-i\omega\widetilde\theta_{\rm HN}}\rangle=\langle e^{-i\omega \tau_p\,S_\alpha\,(\varGamma_\gamma)^{1/\alpha}}\rangle\nonumber\\
&=&\int_0^\infty e^{-(i\omega/\omega_p)^\alpha t}\,g_\gamma(t)\,dt=
\frac{1}{(1+(i\omega/\omega_p)^\alpha)^\gamma}\,.\nonumber
\end{eqnarray}
Therefore, the waiting time $\widetilde\theta_{\rm HN}$ with $0<\alpha,\gamma<1$ represents the HN relaxation pattern and, moreover, in the time domain we have
\begin{equation}
\phi_{\rm HN}(t)=\int_0^\infty \left\{1-G_\gamma\left[\Bigl(\frac{t}{s\tau_p}\Bigr)^\alpha\right]\right\}\,h_\alpha(s)\,ds\,,\label{eqiii15}
\end{equation}
where $G_\gamma(x)=\int_0^x g_\gamma(t)\,dt=1-\Gamma(\gamma,x)/\Gamma(\gamma)$, and $\Gamma(\gamma,x)$ is the upper incomplete gamma function.

On the other hand, the distribution of the random variable $\widetilde\theta_{\rm HN}$  can be identified as the Mittag-Leffler distribution \cite{pillai90} and, hence, the time-domain HN relaxation function is represented by the following series
\begin{eqnarray}
\phi_{\rm HN}(t)&=&1-\sum_{j=0}^\infty\frac{(-1)^j\,\Gamma(\gamma+j)}{j!\Gamma(\gamma)\Gamma(1+\alpha(\gamma+j))}
\left(\frac{t}{\tau_p}\right)^{\alpha(\gamma+j)}\nonumber\\
&=&1-(t/\tau_p)^{\alpha\gamma}E^\gamma_{\alpha,\alpha\gamma+1}\left[-(t/\tau_p)^\alpha\right]\,,\label{eqiii16}
\end{eqnarray}
where $E^\gamma_{\alpha,\beta}(x)$ is the three-parameter Mittag-Leffler function \cite{prab71}.

Formula (\ref{eqiii12}) and hence the time-domain relaxation function (\ref{eqiii15}) take on simpler forms in case of the CD, CC and D responses. For the CD function ($\alpha=1$, $0<\gamma<1$) one gets the gamma waiting-time distribution and
\begin{displaymath}
\phi_{\rm CD}(t)=1-G_\gamma\left(t/\tau_p\right)\,,\quad t>0\,.
\end{displaymath}
The respective response function equals $f_{\rm CD}(t)=g_\gamma(t/\tau_p)/\tau_p$. Similarly, in case of the D function ($\alpha=\gamma=1$) the corresponding waiting time is distributed according to the exponential law and
\begin{displaymath}
\phi_{\rm D}(t)=e^{-t/\tau_p}\,,\qquad t>0\,.
\end{displaymath}

For the CC response ($0<\alpha<1$, $\gamma=1$)  the gamma random variable becomes an exponential one $\varGamma_\gamma=\varGamma_1$ and the series representation (\ref{eqiii16}) simplifies to the one-parameter Mittag-Leffler function
\begin{equation}
\phi_{\rm CC}(t)=\sum_{j=0}^\infty\frac{(-1)^j}{\Gamma(1+\alpha j)}\left(\frac{t}{\tau_p}\right)^{\alpha j}=E_\alpha\left[-(t/\tau_p)^\alpha\right]\,.\label{eqiii17a}
\end{equation}
It is worth noting that all the waiting-time distributions, underlying the considered empirical relaxation responses, are infinitely divisible \cite{feller66,jw00b}.

\subsubsection{Infinite mean cluster sizes}

Let us now study the complex dynamics of clustered systems from the probabilistic point of view \cite{wjj01,jw02,jjw03}. In every complex system capable of responding to an external field, the total number $N$ of entities in the system may be divided into two parts. One of them includes so-called active entities being able to follow changes of the field. Another part consists of inactive neighbors. Even if some entities do not contribute directly to the relaxation dynamics, they may affect the stochastic transition of the active ones. Note, this influence lies in the properties of individual relaxation rates $\beta_{1N},\beta_{2N},\dots$ of the active entities in the system. According to the concept of relaxation rates, the individual rates take the form $\beta_{iN}=\beta_i/A_N$, where $\beta_i$ is independent of $N$, and $A_N$ is the same normalizing constant for each entity. Assume further that the $i$th active entity interacts with $N_i-1$ inactive neighbors forming a cluster of size $N_i$. The unknown number $K_N$ of active entities in the system, random in general, equals to the number of clusters due to the local interactions. The value $K_N$ is determined by the first index $k$ for which the sum $N_1+\dots+N_k$ of the cluster sizes exceeds $N$, the total number of entities. Mathematically this proposition can be written as
\begin{equation}
K_N=\min\left\{k\ :\ \sum_{i=1}^kN_i>N\right\}\,,\label{eqvi1}
\end{equation}
where $\{k:X\}$ implies the value of $k$ such that $X$ holds. Interactions among active entities have a local character because of their surrounding by inactive entities (due to screening effects, for example, \cite{jonscher83}). Therefore,  every evolving active entity may ``feel'' only some of other active neighbors. In these conditions nothing prevents the emergence of cooperative regions (super-clusters) built up from the active entities and their surroundings. Let the random number $L_N$ of such super-clusters be determined by their sizes $M_1,M_2,\dots$. In the same way as (\ref{eqvi1}) we define
\begin{equation}
L_N=\min\left\{l\ :\ \sum_{j=1}^lM_j>K_N\right\}\,,\label{eqvi2}
\end{equation}
where $M_j$ is a number of interacting active entities in the {\it j}\,th super-cluster. A contribution of each super-cluster to the total relaxation rate is the sum of the contributions of all active entities in the super-cluster. Hence, for the {\it j}\,th super-cluster, its relaxation rate, say $\overline{\beta}_{jN}$, is equal to
\begin{equation}
\overline{\beta}_{jN}=\sum_{i=M_1+\dots+M_{j-1}+1}^{M_1+\dots+M_j}\beta_{iN}/A_n\,.\label{eqvi3}
\end{equation}
For $j=1$ it is simply the sum
\begin{equation}
\overline{\beta}_{1N}=\sum_{i=1}^{M_1}\beta_{iN}/A_n\,,\label{eqvi4}
\end{equation}
for $j=2$ it becomes
\begin{equation}
\overline{\beta}_{2N}=\sum_{i=M_1+1}^{M_1+M_2}\beta_{iN}/A_n\label{eqvi5}
\end{equation}
and so on. The effective representation of the system as a whole
is provided by the total relaxation rate $\widetilde\beta_N$ which is the sum of the contributions over all super-clusters
\begin{equation}
\widetilde\beta_N=\sum_{j=1}^{L_N}\overline{\beta}_{jN}\,.\label{eqvi7}
\end{equation}
In fact, considering relaxation phenomena, one usually deals with systems consisting of a large number of relaxing entities so that the weak limit
\begin{equation}
\widetilde\beta\stackrel{d}{=}\lim_{N\to\infty}\widetilde\beta_N\label{eqvi8}
\end{equation}
can describe the entire system, and, hence the relaxation function reads
\begin{displaymath}
\phi(t)=\left\langle\exp(-\widetilde\beta t)\right\rangle\,.
\end{displaymath}
In general, all the quantities $N_i$, $M_j$, $\beta_{iN}$, together with those defined by them, must be considered as random variables. The point is that the number of relaxing entities directly engaged in the relaxation process, their locations, as well as their ``birth'' and ``death'', are random. Obviously, their stochastic features would determine the properties of the total relaxation rate $\widetilde\beta_N$ if they were known. But they are rather not known. Nevertheless, on the basis of the limit theorems of probability theory, it is possible to define the distribution of the limit $\widetilde\beta$ representing a macroscopic relaxing system, even with rather limited knowledge about the distributions of micro/mesoscopic random quantities used in the model.

Assume that $M =\{M_j,\ j = 1, 2, \dots\}$, $N = \{N_i,\ i = 1, 2,\dots\}$, and $\beta = \{\beta_i,\ i = 1, 2, \dots\}$ are independent sequences, each of which consists of iid positive random variables, $M_j$ and $N_i$ being integer-valued. For the sake of simplicity, let us introduce the following notations
\begin{eqnarray*}
{\cal S}_X(0)=0\,,\quad {\cal S}_X(k)=\sum_{i=1}^kX_i\,,\\
\nu(n)=\min\left\{k\ :\ {\cal S}_X(k)>n\right\}
\end{eqnarray*}
for $k = 1, 2, \dots$, and $X$ corresponding to one of sequences $M$, $N$, or $\beta$. Denoting the sum of the contributions over all super-clusters (\ref{eqvi7}) for convenience as $\widetilde\beta_n$, the latter takes the form of a normalized random sum
\begin{displaymath}
\widetilde\beta_n={\cal S}_\beta(\Sigma_n)/a_n\,,
\end{displaymath}
with the random index $\Sigma_n={\cal S}_M(\nu_M(\nu_N(n)))$ independent of the components $\beta_j$, and $a_n$ as a sequence of the normalizing constants. If the distribution of a positive random variable $X_j$ has a heavy tail as defined in (\ref{eqii4}), then asymptotic properties of $\widetilde\beta_n$ (as $n\to\infty$) can be found exactly.

Assume that both $N_j$ and $\beta_j$ have heavy-tailed distributions with the same exponent $\alpha\in(0,1]$. Following \cite{feller66}, we have
\begin{eqnarray}
\frac{{\cal S}_N(n)}{n^{1/\alpha}}\stackrel{d}{\rightarrow}\left(c_1\Gamma(1-\alpha)\right)^{1/\alpha}S_\alpha\,,\nonumber\\
\frac{{\cal S}_\beta(n)}{n^{1/\alpha}}\stackrel{d}{\rightarrow}\left(c_2\Gamma(1-\alpha)\right)^{1/\alpha}Z_\alpha\,,\label{eqvi8a}
\end{eqnarray}
where random variables $S_\alpha$ and $Z_\alpha$ are distributed with completely asymmetric $L\alpha S$ distributions, $c_1$ and $c_2$ being positive constants. Here $\stackrel{d}{\rightarrow}$ means ``tends in distribution''. Since $\nu_N(n)$ and ${\cal S}_N(k)$ are connected by the relation
\begin{displaymath}
\{\nu_N(n)>k\}=\{{\cal S}_N(k)\leq n\}
\end{displaymath}
for any $n,k=1,2,\dots$, then
\begin{equation}
\frac{\nu_N(n)}{n^\alpha}\stackrel{d}{\rightarrow}\frac{1}{c_1\Gamma(1-\alpha)}\left(\frac{1}{S_\alpha}\right)^\alpha\,.\label{eqvi8b}
\end{equation}
If the distribution of $M_j$ has also a heavy tail, according to Eq.(\ref{eqii4}), with $\gamma\in(0,1]$, as has been shown in \cite{feller66}, the normalized random sum
\begin{equation}
\frac{{\cal S}_M(\nu_M(n))}{n}\stackrel{d}{\rightarrow}\frac{1}{{\cal B}_\gamma}\label{eqvi8c}
\end{equation}
for $n\to\infty$ behaves as a reciprocal to the beta-distributed random variable ${\cal B}_\gamma$ governed by the following pdf
\begin{displaymath}
f_\gamma(x)=\begin{cases}\frac{1}{\Gamma(\gamma)\Gamma(1-\gamma)}\,x^{\gamma-1}\,(1-x)^{-\gamma} & {\rm for}\ 0<x<1\,,\\
\qquad\qquad\qquad 0 & {\rm otherwise}\,,\end{cases}
\end{displaymath}
where $\Gamma(\cdot)$ is the gamma function. It should be pointed out that the distribution is a particular case of the beta-distribution \cite{john70}. As the random sequences $\{{\cal S}_M(\nu_M(n))\}$ and $\{\nu_N(n))\}$ are independent, it follows from \cite{dobr55} that the results (\ref{eqvi8b}) and (\ref{eqvi8c}) allow us to get
\begin{equation}
\frac{\Sigma_n}{n^\alpha}\stackrel{d}{\rightarrow}\frac{1}{c_1\Gamma(1-\alpha)}\left(\frac{1}{S_\alpha}\right)^\alpha\frac{1}{{\cal B}_\gamma}\,.\label{eqvi8d}
\end{equation}
Using the relations (\ref{eqvi8a}) and (\ref{eqvi8d}), we find the tendency of the normalized random sum $\widetilde\beta_n$ for $n\to\infty$, namely
\begin{displaymath}
\frac{{\cal S}_\beta(\Sigma_n)}{n}\stackrel{d}{\rightarrow}\frac{c_2^{1/\alpha}}{c_1^{1/\alpha}}\frac{Z_\alpha}{S_\alpha}\left(\frac{1}{{\cal B}_\gamma}\right)^{1/\alpha}\,.
\end{displaymath}
Derivation of $\lim_{n\to\infty}\widetilde\beta_n$ has been presented in \cite{jurl03}.

\begin{figure}
\centerline{\includegraphics[clip,width=1.1\columnwidth]{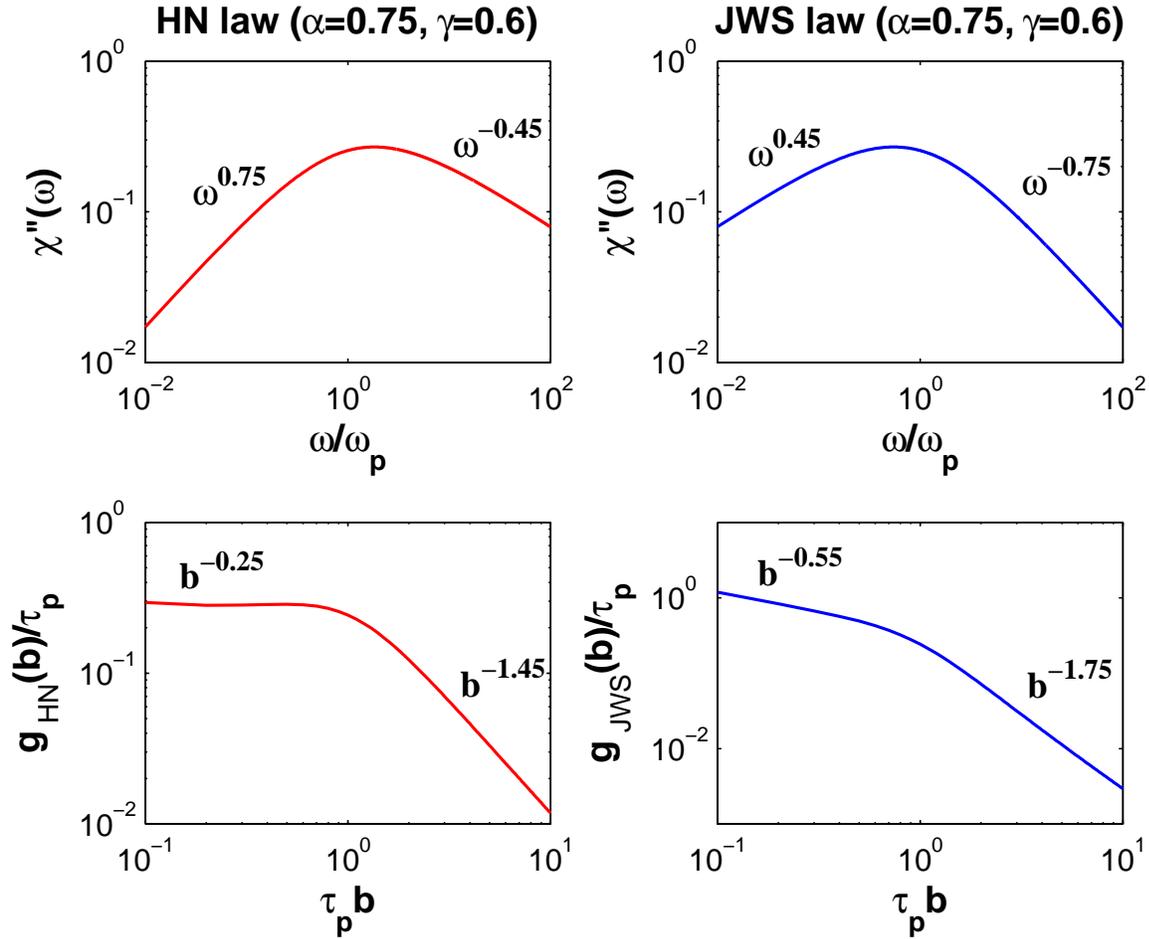}}
\caption{(Colour on-line). Log-log plots of the frequency-domain relaxations functions (corresponding to the HW law in the top left panel, and the JWS relationship in top right panel) and the effective relaxation rate density functions: $g_{\rm HN}(b)$ (bottom left panel) and $g_{\rm JWS}(b)$ (bottom right panel).}
\label{HN-JWS}
\end{figure}

Recall, the relaxation response can be associated with $\widetilde\theta$, the system's waiting time for its transition from the initially imposed state, and $\widetilde\beta$, the effective relaxation rate. The random variables $\widetilde\theta$ and $\widetilde\beta$ are strictly connected with each other, as $\phi(t)=\Pr(\widetilde\theta\geq t) = {\cal L}(\widetilde\beta; t)$, where ${\cal L}(X;t)=\left\langle\exp(-Xt)\right\rangle$ denotes the Laplace transform of the pdf of a random variable $X$. As it has been shown above (see Eqs.(\ref{eqiii12})-(\ref{eqiii16})), the relaxation function corresponding to the HN law is $\phi_{\rm HN}(t)=\Pr(\tau_p\,S_\alpha(\varGamma_\gamma)^{1/\alpha}\geq t)$. By direct calculations \cite{jurl03} one can easily find
\begin{equation}
{\cal L}(1/{\cal B}_\gamma; t)=\int_0^1 e^{-t/x}\,f_\gamma(x)\,dx=\Pr(\varGamma_\gamma\geq t)\,.\label{eqvi8e}
\end{equation}
Using formula (\ref{eqiii13}) for the Laplace transform of completely asymmetric $L\alpha S$ random variables (\ref{eqvi8a}), i.\,e. ${\cal L}(S_\alpha; t)={\cal L}(Z_\alpha; t)=e^{-t^\alpha}$, together with (\ref{eqvi8e}), we come to
\begin{displaymath}
{\cal L}\Bigg(\tau_p\frac{Z_\alpha}{S_\alpha}\left(\frac{1}{{\cal B}_\gamma}\right)^{1/\alpha}; t\Bigg)=\Pr\left(\tau_p\,S_\alpha(\varGamma_\gamma)^{1/\alpha}\geq t\right)\,.
\end{displaymath}
Thus, the effective relaxation rate of the clustered system considered above is
\begin{equation}
\widetilde\beta_{\rm HN}\stackrel{d}{=}\frac{1}{\tau_p}\left({\cal B}_\gamma\right)^{-1/\alpha}\frac{Z_\alpha}{S_\alpha}\,,\quad 0<\alpha,\gamma\leq 1\,.\label{eqvi8f}
\end{equation}
So, the HN relaxation function can be expressed in the form of a weighted average (\ref{eqiii5}) of an exponential decay $e^{-bt}$ with respect to the distribution $g_{\rm HN}(b)$ of the effective relaxation rate $\widetilde\beta_{\rm HN}$. In this case we obtain
\begin{displaymath}
g_{\rm HN}(b)=\begin{cases}\frac{\sin(\gamma\psi(b))(\pi b)^{-1}}{\left[(\tau_pb)^{2\alpha}+2(\tau_pb)^\alpha\cos(\pi\alpha)+1\right]^{\gamma/2}} &
{\rm for}\ b>0\,,\\ \qquad\qquad\qquad 0 & {\rm for}\ b\leq 0\,,\end{cases}
\end{displaymath}
where $\psi(b)=\frac{\pi}{2}-\arctan\left(\frac{(b\tau_p)^{-\alpha}+\cos(\pi\alpha)}{\sin(\pi\alpha)}\right)$ (Figure~\ref{HN-JWS}). At this point we have to stress that such a representation, following the most natural attempt to non-exponential relaxation \cite{bb78,jonscher83}, it is not possible to derive the HN relaxation function with $\gamma>1$. To describe the atypical relaxation data represented in Figure~\ref{diag-1}, the proposed scheme with random variables $N_i$, $M_j$, and $\beta_{iN}$ should be modified. Instead of the overestimating the number of clusters and super-clusters (see Eqs.(\ref{eqvi1}) and (\ref{eqvi2})), the procedure of underestimating these numbers should be involved.

\begin{figure}
\centerline{\includegraphics[clip,width=1.\columnwidth]{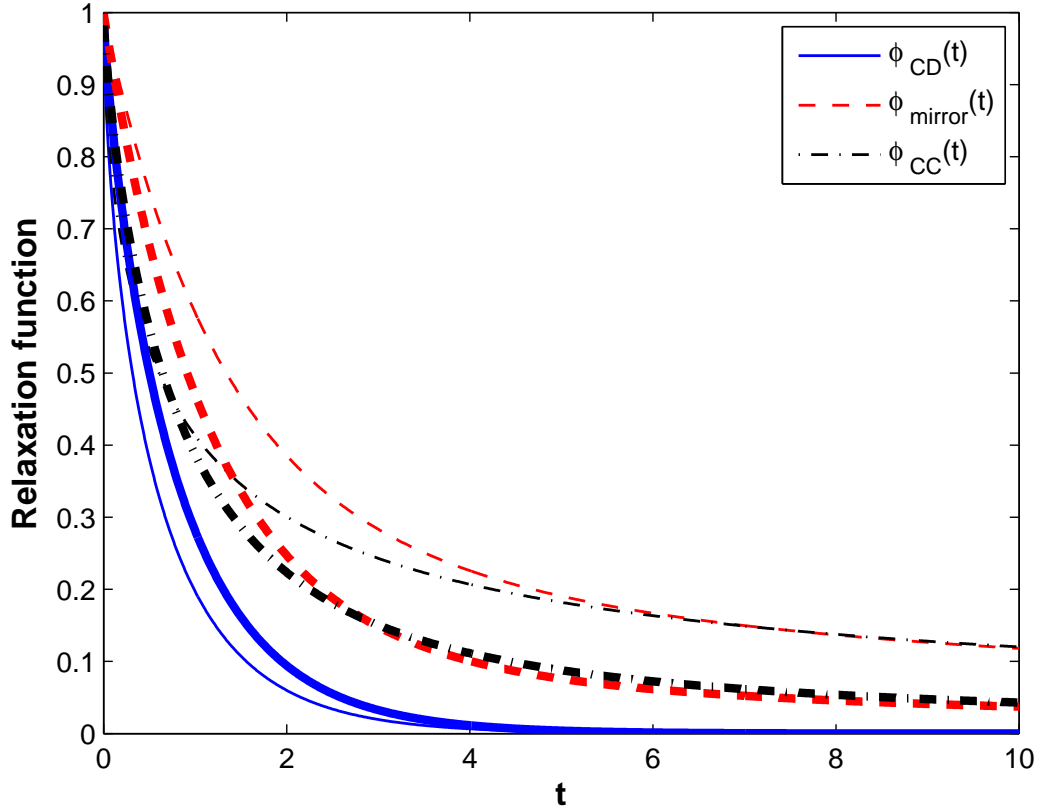}}
\caption{(Colour on-line). Relaxation functions corresponding to the CD law and its mirror case with the index $\gamma$ = 0.6 (thin line) as well as the CC case with $\alpha=0.6$\,. The thick lines indicate the cases with $\gamma$ = $\alpha$ = 0.8\,.}
\label{relfun_CDmirCC}
\end{figure}

Now the number $K_N$ of active entities in the system satisfies the relation
\begin{equation}
K_N=\max\left\{k\ :\ \sum_{i=1}^k(N_i+1)\leq N\right\}\,,\label{eqvi8h}
\end{equation}
and the number $L_N$ of super-clusters is defined by another dependence
\begin{equation}
L_N=\max\left\{l\ :\ \sum_{j=1}^lM_j\leq K_N\right\}\,.\label{eqvi8g}
\end{equation}
In this case we analyze (\ref{eqvi7}) in a way analogous to that using the overestimating scheme. The study provided above makes the similar derivations unnecessary, so we omit them and at once write the relaxation rate of the clustered system corresponding to the atypical relaxation data, namely we have
\begin{equation}
\widetilde\beta_{\rm JWS}\stackrel{d}{=}\frac{1}{\tau_p}\left({\cal B}_\gamma\right)^{1/\alpha}\frac{Z_\alpha}{S_\alpha}\,.\label{eqvi8j}
\end{equation}
The subscript JWS has been used to show a link to the relaxation function for atypical relaxation data, derived in the diffusion framework by Jurlewicz, Weron and Stanislavsky (JWS) \cite{jw08,wjmwt10,swt10}. The relaxation function, corresponding to the JWS law, has a different form than the HN one \cite{jwt08}. It reads
\begin{displaymath}
\phi_{\rm JWS}(t)={\mathcal L}\Bigg(\tau_p\,\frac{Z_\alpha}{S_\alpha}\left({\mathcal B}_\gamma\right)^{1/\alpha}; t\Bigg)\,.
\end{displaymath}
Then the pdf of $\tilde\beta_{\rm JWS}$ (giving the relaxation function in the form of a weighted average of an exponential decay $e^{-bt}$) takes the very similar (but not the same as $g_{\rm HN}(b)$) form
\begin{displaymath}
g_{\rm JWS}(b)=\begin{cases}\frac{\sin(\gamma\psi(b))(\pi b)^{-1}}{\left[(\tau_pb)^{-2\alpha}+2(\tau_pb)^{-\alpha}\cos(\pi\alpha)+1\right]^{\gamma/2}} &
{\rm for}\ b>0\,,\\ \qquad\qquad\qquad 0 & {\rm for}\ b\leq 0\,,\end{cases}
\end{displaymath}
where $\psi(b)=\frac{\pi}{2}-\arctan\left(\frac{(b\tau_p)^\alpha+\cos(\pi\alpha)}{\sin(\pi\alpha)}\right)$ (see Figure~\ref{HN-JWS}). In relation to Eqs.(\ref{eqii8}) and (\ref{eqiii5}), for the two-power law dielectric susceptibilities (see Eqs.(\ref{eqi1}) and (\ref{eqi2})), by Tauberian theorem \cite{feller66}, we have the following asymptotic properties of effective relaxation rate pdf
\begin{displaymath}
g(b)\propto\begin{cases} b^{m-1} &
{\rm for}\ b\to 0\,,\\  b^{n-2} & {\rm for}\ b\to\infty\,,\end{cases}
\end{displaymath}
see the bottom panels in Figure~\ref{HN-JWS}. Let us note that the Laplace transform of the generalized arcsine pdf $f_\gamma(x)$
%\begin{eqnarray}
%{\cal L}(B_\gamma; t)&=&\int_0^1 e^{-tx}\,f_\gamma(x)\,dx\nonumber\\
%&=&\Pr(\Xi_\gamma\geq t)=M(\gamma,1,-t)\,,\nonumber
%\end{eqnarray}
\begin{displaymath}
{\cal L}({\cal B_\gamma}; t)=\int_0^1 e^{-tx}\,f_\gamma(x)\,dx=\Pr(\varXi_\gamma\geq t)=M(\gamma,1,-t)\,,\nonumber
\end{displaymath}
where $M(a,b,x)$ is the Kummer (confluent) function \cite{abr65}, describes a mirror reflection of the CD law in frequency domain as Fig.~2 in \cite{sw12}. Hence, the random effective waiting time $\widetilde\theta_{\rm JWS}$ is given by a mixture of random variables $\tau_p\,S_\alpha\,(\varXi_\gamma)^{1/\alpha}$, and the relaxation function takes the form
\begin{displaymath}
\phi_{\rm JWS}(t)=\Pr(\tau_p\,S_\alpha\,(\varXi_\gamma)^{1/\alpha}\geq t)=\Pr(\widetilde\theta_{\rm JWS}\geq t)\,.
\end{displaymath}
This allows us to find the frequency-domain shape function
%\begin{eqnarray}
%\phi^*_{\rm JWS}(\omega)&=&\langle e^{-i\omega \tau_p\,S_\alpha(\Xi_\gamma)^{1/\alpha}}\rangle\nonumber\\
%&=&\int_0^\infty e^{-(i\omega\tau_p)^\alpha t}\,v_\gamma(t)\,dt\,,\nonumber
%\end{eqnarray}
\begin{displaymath}
\phi^*_{\rm JWS}(\omega)=\langle e^{-i\omega \tau_p\,S_\alpha(\varXi_\gamma)^{1/\alpha}}\rangle
=\int_0^\infty e^{-(i\omega\tau_p)^\alpha t}\,\varrho_\gamma(t)\,dt\,,
\end{displaymath}
where $\varrho_\gamma$ is the pdf of the random variable $\Xi_\gamma$, i.\,e.
\begin{displaymath}
\varrho_\gamma(x)=\frac{1}{\Gamma(\gamma)\Gamma(1-\gamma)}\int_0^1 e^{-xz}\,z^\gamma\,(1-z)^{-\gamma}\,dz
\end{displaymath}
for $x>0$. As a result, we get the following expression
\begin{equation}
\phi^*_{\rm JWS}(\omega)=1-\frac{(i\omega/\omega_p)^{\alpha\gamma}}{[1+(i\omega/\omega_p)^\alpha]^\gamma}\,.\label{eqvi8jws}
\end{equation}
It is useful to compare probabilistic properties of two different random variables, $\varGamma_\gamma$ and $\varXi_\gamma$, appearing in this context. The distribution of the first of them is characterized by all finite non-zero moments, whereas in the second case all the integer moments become equal to zero. Note that the relaxation patterns very close to the mirror CD law are observed in neo-hexanol ($m=0.72$ and $1-n=0.95$, see Fig. 5.27 in \cite{jonscher83}) as well as in gallium (Ga)-doped mixed crystals \cite{twpp08}. The difference among CD, CC and mirror relaxations with the same parameter are shown in Figure~\ref{relfun_CDmirCC}.

\begin{figure}
\centerline{\includegraphics[clip,width=1.15\columnwidth]{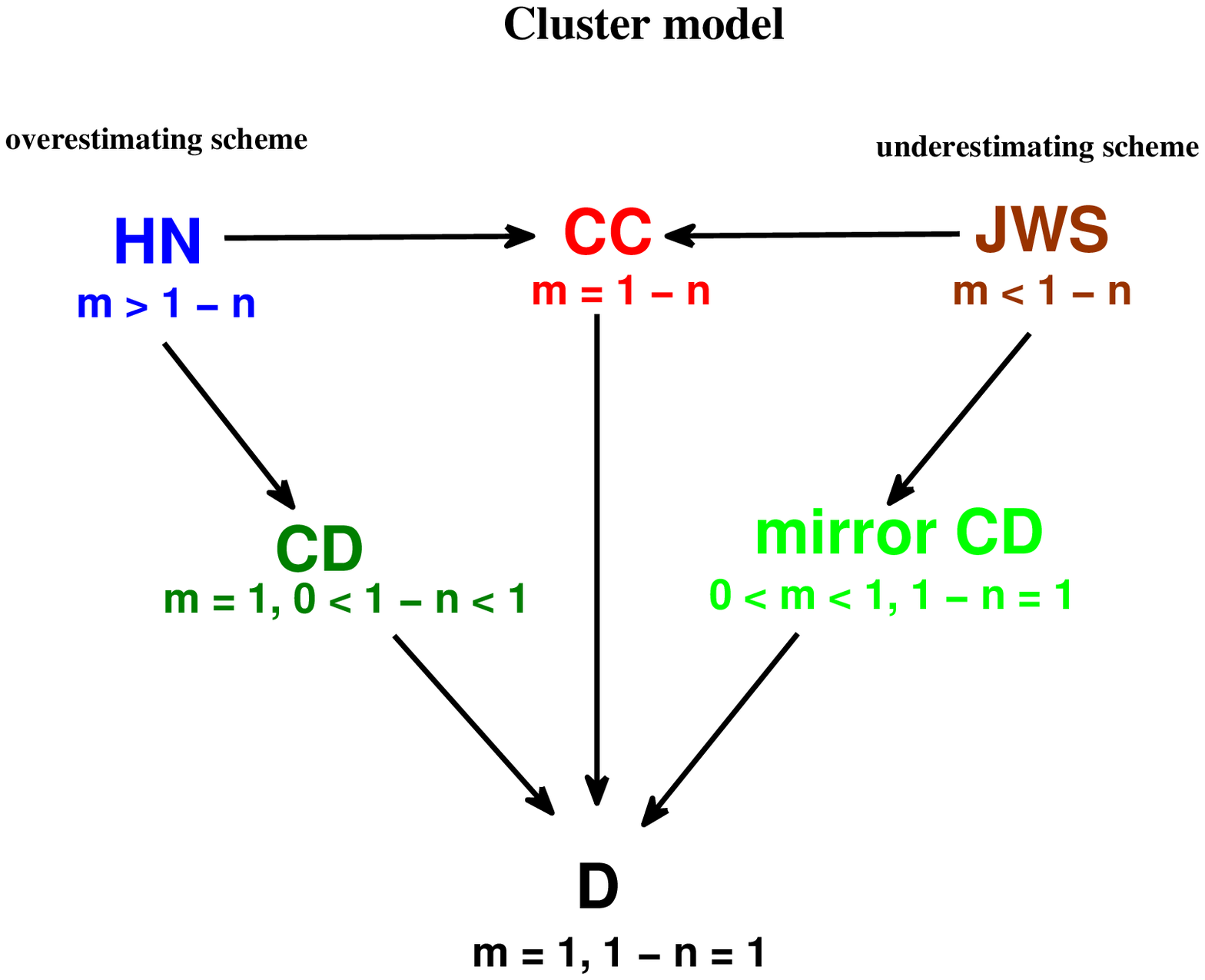}}
\caption{(Colour on-line). Space representations of clusterization in relaxation systems. In this graph, as in Figure~\ref{diag-1}, the indices $1-n$ and $m$ denote the high- and low-frequency power-law exponents, respectively.}
\label{space_scheme}
\end{figure}

The finiteness of the expected value and long-tail property (\ref{eqii4}) can be presented only on three different levels: active entity $\Rightarrow$ cluster $\Rightarrow$ cooperative region (super-cluster) of the complex system. To sum up, Table~\ref{tab_inter} shows the connection between the internal properties of complex system's dynamics and the empirical relaxation responses, as well as, the physical sense of the
parameters characterizing the responses. The proposed approach leads to a very general scenario of relaxation, from the stochastic nature of microscopic dynamics through the hierarchical structure of parallel multi-channel processes to the empirical macroscopic laws of relaxation (see Figure~\ref{space_scheme}).

\begin{table} \caption{Relationship between internal random properties of clustered
  complex systems and their relaxation responses. The indices
  of the column ``Power-law exponents'' correspond to the values $0<\alpha,\gamma\leq 1$ in Eqs.(\ref{eqvi8f}) and (\ref{eqvi8j}), $m$ and $n$ are related to Eqs.(\ref{eqi1}) and (\ref{eqi2}).\\ }
  \label{tab_inter}
	\centering{
  \begin{tabular}{|c|c|c|c|c|c|}\hline
    & \multicolumn{2}{|c|}{Power-law} &  &  & \\
    Law & \multicolumn{2}{|c|}{exponents} & $N_i$ & $M_j$ & $\beta_{iN}$  \\  \cline{2-3}
    & $m$  & $1-n$  &  &  & \\   \hline
    &  &  &  &  & \\
    D & 1 & 1 & $\langle N_i\rangle <\infty$ & $\langle M_j\rangle <\infty$ & $\langle\beta_{iN}\rangle <\infty$ \\
    &  &  &  &  & \\ \hline
    &  &  &  &  & \\
    &  &  &  & long tail & \\
    CD & 1 & $\gamma$ & $\langle N_i\rangle <\infty$ & $\gamma$ &
    $\langle\beta_{iN}\rangle <\infty$ \\
    &  &  &  &  & \\  \hline
    &  &  &  &  & \\
    &  &  & long tail &  &long tail \\
    CC & $\alpha$ & $\alpha$ & $\alpha$ & $\langle M_j\rangle <\infty$ & $\alpha$ \\
    &  &  &  &  & \\ \hline
    &  &  &  &  & \\
    mirror &  &  &  & long tail & \\
    CD & $\gamma$ & 1 & $\langle N_i\rangle <\infty$ & $\gamma$ & $\langle\beta_{iN}\rangle <\infty$ \\
    &  &  &  &  & \\  \hline
    &  &  &  &  & \\
    &  &  & long tail & long tail & long tail \\
    HN & $\alpha$ & $\alpha\gamma$ & $\alpha$ & $\gamma$ & $\alpha$ \\
    &  &  &  &  & \\  \hline
    &  &  &  &  & \\
    &  &  & long tail & long tail & long tail \\
    JWS & $\alpha\gamma$ & $\alpha$ & $\alpha$ & $\gamma$ & $\alpha$ \\
    &  &  &  &  & \\  \hline
  \end{tabular}}
\end{table}

The fundamental consequence of property (\ref{eqi1}) is that for large $\omega$ the ratio of the imaginary to real term of the complex susceptibility $\chi(\omega)=\chi'(\omega)-i\chi''(\omega)$ is a constant, dependent only on the exponent $n$
\begin{equation}
\frac{\chi''(\omega)}{\chi'(\omega)}=\cot\left(n\frac{\pi}{2}\right)\qquad{\rm for\ }\omega\gg\omega_p\,.\label{eqiii18}
\end{equation}
The physical significance of this simple property is that at high frequencies the ratio of the macroscopic energy lost per radian to the energy stored at the peak is independent of frequency \cite{jonscher83}. However, the D response does not have this property.

A. K. Jonscher \cite{jonscher83} has advanced a hypothesis that the fractional power law (\ref{eqi1}) and the energy criterion (\ref{eqiii18}) are inescapably connected with the fact that the energy loss in every microscopic reversal is independent of the rate of reversals in the corresponding frequency range. He assumed that since in any dielectric system the total polarization is a sum of individual microscopic polarizations and the total loss is the sum of individual microscopic losses, the microscopic relationship also must have the property of energy lost to energy stored being independent of frequency. The fact is based on the identical property of individual structural elements of the systems. This explains the universality in the large scale behavior of complex systems.

Note, the above-mentioned limiting form (\ref{eqvi8}) basically is determined by the
tail behavior of $F_{\beta_i}(b)$ for large $b$, i.\,e.\ by
asymptotic properties of $F_{\beta_i}(b)$. The detailed knowledge
of its other properties is not necessary. The distribution function $F_{\beta_i}(b)$
belongs to the domain of attraction of the
$L\alpha S$ law with the index of stability
$0<\alpha< 1$ if and only if \cite{feller66} for each $x>0$
\begin{equation}
\lim_{b\to\infty}\frac{1-F_{\beta_i}(xb)}{1-F_{\beta_i}(b)}=
x^{-\alpha}. \label{eqiii19}
\end{equation}
This condition can be interpreted as a type of self-similarity:
%\begin{eqnarray}
%&&\mathrm{Pr}\,(\beta_i>\mathit{xb})\approx
%x^{-\alpha}\,\mathrm{Pr}\,(\beta_i>\mathit{b})\quad\mathrm{for\ any}\
%\mathit{x}>0,\nonumber\\
%&&\qquad\qquad\qquad\qquad\qquad\qquad\qquad\mathrm{and\ large}\ \mathit{b}. \label{eqiii20}
%\end{eqnarray}
\begin{equation}
\mathrm{Pr}\,(\beta_i>\mathit{xb})\approx
x^{-\alpha}\,\mathrm{Pr}\,(\beta_i>\mathit{b})\quad\mathrm{for\ any}\
\mathit{x}>0\ \mathrm{and\ large}\ \mathit{b}. \label{eqiii20}
\end{equation}
The self-similarity was suggested earlier \cite{ks86,dh89,nikl93} as a fundamental feature
of relaxation phenomena. Let us stress that in the limit-theorem
approach this result is obtained on the pure probabilistic base,
independently of the physical details of systems.

In the framework of the correlated-cluster approach the physical intuition of A. K. Jonscher may be strictly argumentative. Really, the condition (\ref{eqii4}) applied to any relaxation rate $\beta_i$ leads to the scaling property of the relaxation-rate distribution at large $b$ (see also Eq.(\ref{eqiii20})). The asymptotic behavior of the distribution is connected with the short-time asymptotic properties of the associated relaxation function $\phi(t)$, and the response function as its derivative $f(t)=-d\phi(t)/dt$ takes the form
\begin{displaymath}
f(t)\propto t^{\alpha-1}\,U(t)
\end{displaymath}
for $t\to 0$, where $U(t)$ is a slowly varying function so that $\lim_{t\to 0}U(ct)/U(t)=1$ for any constant $c>0$ \cite{sawo97}. It may be easily verified that the short-time behavior of $f(t)$ corresponds to the high-frequency properties of the susceptibility
$\chi(\omega)$:
\begin{displaymath}
\chi(\omega)=\chi'(\omega)-i\chi''(\omega)\propto
(i\omega)^{-\alpha}\,U(1/\omega).
\end{displaymath}
The result yields straightforwardly the energy criterion (\ref{eqiii18}) with $n=1-\alpha$. The long-tail property of micro/meso/macroscopic relaxation rates with the parameter $\alpha$ leads to micro/meso/macroscopic energy criterion with the characteristic constant $1-\alpha$. The analysis of the model shows \cite{wjj01,jw02,jjw03} that in the HN, JWS, CC  and KWW responses the energy criterion holds for all micro/meso/macro levels, and the power-law exponent $n$ for the HN case (as well as the power-law exponent $m$ for the JWS case) is defined not only by the long-tail property of the distribution of cluster sizes, but also of super-cluster sizes. In the CD case the microscopic energy criterion is not fulfilled. The high-frequency power law of this response results only from the long-tail property of the distribution of super-cluster sizes (see \cite{jjw03} for details).

\subsubsection{Finite mean cluster sizes}

Assume that the sequences of random variables $N_{i}$,
$M_{j}$, and $\beta_{iN}$ are stochastically independent; and each
sequence consists of iid non-negative random variables. Assuming moreover finite-average cluster size
$\langle N_i\rangle=\eta_0$ we obtain that $K_N\approx \eta_0^{-1}N$ (with probability~1) for large $N$.
Hence, the random sum $\sum\limits_{j=1}^{L_N}M_{j}$ is asymptotically distributed as
$\mu_{\gamma} \eta_0^{-1}N$, where $\mu_{\gamma}$ is a random variable representing a continuous limit of the random number of randomly sized super-clusters. The random variable $\mu_\gamma$ indicates the random space in which the super-clusters exist. If the distribution of $M_j$ is heavy-tailed (see Eq.(\ref{eqii4})) with the tail exponent $0<\gamma<1$, then the pdf of $\mu_{\gamma}$ is given by
$q_{\gamma}(x) = [\Gamma(\gamma)\Gamma(1-\gamma)]^{-1}x^{-1}(x-1)^{-\gamma}$ for $x>1$, and 0 otherwise.  Here $\Gamma(\cdot)$ is the gamma function. If the super-cluster average size is
finite ($\langle M_j\rangle<\infty$), then $\gamma=1$ and $\mu_1=1$.

Taking $\beta_{iN}=A_{N}^{-1}\beta_{i}$, where $\beta_{i}$ is
independent of the system size $N$ and the normalizing system-size
dependent constant $A_{N}$ is the same for each dipole, one can
write
\begin{displaymath}
\tilde\beta_{N} = \frac{1}{A_N}\sum\limits_{i=1}^{\sum\limits_{j=1}^{L_N}M_{j}} {\beta_{i}}.
\end{displaymath}
For a finite-average distribution of the individual relaxation
rates with $\langle \beta_i\rangle=b_0$, and for $A_N=N$ we obtain
\begin{equation}
\tilde\beta = b_0\mu_{\gamma}\eta_0^{-1}.\label{eqvii14}
\end{equation}
On the other hand, if the distribution of the individual
relaxation rates $\beta_i$ is heavy-tailed with the tail exponent
$0<\alpha\leq 1$ and the scaling constant equals $b_0$, then for the normalizing sequence
$A_N=(\Gamma(1-\alpha)N)^{1/\alpha}$ we get
\begin{equation}
\tilde\beta = b_0(\mu_{\gamma}\eta_0^{-1})^{1/\alpha} S_{\alpha},\label{eqvii15}
\end{equation}
where $S_{\alpha}$ is a completely asymmetric $L\alpha S$
random variable with the index of stability $0<\alpha< 1$, independent
of $\mu_{\gamma}$; formula (\ref{eqvii15}) coincides with Eq.(\ref{eqvii14}), if we take $\alpha=1$ and $S_1=1$.

Now, one can derive the
corresponding relaxation or response functions. In particular, the
case $0<\gamma<1$ and $\alpha=1$ corresponds to the CD relaxation
pattern with $\tau_p=\eta_0b_0^{-1}$, while the case
$\gamma=1$ and $0<\alpha<1$ is related to the KWW relaxation
function with $\tau_p=\eta_0^{1/\alpha}b_0^{-1}$. Let us
remind that $\alpha=1$ refers to $\langle
\beta_i\rangle=b_0<\infty$, and $\gamma=1$ to $\langle
M_j\rangle<\infty$. For $0<\gamma<1$ and $0<\alpha<1$ one obtains
the Generalized Gamma (GG) relaxation for which the relaxation
function can be represented as
\begin{displaymath}
\phi_{\rm GG}(t)=\Pr[\varGamma_{\gamma}\geq (t/\tau_p)^{\alpha}],
\end{displaymath}
where $\varGamma_{\gamma}$ is the gamma distributed random variable with
the shape parameter $\gamma$ and the scale parameter equal to 1,
and $\tau_p=\eta_0^{1/\alpha}b_0^{-1}$. The corresponding response
function $f_{GG}(t)$ takes hence the form of GG pdf \cite{john70}
\begin{equation}
f_{\rm GG}(t)=\frac{\alpha}{\tau_p\Gamma(\gamma)}(t/\tau_p)^{\alpha\gamma-1}\exp\left [-\left(t/\tau_p\right )^{\alpha}\,\right ].\label{eqvii16}
\end{equation}
It should be noticed that this relaxation function is supported in experimental data \cite{kkbnr10} (propylene glycol and 2-picoline in tri-styrene).
As one can see, the GG response results from
the heavy-tail properties of both the active-dipole
relaxation-rate ($\beta_{iN}$) and the super-cluster-size ($M_j$)
distributions. Its parameters $0<\alpha,\gamma<1$ are equal to the
respective tail exponents. The special case of the CD pattern
($\alpha=1$) corresponds to the finite-average $\beta_{iN}$
distributions taken instead of the heavy-tailed one. Similarly,
the KWW response ($\gamma=1$) refers to the finite-average $M_j$
distribution. As the generalized model has two spatial scales
(clusters and super-cluster regions), one could expect that the
latter scale corresponds to larger relaxation times than the
scales being related to the clusters. But the clusters of
super-cluster regions are dynamically constrained. Therefore,
their parameters $\alpha$ and $\gamma$ only influence on
short-time behavior of this relaxation via active dipoles. When
such a constraint is absent or weak, the clusters and
super-cluster regions are responsible for different time scales.
Consequently, the super-cluster evolution can determine the
relaxation trend in lower frequencies. To sum up, Table~\ref{tab_inter2} demonstrates the connection between the internal properties of such complex system's dynamics and the parameters characterizing the empirical relaxation responses.

\begin{table} \caption{Relationship between the internal random properties of
  clustered complex systems and their fractional short-time power-law relaxation responses. The indices
  of the column ``Short-time power-law exponent'' corresponds to Eqs.
  (\ref{eqii10}), and $0<\alpha,\gamma\leq 1$ are related to Eq.(\ref{eqvii16}).\\ }
  \label{tab_inter2}
	\centering{
  \begin{tabular}{|c|c|c|c|c|}\hline
    & Short-time &  &  &  \\
 Law & power-law & $N_i$ & $M_j$ & $\beta_{iN}$  \\
    & exponent $n$  &  &  &  \\
 \hline
    &  &  &  &  \\
    &  & & long tail & \\
    CD & $1-\gamma$ & $\langle N_i\rangle <\infty$ & $\gamma$ &
    $\langle\beta_{iN}\rangle <\infty$ \\
    &  &  &  &  \\  \hline
    & & & & \\
    KWW & $1-\alpha$ & $\langle N_i\rangle <\infty$ & $\langle M_j\rangle
    <\infty$& long tail \\
		&  &  &  & $\alpha$  \\
    &  &  &  &  \\ \hline
    &  &  &  &  \\
    &  & $\langle N_i\rangle <\infty$ & long tail & long tail \\
    GG & $1-\alpha\gamma$ &  & $\gamma$ &
    $\alpha$ \\
    &  &  &  &  \\ \hline
  \end{tabular}}
\end{table}

\begin{figure}
\centerline{\includegraphics[clip,width=1.\columnwidth]{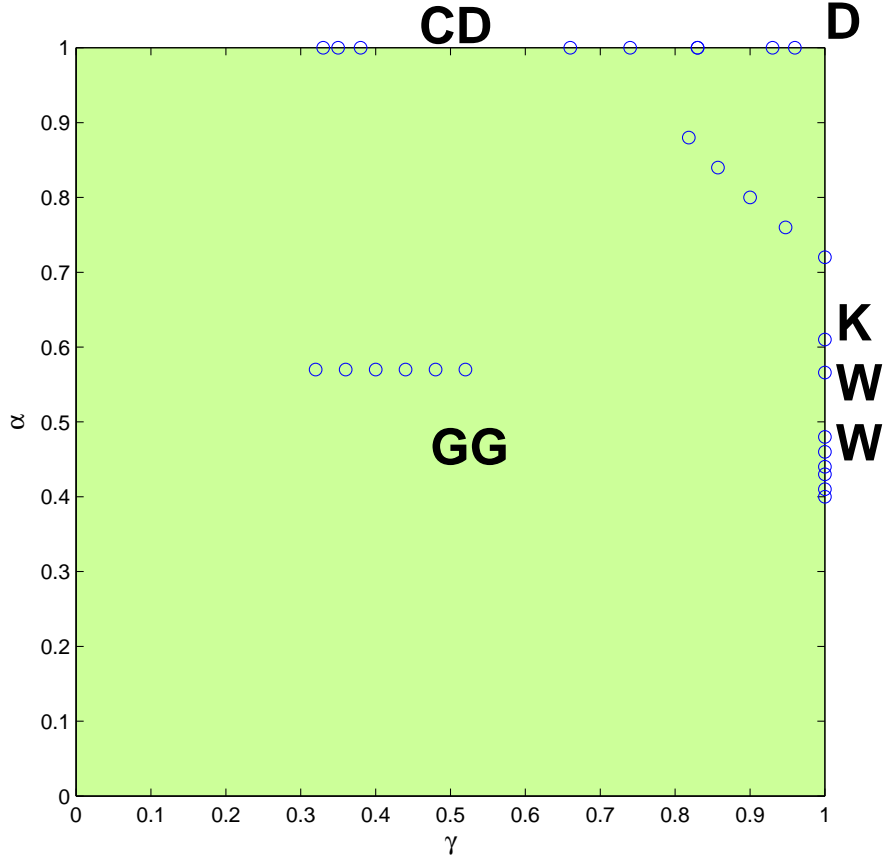}}
\caption{(Color on-line) Relaxation diagram positioning different
one-power (short-time) laws of relaxation. The circles are experimental points (for
various materials) taken from \cite{jonscher83,kkbnr10,bfj01}. The abbreviations have the following meanings: D -- Debye; KWW -- Kohlrausch-Williams-Watts; CD -- Cole-Davidson; GG -- Generalized Gamma. The D relaxation is considered as a special case ($\alpha=1$ and $\gamma=1$).}
\label{diag-2}
\end{figure}

Now we  observe  that the GG function,
as well as KWW and CD ones (see Figure~\ref{diag-2}), exhibit the short-time power law
\begin{equation}
f_{\rm GG}(t)\propto (t/\tau_p)^{-n}\;\;\;\;\;{\rm for}\;t\ll\tau_p,\label{eqvii17}
\end{equation}
where $n=1-\alpha\gamma$. For the long-time
limit both the GG and KWW functions decay stretched exponentially
with the exponent $\alpha<1$, while the CD function decays simply
exponentially. The time-domain limiting properties of the GG function correspond
to those  of the frequency-domain response given by
\begin{eqnarray}
&\phi^{*}_{\rm GG}(\omega)=\frac{\alpha}{\tau_p\Gamma(\gamma)}\int\limits^{\infty}_{0}e^{-{\rm i}\omega t} (t/\tau_p)^{\alpha\gamma-1}\exp[- (
t/\tau_p)^{\alpha} ]\,dt&\nonumber\\
&=\frac{\alpha}{\Gamma(\gamma)}({\rm i}\omega/\omega_p)^{-\alpha\gamma}\, {_1\!\Psi_0}[-({\rm i}\omega/\omega_p)^{-\alpha}\ |\ (\alpha\gamma,\alpha)]\,,&\label{eqvii18}
\end{eqnarray}
where ${_1\!\Psi_0}[z\ |\ (\alpha\gamma,\alpha)]=\sum^\infty_{n=0}\Gamma(\alpha(\gamma+n))\,z^n/n!$ is a special case of the Fox-Wright Psi function \cite{miller}.

Substituting $s=\omega t$ in (\ref{eqvii18}), we get
\begin{equation}
\lim\limits_{\omega\rightarrow\infty}\frac{\phi^{*}_{\rm GG}(\omega)
}{ ({\rm i}\omega/\omega_{p})^{-\alpha\gamma}}=\frac{\alpha \Gamma(\alpha\gamma)}{\Gamma(\gamma)}.\label{eqvii19}
\end{equation}
The dielectric susceptibility exhibits
hence the fractional high-frequency power law with
different fractional exponents for the GG, CD and KWW functions. In the low-frequency range we come to
\begin{displaymath}
\lim\limits_{\omega\rightarrow 0+}\frac{-{\rm Im} [\phi^{*}_{\rm GG}(\omega)]
}{\omega/\omega_{p}}=\frac{\Gamma(\gamma+1/\alpha)}{\Gamma(\gamma)}\,,
\end{displaymath}
while
\begin{displaymath}
\lim\limits_{\omega\rightarrow 0}\frac{{\rm Re}[\phi^{*}_{\rm GG}(0)]-{\rm Re}[\phi^{*}_{\rm GG}(\omega)]
}{(\omega/\omega_{p})^{2}}=
\frac{\Gamma(\gamma+2/\alpha)}{2\Gamma(\gamma)}.
\end{displaymath}
Therefore, in the low-frequency limit for the GG pattern (with its
special KWW and CD cases) one observes linear dependence on
frequency in the absorption term. The results have been obtained in \cite{wjps13}. It should be noticed that the attempt to fit the experimental data analyzed in \cite{kkbnr10} by two different Havriliak-Negami relaxation functions requires seven various parameters \cite{mpvdba11} instead of the three ones as in the case \cite{kkbnr10} based on the generalization of CD and KWW functions. In this context the GG function is more preferable.

\subsection{Spatial randomness}

The Bernoulli binomial distribution \cite{feller66}
\begin{displaymath}
\Pr\left(X = k\right) = \binom{N}{k}\,p^k\,(1 - p)^{N-k}\ {\rm for}\ k = 0,1,2,3,\dots,N
\end{displaymath}
is often applied for a finite area study as the exact model with spatial randomness.
It is easy to show that when $N\to\infty$ and $p\to0$ so that $pN$ is a constant,
the Bernoulli distribution becomes the Poisson distribution, that describes
random patterns in infinitely large areas. Both Bernoulli and
Poisson distributions suppose that the probability that an
individual of species is found in a given area is independent of
the presence of other individuals in the same area.  If a space cell already contains an
individual, then the cell will be more likely to keep more
individuals, whereas empty cells are apt to remain empty. It cannot be
considered strictly random from the spatial point of view. The
important feature relates to aggregated patterns. To describe such patterns, we use the negative
binomial distribution \cite{john70}. The latter is often
considered as a flexible alternative to the Poisson model for
count data. The negative binomial distribution is a substitute
for the Poisson distribution, when it is doubtful whether the strict
requirements, particularly independence, for the Poisson distribution
will be satisfied. In this case the number of entities taking
part in the relaxation process is not necessarily fixed with
respect to all $N$ dipoles forming the system.
The transition process starts with a random initial number $\nu_M$
of ordering dipole orientations and then runs due to the individual transitions of
dipole orientations at random instants of time
$\theta_{1},\theta_{2},\ldots,\theta_{N}$. The number $\nu_M$ is
an integer-valued random variable depending on the size $N$ of the
system. The effective relaxation rate $\widetilde\beta$ is given by
the summation of individual rates $\beta_i$ over all $\nu_N$
possible routes for its realization, i.\,e.
\begin{equation}
\widetilde\beta_N=\sum^{\nu_N}_{i=1}\beta_i/A_N\,.\label{eqiii38a}
\end{equation}
This form of $\widetilde\beta_N$, instead of that defined in Eq.(\ref{eqiii11}), yields now a mixed \cite{wk97} effective relaxation rate $\widetilde\beta$. The negative binomial distribution for
$\nu_N$ is written as
\begin{equation}
{\rm Pr}\Bigl(\nu_N=n\Bigr)=  \frac{\Gamma(\gamma+n-1)}{\Gamma(\gamma)
\,(n-1)!}\Bigl[\frac{1}{N}\Bigr]^{\gamma}\Bigl[1-\frac{1}{N}\Bigr]^{n-1}\,,
\label{eqiii39a}
\end{equation}
with $n=1,2,\,\dots$ and the parameter $\gamma>0$. It follows from the book of Johnson and
Kotz \cite{john70} that in the limit $N\to\infty$
Eq.(\ref{eqiii39a}) transforms into the gamma distribution with pdf
$g_\gamma(y)=y^{\gamma-1}e^{-y}/\Gamma(\gamma)$ for $y>0$. Then
the survival probability of the entire system ${\rm Pr}\bigl(\widetilde\theta\geq t\bigr)=\left\langle e^{-\widetilde\beta t}\right\rangle$, where $\widetilde\beta$ is the limiting random variable of $\widetilde\beta_N$ given by (\ref{eqiii38a}), can be expressed as a mixture of distributions (see Subsection~\ref{subsiicn})
\begin{eqnarray}
\phi(t)&=&{\rm Pr}\bigl(\widetilde\theta\geq t\bigr)\nonumber\\
&=&\int^\infty_0\left(\int^\infty_0
e^{-At(y/\gamma)^{1/\alpha}x}\,h_\alpha(x)\,dx\right)\,g_\gamma(y)\,dy
\nonumber\\
&=&\frac{1}{[1+(1/\gamma)(At)^\alpha]^\gamma}\,,\quad 0<\alpha,\alpha\gamma\leq1\,,\label{eqiii40}
\end{eqnarray}
what can be identified as the tail of Burr distribution \cite{john70,wk97}.
The parameter $\gamma$ has the same physical (or chemical) sense in both gamma
distribution and negative binomial one, i.\,e. it shows a measure
of aggregation in the system. It should be pointed out that the
above probabilistic schemes start to work even with $N\approx
10^5-10^6$. If the aggregation is lack, then
\begin{displaymath}
\frac{1}{[1+(1/\gamma)(At)^\alpha]^\gamma}\stackrel{\gamma\to\infty}{\longrightarrow}
\exp\bigl[-(At)^\alpha\bigr]\,.
\end{displaymath}
The limiting case ($\gamma\to\infty$) describes the deterministic number of the $L\alpha S$ contributions (see Eq.(\ref{eqiii25})). The response function $f(t)$, corresponding to Eq.(\ref{eqiii40}), exhibits the two-power-law asymptotics, namely
\begin{equation}
f(t)\propto
\begin{cases} (At)^{\alpha-1} &
{\rm for}\quad At \ll 1\\ (At)^{-\alpha\gamma-1} & {\rm
for}\quad At \gg 1\,.\end{cases}\label{eqiii41}
\end{equation}
The results are supported in experimental data \cite{kwk01,gnpawets09,otbcdfgnmnos14}. Note that by direct calculations the relaxation function (\ref{eqiii40}) leads to the transition rate
\begin{equation}
r_{\rm bin}(t)= \frac{\alpha A^\alpha t^{\alpha-1}}{1+1/\gamma(At)^\alpha}\,,\label{eqiii41a}
\end{equation}
so that for $\gamma\to\infty$ it tends to $\alpha A^\alpha t^{\alpha-1}$, the transition rate of KWW relaxation.

Now we discuss interpretation of the obtained model in more details. As it has been established in the framework of the limit theorems of probability theory, the stretched exponential law (\ref{eqiii25}) is the only form of the relaxation decay realized in random distributions (such as Bernoulli and Poisson ones) of species as a null model for the species-area relationship. The hyperbolic law (\ref{eqiii40}) corresponds to spatial aggregations described by the negative binomial distribution. It can arise from a variety of random processes. One of such well-known processes is immigration/birth/death-like scheme. In this case the birth (and death) events are not an independent but a contagious process, meaning that a birth has the tendency to induce more births and a death to induce more deaths. As applied to dielectric relaxation, the birth is a merger of ordered dipole orientations, and the death is their breakup. The macroscopic picture of the dipole evolution is an echo of the spatial distributions. There are parameters standing for both, microscopic and macroscopic dynamics. We have denoted them as $\alpha$ and $\gamma$. The parameter $\alpha$ characterizes a L\'evy stable (scaling on different levels) character of random processes participating in the development of such systems. Any complex system consists of many objects. Their statistics is assumed to be a summation procedure of many random variables. The variables and their sum (full or partial) belong to the same probability distribution. This means just its stability. If $\alpha\to1$, then the transition of the system from an excited state to an equilibrium one looks like nothing else but an exponential decay. Otherwise, when the expected value does not exist, the decay is stretched exponential. The relaxation rates take a continuous value from zero to infinity. Its probability density has a power tail denoting that one cannot avoid very large values of the rates. They make a significant contribution to the relaxation evolution to change its behavior with strongly exponential features into stretched ones. The parameter $\gamma$ detects the dipoles as clustered. The clustering is stochastically independent on the stable character of random processes leading to the stretched decay mentioned above. Therefore, knowing the values of $\alpha$ and $\gamma$, we can answer whether there are clusters or not, as well as clarify the probabilistic character for individual relaxation rates. If $\gamma\to\infty$ and $\alpha\to1$ (no clusters and no stable distribution of rates), then the relaxation decay is only exponential. In this context the dependence (\ref{eqiii40}) is more general than the stretched exponential output.

\subsection{Probabilistic vs deterministic modeling}

The concept of time-dependent transition rate $r(t)$ in the study of non-exponential relaxation is not always convenient because of its overloaded form in force of $r(t)=-\frac{d}{dt}\ln\phi(t)$ (see Eqs.(\ref{eqiii34})- (\ref{eqiii35}), and Eq.(\ref{eqiii41a})), where $\phi(t)$ is the relaxation function under consideration. Moreover, $r(t)$ of the HN case (\ref{eqiii15}) becomes very cumbersome, and it does little in understanding of the relaxation mechanisms. Even in the case of the CC relaxation function the transition rate $r(t)$ is of a special form \cite{wk00}. It is expressed in terms of a ratio of Mittag-Leffler functions %(see (\ref{eqiii17a}) and \cite{pillai90}). %$\phi'_{\rm CC}(t)$ and $\phi_{\rm CC}(t)$ that is written as (\ref{eqiii17a}).
%If we express $\phi_{\rm CC}(t)=E_\alpha(-(t/\tau_p)^\alpha)$ in terms of the Mittag-Leffler function ($0<\alpha\leq1$), then the CC relaxation rate is defined as
\begin{displaymath}
r_{\rm CC}(t)= \frac{t^{\alpha-1} E_{\alpha,\alpha}\left[-(t/\tau_p)^\alpha\right]}{\tau_p^\alpha E_\alpha\left[-(t/\tau_p)^\alpha\right]}\,,
\end{displaymath}
where $E_{\alpha,\beta}(x)=\sum_{j=0}^\infty x^j/\Gamma(\alpha j+\beta)$, and $E_\alpha(x)=E_{\alpha,1}(x)$. Although in the kinetic equation (\ref{eqii7}), the relaxation rate $r(t)$ contains some information about stochastic features of the given relaxing system, this approach is not unique.
%\begin{displaymath}
%\frac{d}{dt}\,E_\mu(-\lambda t^\mu)=-\lambda t^{\mu-1}E_{\mu,\mu}(-\lambda t^\mu)\,.
%\end{displaymath}
There exists another view on the kinetic equations. If one gives up the kinetic description based on the derivative of first order and accepts the transition rate as a constant, equal to the material one $\tau_p$, then the kinetic equation can be rewritten into another  representation (using the fractional operators). For example, the CC relaxation equation takes the following fractional form \cite{main96,hilfer00}
\begin{equation}
\left(\frac{\partial^\alpha}{\partial t^\alpha}+
\tau_p^{-\alpha}\right)\phi_{\rm CC}(t)=\frac{t^{-\alpha}}{\Gamma(1-\alpha)}\label{eqiii41cc}
\end{equation}
with the initial condition $\phi_{\rm CC}(0)=1$, where $\partial^\alpha/\partial t^\alpha$ is the Riemann-Liouville fractional derivative \cite{oldham74}. It should be stressed that appearance of the fractional derivative in kinetic equations is caused by the completely asymmetric $L\alpha S$ law, describing the major features of stochastic processes in the relaxation of complex systems. In fact, this suggests the probabilistic interpretation of the fractional calculus. The corresponding pseudo-differential equation for the HN relaxation will be considered below.

%%%%%%%%%%%%%%
%%%%%%%%%%%%%%
\section{Relaxation in two-state systems}
%%%%%%%%%%%%%%
%%%%%%%%%%%%%%
Relaxation, following the D law, may be simply explained by means of a two-state system \cite{repke}. Let $N$ be the common number of entities in a complex system. If $N_\uparrow$ is the number of entities in the state
$\uparrow$, $N_\downarrow$ is the number of entities in the state $\downarrow$ so that
$N=N_\uparrow+N_\downarrow$. Assume that for $t=0$ the system is stated in such an order so that
the states $\uparrow$ dominate, namely
\begin{displaymath}
\frac{N_\uparrow(t=0)}{N}=n_\uparrow(0)=1,\quad
\frac{N_\downarrow(t=0)}{N}=n_\downarrow(0)=0\,,
\end{displaymath}
where $n_\uparrow$ is the part of entities in the state $\uparrow$, $n_\downarrow$ the
part in the state $\downarrow$. Denote the transition rates by $B$ defined from
microscopic properties of the system (for instance, according to the given Hamiltonian of
interaction and the Fermi's golden rule). In this case the kinetic equation describing the
ordinary relaxation (D relaxation) takes the form
\begin{equation}
\begin{cases}\dot n_\uparrow(t)-B\,\{n_\downarrow(t)-
n_\uparrow(t)\}=0\\ \dot n_\downarrow(t)-B\,\{n_\uparrow(t)-
n_\downarrow(t)\}=0,\end{cases}\label{eqiv1}
\end{equation}
where, as usual, the dotted symbol  means the first-order time
derivative. The relaxation function for the simple two-state system is
\begin{equation}
\phi_{\rm D}(t)=1-2n_\downarrow(t)=2n_\uparrow(t)-1=\exp(-2Bt)\,,\label{eqiv1a}
\end{equation}
where $2B=\tau_p^{-1}$ is the characteristic material constant.
It is easy to see that the steady state of the system corresponds to equilibrium with
$n_\uparrow(\infty)=n_\downarrow(\infty)=1/2$. Clearly its response has an exponential
character. However, this happens to be the case when entities relax irrespectively of
each other and of their environment. If the entities interact with their environment, and
the interaction is complex (or random), their contribution in relaxation already will not
result in any exponential decay.

\subsection{Stochastic arrow of time}

The system of equations (\ref{eqiv1}) is a particular case of the general kinetic equation used for the description of evolution of Markov random processes \cite{kampen84}. It is governed by the deterministic array of time. As it was shown in \cite{st03,stan03,stan04}, if one accounts for the subordination of $n_\uparrow(\tau)$ and $n_\downarrow(\tau)$ by an appropriate random process, the kinetic equation (\ref{eqiv1}) changes its form in dependence of chosen subordinators. Recall, a subordinator itself is a stochastic process (with independent, stationary, non-negative increments) describing the evolution of time within another stochastic process (see Section~\ref{pariid}). Such a subordinator introduces a new time clock (stochastic time arrow). In fact, the notion of a subordinator allows determining the random number of ``time steps''. This concept is very helpful to account for appearance of traps in the evolution of relaxing entities.

Let us consider time evolution of the number of entities in the states $\downarrow$ and $\uparrow$ as parent random
processes in the sense of subordination. Assuming that they may be subordinated by another random process with a pdf, say $p(t,\tau)$. Then the entity ratio $m_\uparrow(t)$ in the state $\uparrow$ and the entity ratio $m_\downarrow(t)$ in the state $\downarrow$ under the temporal subordination are determined by the following integral relations
\begin{eqnarray}
m_\uparrow(t)=\int^\infty_0 n_\uparrow(\tau)\,p(t,\tau)\,d\tau\,,\nonumber\\
m_\downarrow(t)=\int^\infty_0 n_\downarrow(\tau)\,p(t,\tau)\,d\tau\,.\label{eqi1a}
\end{eqnarray}
To derive the above equation, it is necessary to choose a subordinator in the clear form.

\begin{figure}
\centerline{\includegraphics[clip,width=1.\columnwidth]{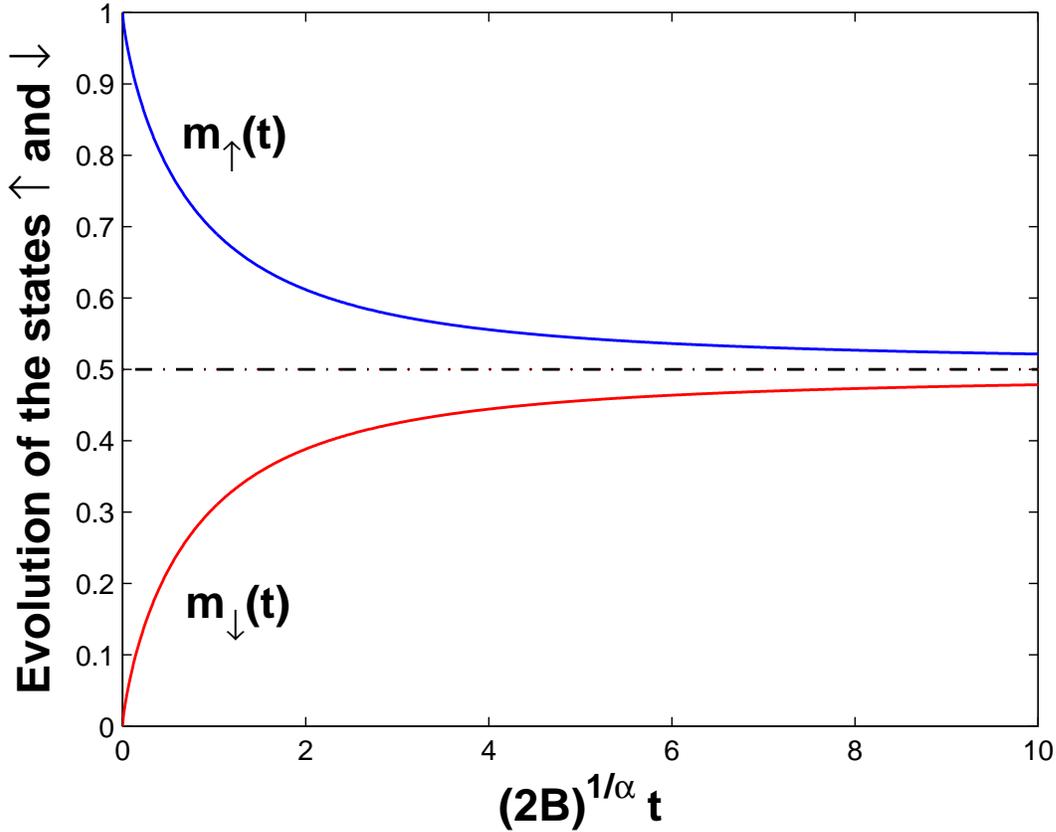}}
\caption{(Color on-line) Relaxation of the part $m_\uparrow$ of entities in the state $\uparrow$ and
the part $m_\downarrow$ of entities in the state $\downarrow$ (for $\alpha$ = 0.8).}
\label{two-states}
\end{figure}

If a new time clock is the inverse $L\alpha S$ subordinator, then the equation of the two-state system (\ref{eqi1a}) takes the following form
\begin{equation}
\begin{cases}D^\alpha m_\uparrow(t)-B\,\{m_\downarrow(t)-m_\uparrow(t)\}=0\\
D^\alpha m_\downarrow(t)-B\,\{m_\uparrow(t)-m_\downarrow(t)\}=0\,,\end{cases}\label{eqiv2}
\end{equation}
where $D^\alpha$ is the $\alpha$-order fractional derivative with respect to time, and $0<\alpha\leq 1$. Here
we use the Caputo derivative \cite{cap79}, namely
\begin{displaymath}
D^\alpha x(t)=\frac{1}{\Gamma(n-\alpha)}\int^t_0\frac{x^{(n)}(\tau)}
{(t-\tau)^{\alpha+1-n}}\,d\tau,\ n-1<\alpha<n,
\end{displaymath}
where $x^{(n)}(t)$ means the $n$-derivative of $x(t)$. In this case the relaxation
function reads
\begin{displaymath}
\phi_{\rm CC}(t)=1-2m_\downarrow(t)=2m_\uparrow(t)-1=E_\alpha(-2Bt^\alpha),
\end{displaymath}
where $E_\alpha(z)$ is the one-parameter
Mittag-Leffler function \cite{erd55}. Evolution of the states $\downarrow$ and $\uparrow$ is shown in Figure~\ref{two-states}.
Following Eq.(\ref{eqii8}), the frequency-domain CC function is
\begin{equation}
\phi^*_{\rm
CC}(\omega)=\frac{1}{1+(i\omega/\omega_p)^\alpha}\,,\quad
0<\alpha\leq 1\,. \label{eqiv4}
\end{equation}
With reference to the theory of subordination \cite{mw06b} the CC law shows that the entities tend to equilibrium via
motion alternating with stops so that the temporal intervals between them are random. The
random values are just governed by the inverse $\alpha$-stable subordinator.

Notice, by a slight change in the method we can easily obtain not only the CC law of relaxation. The approach has a wide potential that will be presented in the next subsection.

\subsection{Memory function}

If the distribution of a nonnegative stochastic process $U(\tau)$ is infinitely divisible, then its Laplace transform takes the form
\begin{equation}
\langle e^{-s\,U(\tau)}\rangle=e^{-\tau\Psi(s)}\,,\label{eqiv5}
\end{equation}
where $\Psi(s)$ is the Laplace  exponent according to the
L\'evy-Khintchine formula \cite{feller66} (called also as the L\'evy exponent). Many
examples of such distributions are well known \cite{john70}. Among them are such as Gaussian,
inverse Gaussian, $\alpha$-stable, tempered $\alpha$-stable,
exponential, gamma, compound Poisson, Pareto, Linnik,
Mittag-Leffler and others, including completely asymmetric $\alpha$-stable
distributions.  Let $f(\tau,t)$ be the pdf of
$U(\tau)$. Then the mean $\langle e^{-sU(\tau)}\rangle$ is the
Laplace transform of $f(\tau,t)$ equal to
\begin{displaymath}
\langle e^{-sU(\tau)}\rangle=\int_0^\infty e^{-st}\,f(\tau,t)\,dt\,.
\end{displaymath}
Knowing $f(\tau,t)$, it is not difficult to find the pdf
$g(t,\tau)$ of its inverse process. Based on the Laplace transform of $g(t,\tau)$ with respect to $t$ we come to
\begin{equation}
\tilde{g}(s,\tau)=\frac{\Psi(s)}{s}\,e^{-\tau\Psi(s)}\,.\label{eqiv6}
\end{equation}
It should be pointed out that the exponential term in the Laplace
image $\tilde{g}(s,\tau)$ allows one to simplify further
calculations by reducing them to algebraic transformations.

Let us now come back to the two-state system where interactions of entities with their environment are taken into account with help of the temporal subordination under an infinitely divisible processes mentioned above. We will analyze the time evolution of the number of entities in the states $\downarrow$ and $\uparrow$. They are parent
random processes in the same manner as in the case of the previous subsection. To account for
interaction of entities with the environment, we subordinate the
latter processes by another random process with a probability
density $g(t,\tau)$ according to the relations (\ref{eqi1a}).
In the steady state the system supports the relation
$m_\uparrow(\infty)=m_\downarrow(\infty)=1/2$ conventionally.
Then the relaxation function can be written as
%\begin{eqnarray}
%\phi(t)&=&\int^\infty_0\phi_{\rm D}(\tau)\,g(t,\tau)\,d\tau\nonumber\\
%&=&\int^\infty_0\exp(-2B\tau)\,g(t,\tau)\,d\tau.\label{eqiv6a}
%\end{eqnarray}
\begin{equation}
\phi(t)=\int^\infty_0\phi_{\rm D}(\tau)\,g(t,\tau)\,d\tau=\int^\infty_0\exp(-2B\tau)\,g(t,\tau)\,d\tau.\label{eqiv6a}
\end{equation}
Using the Laplace image $\tilde{g}(s,\tau)$ as
applied to infinitely divisible random processes (\ref{eqiv6}), we find the
frequency-domain response in a simple form
\begin{equation}
\phi^*(\omega)=\frac{1}{1+\Psi(i\omega)/(2B)}\,,\label{eqiv7}
\end{equation}
where $B$ is a constant transition rate. The latter equation clearly
shows that the relaxation behavior will be determined by the
Laplace exponent $\Psi(s)$.

In the subordination framework, relaxation of the two-state system is described by the following equations
%\begin{displaymath}
%\begin{cases}m_\uparrow(t)&= m_\uparrow(0)\\
%& + B\int^t_0M(t-\tau)\{m_\downarrow(\tau)-m_\uparrow(\tau)\}\,d\tau\,,\\ m_\downarrow(t) & = m_\downarrow(0)\\
%& + B\int^t_0M(t- \tau)\{m_\uparrow(\tau)-m_\downarrow(\tau)\}\,d\tau\,,\end{cases}%\label{eqiv8}
%\end{displaymath}
\begin{displaymath}
\begin{cases}m_\uparrow(t)= m_\uparrow(0) + B\int^t_0M(t-\tau)\{m_\downarrow(\tau)-m_\uparrow(\tau)\}\,d\tau\,,\\
m_\downarrow(t) = m_\downarrow(0) + B\int^t_0M(t- \tau)\{m_\uparrow(\tau)-m_\downarrow(\tau)\}\,d\tau\,,\end{cases}%\label{eqiv8}
\end{displaymath}
where the kernel $M(t)$ plays the role of a memory function. The time-dependent function $M(t)$ can be written as an inverse Laplace transform ${\cal L}_t^{-1}$, i.\,e.
\begin{equation}
M(t)=\frac{1}{2\pi i}\int^{c+i\infty}_{c-i\infty}\frac{e^{st}}{\Psi(s)}\,ds={\cal L}^{-1}_t\frac{1}{\Psi(s)}\,,\label{eqiv8a}
\end{equation}
where $c$ is large enough that $1/\Psi(s)$ is defined for ${\it Re}\
s\geq c$, and $i^2=-1$. Then the relaxation function satisfies
the following equation
\begin{equation}
\phi(t)=1-2B\int^t_0M(t-\tau)\,\phi(\tau)\,d\tau\,.\label{eqiv9}
\end{equation}
Using Eq.(\ref{eqiv8a}), we get
\begin{displaymath}
\phi(t)=1-2B\int^t_0{\cal L}^{-1}_\tau\,\frac{1}{2w+\Psi(s)}\,d\tau\,.
\end{displaymath}
The relaxation response $f(t)=-\frac{d\phi(t)}{dt}$ obeys another
integral equation
\begin{equation}
f(t)=2BM(t)-2B\int^t_0M(t-\tau)\,f(\tau)\,d\tau\,.\label{eqiv10}
\end{equation}
From the above formula it follows immediately an interesting relationship between $f(t)$
and $M(t)$:
\begin{displaymath}
\lim_{B\to 0}\frac{f(t)}{2B}=M(t)\,.
\end{displaymath}
Note that every infinitely divisible stochastic process is defined by
its Laplace (or L\'evy) exponent $\Psi(s)$. There exists one-to-one
connection between the exponent and the corresponding memory
function. Thus, the memory function $M(t)$ is the same important
characteristics for relaxing systems as their time-domain relaxation responses. However, it is not easy to detect $M(t)$ in experiments. The physical significance of the memory function consists in its asymptotic
properties. Power-law tails of memory functions in short or/and long times reflect directly similar properties in the asymptotic behavior of the corresponding relaxation responses.

\subsection{Memory function formalism for empirical laws of relaxation}

\begin{figure}
\centerline{\includegraphics[clip,width=1.\columnwidth]{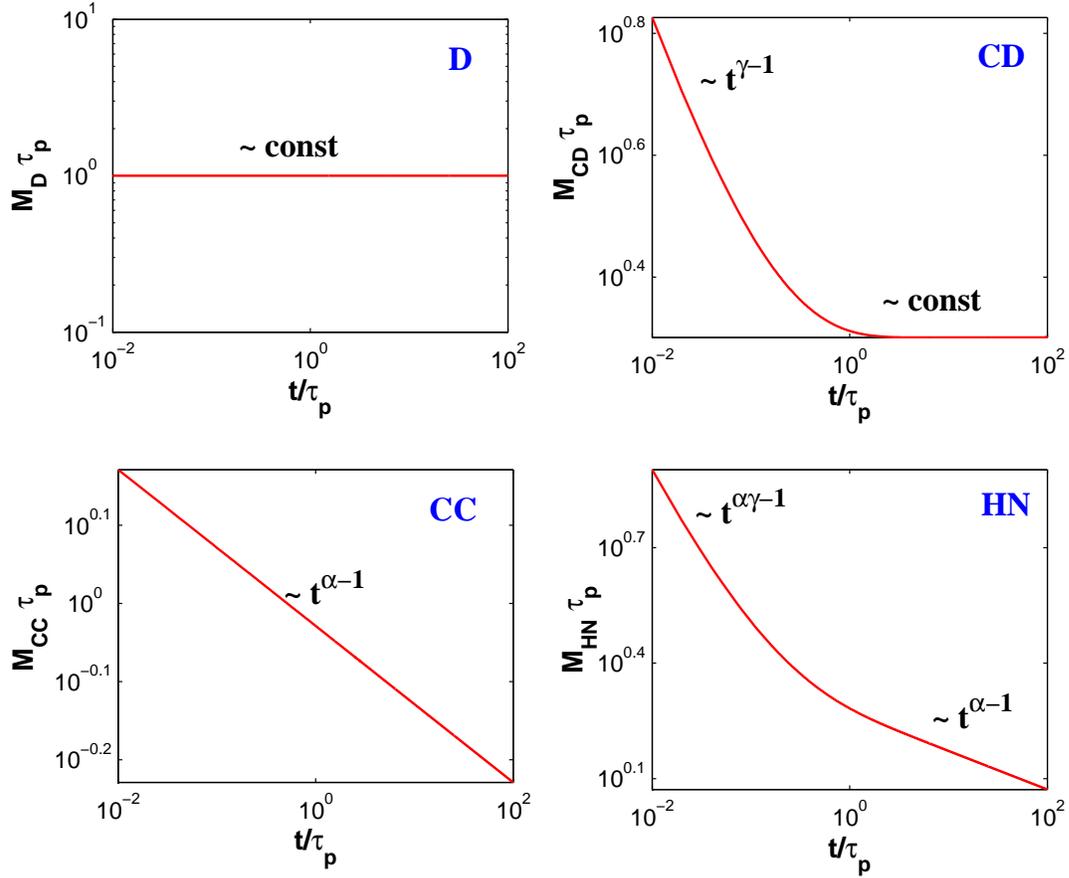}}
\caption{(Color on-line) Memory functions corresponding to the
empirical laws of relaxation ($\alpha$ = 0.9, $\gamma$ = 0.5).}
\label{memory}
\end{figure}

A significant amount of experimental data on relaxation of the disordered systems supports some types of empirical laws (it is about D, CD, CC and HN) \cite{jonscher83,jonscher96}.
According to \cite{sww15}, their memory functions take the following forms
\begin{eqnarray}
M_{\rm D\ }(t)&=&\tau_p^{-1}\,\theta(t)\,,\nonumber\\
M_{\rm CC}(t)&=&\tau_p^{-\alpha} t^{\alpha-1}\,\theta(t)/\Gamma(\alpha)\,,\nonumber\\
M_{\rm CD}(t)&=&e^{-\tau_p^{-1}t}\,\tau_p^{-\gamma}\,t^{\gamma-1}\,E_{\gamma,\gamma}
(\tau_p^{-\gamma} t^\gamma)\,\theta(t)\,,\nonumber
\end{eqnarray}
where $0<\alpha,\gamma\leq 1$, $\theta(t)$ denotes the Heaviside step function, and $E_{\alpha,\beta}(x)=\sum_{j=0}^\infty\,x^j/\Gamma(\alpha j+\beta)$, with $\alpha,\beta>0$, defines the well-known two-parameter Mittag-Leffler function \cite{erd55,gll02}. The memory function connected with the HN law takes a cumbersome form
%\begin{eqnarray}
%&&M_{\rm HN}(t)=\theta(t)\nonumber\\
%&&\times\,\sum^\infty_{j=0}\tau_p^{-\alpha\gamma(j+1)}\,t^{\alpha\gamma(j+1)-1}\,
%E^{\gamma(j+1)}_{\alpha,\alpha\gamma(j+1)}(-\tau_p^{-\alpha} t^\alpha)\,,\nonumber
%\end{eqnarray}
\begin{displaymath}
M_{\rm HN}(t)=\theta(t)\,\sum^\infty_{j=0}\tau_p^{-\alpha\gamma(j+1)}\,t^{\alpha\gamma(j+1)-1}\,
E^{\gamma(j+1)}_{\alpha,\alpha\gamma(j+1)}(-\tau_p^{-\alpha} t^\alpha)\,,
\end{displaymath}
expressed in terms of the three-parameter Mittag-Leffler function \cite{prab71,mh08,MatSaxHaub09}, which has the following Taylor series representation
\begin{displaymath}
E_{\alpha,\beta}^\rho(x)=\sum_{j = 0}^\infty\frac{\Gamma(\rho+j)
}{\Gamma(\rho)\Gamma(\alpha j+\beta)j!}\,x^j\,,\quad\alpha,\beta>0\,.
\end{displaymath}
The shapes of all
these (D, CC, CD and HN) memory functions are shown in
Figure~\ref{memory}. They clearly indicate that the memory function
$M_{\rm HN}(t)$ is the most general case leading to particular
cases $M_{\rm D\ }(t)$, $M_{\rm CC}(t)$ and $M_{\rm CD}(t)$ in
limit values of parameters ($\alpha\to$ 1 and/or $\gamma\to$ 1).
Really, based on relation $E^\mu_{1,\mu}(y)=e^y/\Gamma(\mu)$, the
HN memory function easily transforms into the CD
($\alpha=1$) and D ($\alpha=\gamma=1$) memory ones. When
$\gamma=1$, we obtain the CC memory function using the sum
$\sum^\infty_{j=0}\tau_p^{-\alpha j}\,t^{\alpha
j}\,E^{j+1}_{\alpha,\alpha(j+1)}(-\tau_p^{-\alpha}
t^\alpha)=1/\Gamma(\alpha)$.

%%%%%%%%%%%%%%
%%%%%%%%%%%%%%
\section{Anomalous diffusion approach. Stochastic processes}
%%%%%%%%%%%%%%
%%%%%%%%%%%%%%
Any relaxation process is seen as a change in time (growth or decay) of a macroscopic  physical magnitude  characteristic for the observed system (e.g.  polarization-depolarization of a dielectric material) is accompanied by diffusion of a corresponding physical variable. In particular,  the decay  or growth of polarization is accompanied by diffusion of dipole orientations termed as diffusion of an excitation mode. From the theoretical point of view, the models should be adequate to the dual description of the kinetic process leading however to the same results, i.e. to the empirical evidence. Taking into account the stochastic nature of physical mechanisms underlying  the observed  time evolution of the system, the models from one side should yield a rigorous definition of  the probability of changing or not the initial state of the system and, from the other side,  the models should yield a rigorous stochastic process representing  the underlying diffusion.  Only this dual description can give full information on anomalous dynamical properties of various systems.

In this context the non-exponential relaxation of complex systems can be modeled from the idea based on an excitation undergoing diffusion in the systems \cite{gy95,fy95}. Then the relaxation function $\phi(t)$ is determined by the temporal decay of a given mode $k$, and in the framework of the one-dimensional continuous time random walks (CTRWs) it is given by the following Fourier transform (supported on whole the real axis)
\begin{equation}
\phi(t)=\langle e^{ik\widetilde R(t)}\rangle\,,\label{eqv1}
\end{equation}
where $k>0$ has the physical sense of a wave number, and $\widetilde R(t)$ denotes the diffusion front (scaling limit of the CTRW) under consideration. If $\widetilde R(t)$ is supported on the positive half-line, its Fourier transform is replaced by the Laplace transform \cite{wk96}. Thus we obtain
\begin{equation}
\phi(t)=\langle e^{-k\widetilde R(t)}\rangle\,.\label{eqv1a}
\end{equation}
The details will be given in the next subsections.

\subsection{Uncoupled continuous time random walk. Diffusion. Cole-Cole relaxation}\label{subsiic}

The idea of the continuous time random walk (CTRW) was firstly proposed by E. W. Montroll and G. H. Weiss in 1965 \cite{mw65}. Although the term ``random walk'' was introduced by K. Pearson in 1905 \cite{pearson}, the formalism of simple random walks was known else in the XVII-th century. The random walk approach is based on assumption that step changes are made through equal time intervals. This was the first (and rough) approximation for modeling many various physical, chemical and economical phenomena \cite{volpe16}. The CTRW model went further to study a random waiting time among subsequent random jumps.

Briefly, the core of the CTRW methodology is the following. Consider
a sequence ${T_i}$, $i=1,2,\dots$ of non-negative
iid random variables which represent waiting-time intervals  between subsequent random jumps $R_i$
of a walker. The random time interval of $n$ jumps equals
\begin{equation}
T(n)=\sum_{i=1}^{n}T_i,\quad T(0)=0\,,\label{eqii4a}
\end{equation}
and the space position of the walker after the $n$ jumps is given by the sum
\begin{equation}
R(n)=\sum_{i=1}^{n}R_i,\quad R(0)=0\,,\label{eqii4b}
\end{equation}
where $R_i$ are real iid variables showing both the length and the direction of the {\it i}-th
jump. The random variables $R_1, R_2, \dots$ are assumed to be independent of $T_1,T_2,\dots$, although this is not obligatory. The variables $R_i$ may be multi-dimensional vectors. Without loss of generality, we consider the walks to be one-dimensional.

The random number $N_t$ of jumps, performed by a walker up to time $t>0$, can be determined
by the largest index $n$ for which the sum of $n$ interjump time intervals does not
exceed the observation time $t$, namely
\begin{equation}
N_t=\max\{n:\ T(n)\leq t\}\,.\label{eqii4c}
\end{equation}
The process $N_t$ is often called the counting process or otherwise as the renewal process.
The total distance attaining  by the walker after the $N_t$ jumps becomes then
\begin{equation}
R(N_t)=\sum_{i=1}^{N_t}R_i,\quad R(0)=0\,.\label{eqii5}
\end{equation}
The cumulative stochastic process (\ref{eqii5}) is just known as the CTRW.

Though the aforesaid walks consist only of discrete time and space steps, the random walk model can be generalized to ``continuous steps'' being, hence, closer to the physical reality. The label {\it continuous} indicates the fact that the index $t$ in the CTRW belongs to a continuous set $[0,\infty)$, but does not imply the continuity of the paths. Assuming that the interjump time intervals $T_i$ belong to the domain of attraction of a completely asymmetric $L\alpha S$ distribution with the index $0<\alpha<1$ ($\beta=1$), i.\,e.
\begin{equation}
{\rm Pr}(T_i\geq t)\sim\Bigg(\frac{t}{\tau_0}\Bigg)^{-\alpha}\quad{\rm as}
\quad t\to\infty\,,\quad\tau_0>0\,.\label{eqv7}
\end{equation}
Then the limit theorems \cite{bing71,ms04} yield the continuous limit of the random sum (\ref{eqii4a}), i.\,e.
\begin{equation}
c^{-1/\alpha}T(\lfloor c\tau\rfloor)\stackrel{d}{\rightarrow}U_\alpha(\tau)\quad{\rm as}\ \ c\to\infty\,,\label{eqv7a}
\end{equation}
where $c^{-1/\alpha}$ is the time-rescaling constant chosen appropriately,  $\lfloor x\rfloor$ denotes the integer part of $x$, and $U_\alpha(\tau)$ is a strictly increasing $L\alpha S$ process \cite{mw06b}. Let the jumps $R_i$ belong to the domain of attraction of a $L\eta S$ distribution $S_{\eta,\beta}(x), 0<\eta\leq 2,|\beta|\leq 1$ so that the continuous limit reads
\begin{displaymath}
c^{-1/\eta}R(\lfloor c\tau\rfloor)\stackrel{d}{\rightarrow}X_\eta(\tau)\quad{\rm as}\ \ c\to\infty\,,
\end{displaymath}
where $X_\eta(\tau)$ is a $L\eta S$ process known as the parent process. If $\eta=2$ ($\beta=0$), the parent process $X_2(\tau)$ becomes the classical Brownian motion. Both, the process $U_\alpha(\tau)$ and the process $X_\eta(\tau)$ are indexed by random operational (internal) time $\tau$.

As $T(n)$ is the random time interval elapsed among $n$ jumps, and $N_t$ is the number of jumps occurred up to  time $t>0$, they are connected with each other by the following formula
\begin{equation}
\left\{T(\lfloor x\rfloor)\leq t\right\}=\left\{N_t \geq x\right\}\,.\label{eqii5c}
\end{equation}
Using Eq.(\ref{eqv7a}), we get
\begin{equation}
c^{-\alpha}N_{ct}\stackrel{d}{\rightarrow}S_\alpha(t)=\inf\{\tau:U_\alpha(\tau) >t\}\,,\label{eqv2}
\end{equation}
where $S_\alpha(t)$ is the inverse $L\alpha S$ subordinator,  relating the internal and the observable times, and $c\to\infty$.  In fact, the continuous limit of the discrete counting process $N_t$ is the hitting time process $S_\alpha(t)=\inf\{\tau\!:\! U_\alpha(\tau)>t\}$ satisfying the relation $S_\alpha(U_\alpha(\tau)) =\tau$ (almost surely and almost everywhere). Therefore, this process can be treated as the inverse to $U_\alpha(\tau)$. Because the process $U_\alpha(\tau)$ is strictly increasing, the process $S_\alpha(t)$ is nondecreasing. The hitting time $S_\alpha(t)$ is called also a first passage time \cite{feller66}. For a fixed time it represents the first passage of the stochastic time evolution above this time level. Note, the random process $S_\alpha(t)$ just depends on the true time $t$, and any sample trajectory of the process $S_\alpha(t)$ can be only increasing. Both processes $X_\eta(\tau)$ and $U_\alpha(\tau)$ depend on $\tau$. To find the position of a walker at the real (true) time in this irregular motion, we derive the continuous limit of Eq.(\ref{eqii5}), i.\,e.
\begin{equation}
(c^\alpha)^{-1/\eta}R(\lfloor c^\alpha
S_\alpha(t)\rfloor)\stackrel{d}{\rightarrow}X_\eta(S_\alpha(t)) \label{eqv3},
\end{equation}
known as the anomalous diffusion process (diffusion front $\widetilde R(t)$) according to, for example, \cite{mbsb02}), directed by the inverse $L\alpha S$ subordinator $S_\alpha(t)$. In this issue the process $S_\alpha(t)$ plays the role of a new time clock (stochastic time arrow) in $X_\eta(\tau)$ instead of $\tau$. In the framework of this approach the stochastic process (\ref{eqv3}) opens perspectives to describe an arbitrary diffusion.

Let us discuss a relationship between the pdf of the position $r_t$ of a walking particle at real time $t$ and the probability densities of random processes $X_\eta(\tau)$ and $S_\alpha(t)$. As applied to $X_\eta(S_\alpha(t))$ defined in Eq.(\ref{eqv3}), we can consider it as a subordination of processes. If the processes $X_\eta(\tau)$ and $U_\alpha(\tau)$ are uncoupled (i.\,e.\ independent on each other), the pdf of $r_t$ with $t\geq 0$ can be written as a weighted integration over the internal time $\tau$ so that
\begin{equation}
p(x,t)= \int^\infty_0f(x,\tau)\,g(t,\tau)\,d\tau, \label{eqii6}
\end{equation}
where $f(x,\tau)$ denotes the pdf to find the parent process $X_\eta(\tau)$ at $x$ on operational time $\tau$ and $g(t,\tau)$ is the pdf to be at the operational time $\tau$ on real time $t$.
Then the Fourier transform $\widehat{p}(k,t)=\langle\exp(ikX_\eta[S_\alpha(t)])\rangle$ and the Laplace transform $\widetilde{p}(k,t)=\langle\exp(-kX_\eta[S_\alpha(t)])\rangle$ take the following forms
\begin{eqnarray}
\widehat{p}(k,t)&=&\int^\infty_0\widehat{f}(k,\tau)\,g(t,\tau)\,d\tau\,,\label{eqii6a}\\
\widetilde{p} (k,t)&=&\int^\infty_0 \widetilde{f}(k,\tau)\,g(t,\tau)\,d\tau\,.\label{eqii6b}
\end{eqnarray}
where $k>0$ is the wave number mentioned above. Consequently this allows one to determine the relaxation function from the probabilistic properties of diffusive processes (see Eqs.(\ref{eqv1}) and (\ref{eqv1a})).

The Laplace image for pdf of a  non-negative $L\alpha S$ variable is
\begin{equation}
\tilde{f}(u)=\exp\left(-\tau u^\alpha\right) \,\label{eqv4},
\end{equation}
where $0 < \alpha < 1$. If $f(\tau,t)$ is the pdf of $U_\alpha(\tau)$, then the pdf $g(t,\tau)$ of its inverse $S_\alpha(t)$ reads
\begin{displaymath}
g(t,\tau)=-\frac{\partial}{\partial\tau}\int_{-\infty}^tf(\tau,t')\,dt'.
\end{displaymath}
Taking the Laplace transform of $g(t,\tau)$ with respect to $t$, we get
\begin{equation}
\tilde{g}(u,\tau)=u^{\alpha-1}\,e^{-\tau u^\alpha}\,.\label{eqv5}
\end{equation}
It follows from Eq. (\ref{eqv5}) that the pdf of the inverse $L\alpha S$ process is
\begin{displaymath}
g(t,\tau)=\frac{1}{2\pi i}\int_{Br}e^{ut-\tau
u^\alpha}\,u^{\alpha-1}\,du=t^{-\alpha}F_\alpha(\tau/t^\alpha)\,,
\end{displaymath}
where $Br$ denotes the Bromwich path \cite{abr65}, and the function $F_\alpha(z)$ is given by the following Taylor series expansion
\begin{displaymath}
F_\alpha(z)=\sum_{j=0}^\infty\frac{(-z)^j}{j!\,
\Gamma(1-\alpha-j\alpha)}\,.
\end{displaymath}
For more details see \cite{main96,stan04}. In this framework, the relaxation function $\phi(t)$ that describes the temporal decay of a given mode $k$, can be expressed through the Fourier transform (\ref{eqv1}) of the diffusion process $X_\eta(S_\alpha(t))$. Consequently, this case leads to the CC relaxation function $\phi_{CC}(t)=E_\alpha\left[- (t/\tau_p)^\alpha\right]$, where $\tau_p \sim |k|^{-\eta/\alpha}$ \cite{stan03,mw06b}. The frequency-domain description of the CC relaxation takes the form (\ref{eqiv4}). If $\alpha=1$, it becomes the classical D relaxation.

\subsection{Coupling between the very large jumps in physical and operational times}

Let us now make use of the fact that the stochastic time evolution $U_\alpha(\tau)$ and its (left) inverse process $S_\alpha(t)$ permits one to underestimate or overestimate the time interval $T(N_t)$ of $N_t$ random steps performed up to the physical time $t$ at which the position of a walker is observed:
\begin{equation}
T(N_t)< t < T(N_t+1)\quad {\rm for}\quad t> 0\,,\label{eqv8}
\end{equation}
what follows directly from the definition (\ref{eqii4a}). In fact, the two processes $T(N_t)$ and $T(N_t+1)$ correspond to underestimating and overestimating the real time $t$ from the random time steps $T_i$ of the CTRWs.

In terminology of the Feller's book \cite{feller66} the variable
\begin{displaymath}
Z_t=T(N_t+1)-t
\end{displaymath}
is the residual waiting time (life-time) at the epoch $t$, and
\begin{displaymath}
Y_t=t-T(N_t)
\end{displaymath}
is the spent waiting time (age of the entity that is alive at time $t$). Importance of these random variables can be explained by one remarkable property. For $t\to\infty$ the variables $Y_t$ and $Z_t$ have a common, proper limiting distribution only if their probability distributions $F(y)$ and $F(z)$ have finite expectations. However, if the distribution $F(x)$ satisfies
\begin{displaymath}
1-F(x)=x^{-\alpha}L(x)\,,
\end{displaymath}
where $0<\alpha<1$ and $L(xt)/L(t)\to 1$ as $x\to\infty$, then according to \cite{dyn61}, the pdf of the normalized variable $Y_t/t$ is given by the generalized arcsine law
\begin{equation}
p_\alpha(x)=\frac{\sin(\pi\alpha)}{\pi}\,x^{-\alpha}(1-x)^{\alpha-1}\,,\label{eqv9}
\end{equation}
while $Z_t/t$ obeys
\begin{equation}
q_\alpha(x)=\frac{\sin(\pi\alpha)}{\pi}\,x^{-\alpha}(1+x)^{-1}\,.\label{eqv10}
\end{equation}
Since $\Sigma_{N_t}=t-Y_t$ and $\Sigma_{N_t+1}=Z_t+t$, the distributions of $\Sigma_{N_t}/t$ and $\Sigma_{N_t+1}/t$ can be obtained from Eqs. (\ref{eqv9}) and (\ref{eqv10}) by a simple change of variables $1-x=y$ and $1+x=z$, respectively.

In this case $T(N_t)/t$ tends in distribution in the long-time limit to random variable $Y$ with density
\begin{equation}
p^Y(x)=\frac{\sin(\pi\alpha)}{\pi}\,x^{\alpha-1}(1-x)^{-\alpha}\,,\quad
0<x<1 \label{eqv11}
\end{equation}
and $T(N_t+1)/t\stackrel{d}{\rightarrow}Z$ with the pdf equal to
\begin{equation}
p^Z(x)=\frac{\sin(\pi\alpha)}{\pi}\,x^{-1}(x-1)^{-\alpha}\,,\quad
x>1.\label{eqv12}
\end{equation}
Both pdfs, $p^Y(x)$ and $p^Z(x)$, are special cases of the well-known beta density. It should be noticed that the density $p^Y(x)$ concentrates near 0 and 1, whereas $p^Z(x)$ does near 1. Near 1 both densities tend to infinity. This means that in the long-time limit the most probable values for $T(N_t)$ occur near 0 and 1, while for $T(N_t+1)$ they tend to be situated near 1.

\begin{figure}
\centerline{\includegraphics[clip=,width=1.1\columnwidth]{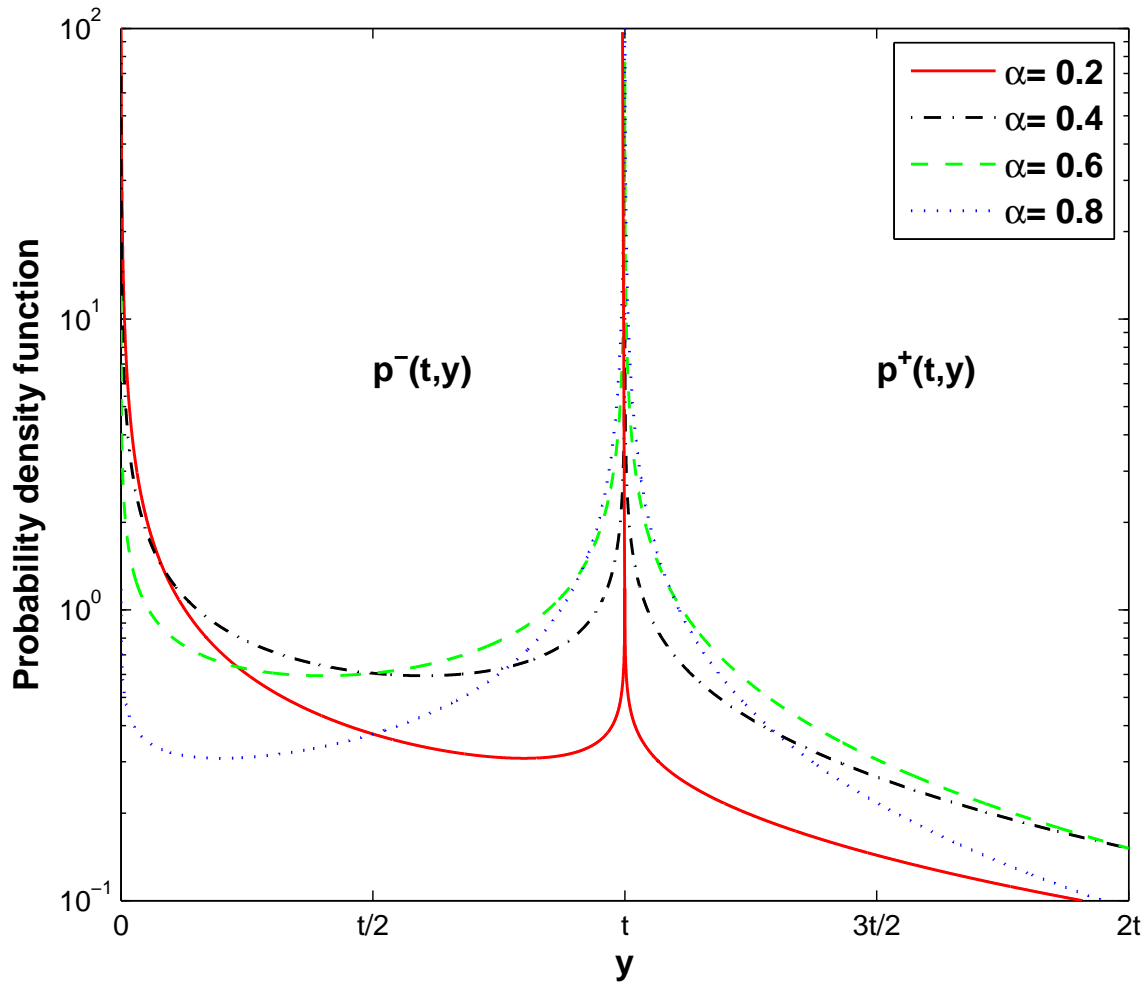}}
\caption{(Colour on-line). The pdf $p^-(y)$ with support on $0<y<t$
and the density $p^+(y)$ with support on $y>t$ for different values of the index $\alpha$.}
\label{dens}
\end{figure}

The nonequality (\ref{eqv8}) can also be represented in a schematic picture of time steps. Next,
a passage from the discrete process $T_i$ to the continuous one $U_\alpha(\tau)$ allows one to reformulate the unequality (\ref{eqv8}) as
\begin{equation}
U^-_\alpha(S_\alpha(t))< t < U_\alpha(S_\alpha(t))\quad {\rm for}\quad t>0\,,\label{eqv13}
\end{equation}
underestimating or overestimating the real time $t$. Note, the difference $U_\alpha(S_\alpha(t))-t$ is the leap-over processes \cite{ek04,klckm07}. The pdfs of $U^-_\alpha(S_\alpha(t))$ and
$U_\alpha(S_\alpha(t))$ take the forms, respectively,
\begin{eqnarray}
p^-(t,y)&=&
\frac{\sin\pi\alpha}{\pi}\,y^{\alpha-1}(t-y)^{-\alpha}\,,
0<y<t\,,\label{eqv14}\\
p^+(t,z)&=&\frac{\sin\pi\alpha}{\pi}\,z^{-1}\,t^\alpha
(z-t)^{-\alpha}\,, z>t\,,\label{eqv15}
\end{eqnarray}
valid for any time $t>0$ (see Figure~\ref{dens}). The moments of $U^-_\alpha(S_\alpha(t))$ and $U_\alpha(S_\alpha(t))$ can be calculated directly from the moments of $Y$ and $Z$ by using the following relations
\begin{displaymath}
U^-_\alpha(S_\alpha(t))\stackrel{d}{=}tY\quad{\rm and}\quad
U_\alpha(S_\alpha(t))\stackrel{d}{=}tZ\,.
\end{displaymath}
Thus, the process $U^-_\alpha(S_\alpha(t))$ has finite moments of any order, while $U_\alpha(S_\alpha(t))$ gives us even no finite the first moment. The process $U_\alpha(S_\alpha(t))>t$ is too long also in the limit formulation. Notice that $p^+(t,y)=y^{-2}p^-(t^{-1},y^{-1})$. In our construction of the compound subordinators $U^-_\alpha(S_\alpha(t))$ and $U_\alpha(S_\alpha(t))$ the processes $U_\alpha(\tau)$ and $S_\alpha(t)$ are clearly coupled.

\subsection{Anomalous diffusion with under- and overshooting subordination}

\begin{figure}
\centerline{\includegraphics[clip=,width=0.6\columnwidth]{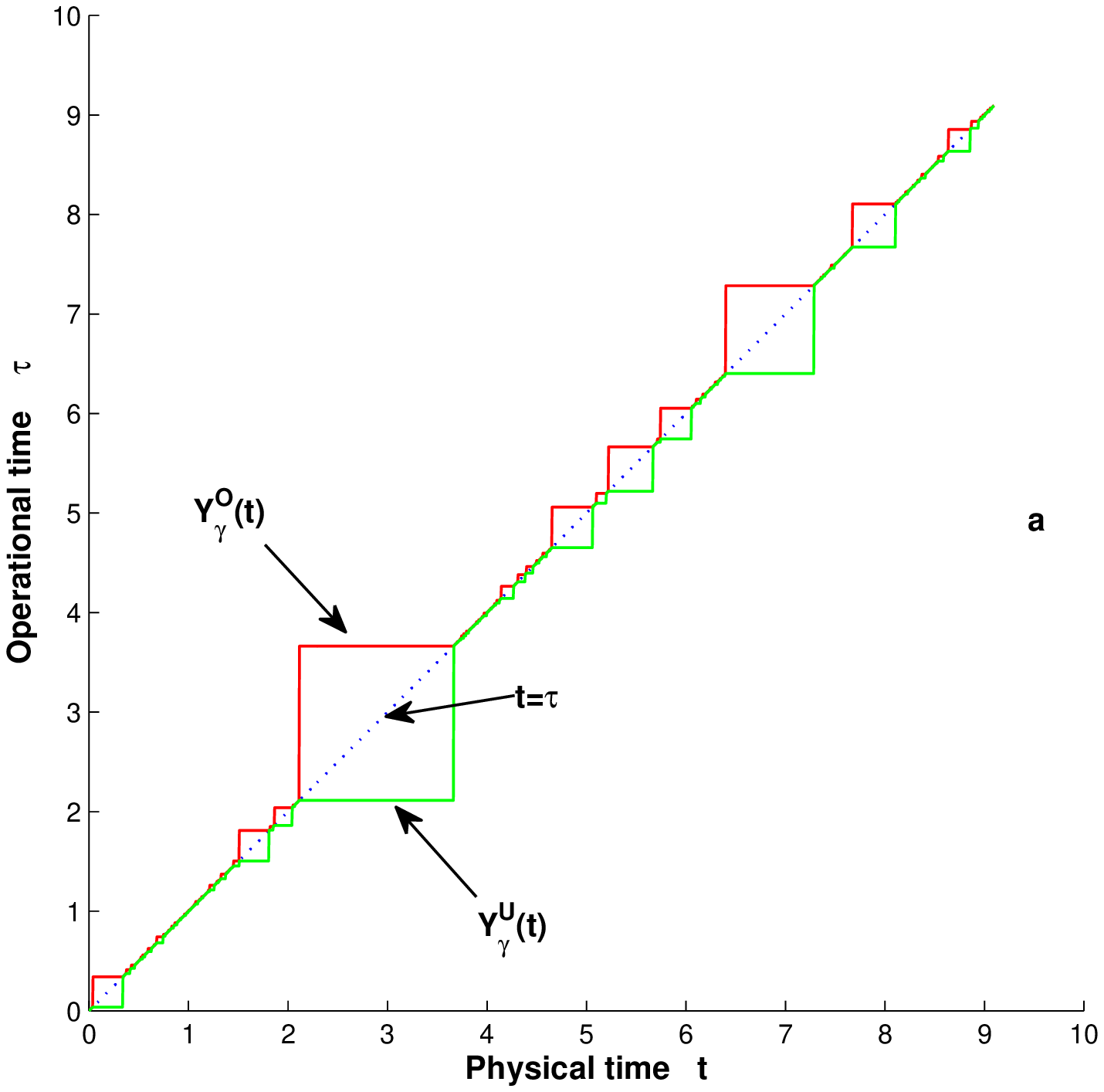}
\includegraphics[clip=,width=0.6\columnwidth]{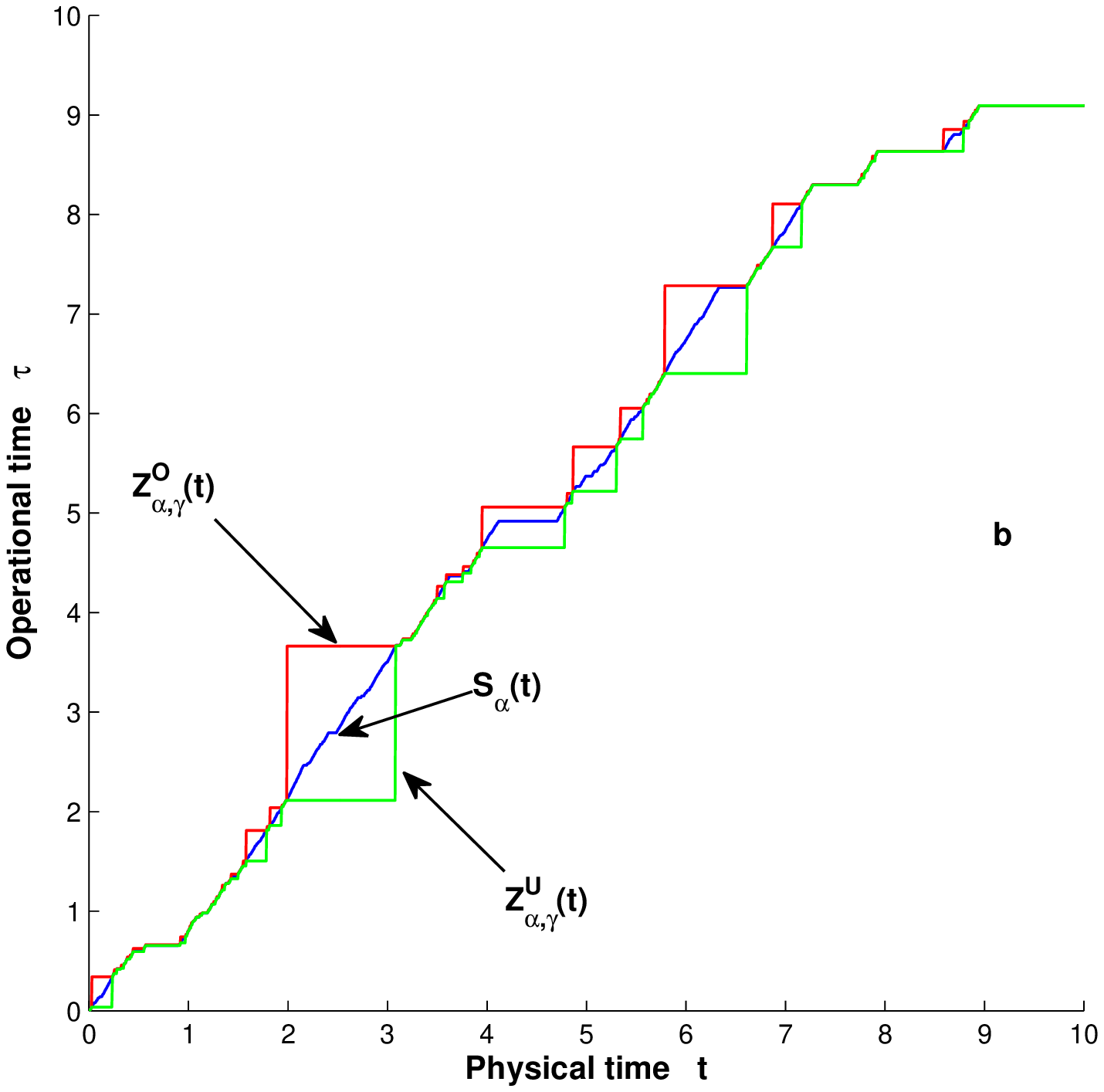}}
%\centerline{\includegraphics[clip=,width=0.5\columnwidth]{fig-oper1b}}
\caption{(Colour on-line). Patterns of operational times
$Y_\gamma(t)$, $S_\alpha(t)$ and $Z_{\alpha,\gamma}(t)=Y_\gamma[S_\alpha(t)]$ under
$\alpha$ = 0.9 and $\gamma$ = 0.8\,. Panel (a): $Y^U_\gamma(t)$ and $Y^O_\gamma(t)$.  Panel (b):
$Z^U_{\alpha,\gamma}(t)$ and $Z^O_{\alpha,\gamma}(t)$.  }
\label{oper-time}
\end{figure}

According to studies presented in \cite{wjmwt10}, the widely observed fractional two-power relaxation dependencies (\ref{eqi1}) and (\ref{eqi2}) are closely connected with the under- and overshooting subordination
\begin{displaymath}
Z^U_{\alpha,\gamma}(t)< S_\alpha(t) < Z^O_{\alpha,\gamma}(t)\quad
{\rm for}\quad t> 0\,,
\end{displaymath}
where
%\begin{eqnarray*}
%Z^U_{\alpha,\gamma}(t)&=&Y^U_\gamma[S_\alpha(t)]\,,\\
%Z^O_{\alpha,\gamma}(t)&=&Y^O_\gamma[S_\alpha(t)]\,,
%\end{eqnarray*}
\begin{displaymath}
Z^U_{\alpha,\gamma}(t)=Y^U_\gamma[S_\alpha(t)]\,,\quad Z^O_{\alpha,\gamma}(t)=Y^O_\gamma[S_\alpha(t)]\,,
\end{displaymath}
the processes $Y^U_\gamma(t)$ and $Y^O_\gamma(t)$ being nothing else as $U^-_\gamma(S_\gamma(t))$ and $U_\gamma(S_\gamma(t))$ with the index $\gamma$ (Figure~\ref{oper-time}a). Their main feature is that they are subordinated by an independent inverse $L\alpha S$ process $S_\alpha(t)$ forming the compound subordinators $Z^U_{\alpha,\gamma}(t)$ and $Z^O_{\alpha,\gamma}(t)$, respectively (Figure~\ref{oper-time}b). The approach enlarges the class of diffusive scenarios in the framework of the CTRWs. Assuming a heavy-tailed cluster-size distribution with the tail exponent $0<\gamma<1$, the coupling between jumps and interjump times tends to the compound operational times $Z^U_{\alpha,\gamma}(t)$ and $Z^O_{\alpha,\gamma}(t)$ as under- and overshooting subordinators, respectively.

The overshooting subordinator yields the anomalous diffusion scenario leading to the well-known HN relaxation pattern (\ref{eqi0a}), and the undershooting subordinator leads to a new relaxation law (\ref{eqvi8jws}). Derivation of both cases is presented below (see Section~\ref{parve}). These results are in agreement with the idea of a superposition of the classical (exponential) Debye relaxations \cite{bb78}. Let $X_\eta(\tau)$  with $0<\eta\leq2$ be the parent process that is subordinated either by $Z^U_{\alpha,\gamma}(t)$ or $Z^O_{\alpha,\gamma}(t)$. Then the subordination relation, expressed by means of a mixture of pdfs, takes the form
\begin{equation}
p^{\,r}(x,t)=\int_0^\infty\int_0^\infty
p^X(x,y)\,p^\pm(y,\tau)\,p^S(t,\tau)\,dy\,d\tau\,,\label{eqv18}
\end{equation}
where $p^{\,r}(x,t)$ is the pdf of the subordinated process $X_\eta[Z^U_{\alpha,\gamma}(t)]$ (or $X_\eta[Z^O_{\alpha,\gamma}(t)]$) with respect to the coordinate $x$ and time $t$, $p^X(x,\tau)$ the pdf of the parent process, $p^\pm(y,\tau)$ the pdf of $U^-_\gamma(S_\gamma(t))$ and $U_\gamma(S_\gamma(t))$ respectively, and $p^S(t,\tau)$ the pdf of $S_\alpha(t)$. Note, the processes $X_\eta[Z^U_{\alpha,\gamma}(t)]$  and $X_\eta[Z^O_{\alpha,\gamma}(t)]$ can be factorized in such a way, that if $X_2(\tau)$ is the Brownian process, then, e.\,g. (see Eq.(\ref{eqv3})), the diffusion front $X_2(S_\alpha(t))$ becomes a mixture of the Gaussian and completely asymmetric $L\alpha S$ laws. This issue will be considered in more details in Section~\ref{secvii} (see also \cite{wk96,wjmwt10}).

The subordinator $U_\gamma(S_\gamma(t))$ results in stretching of the real time $t$. It will underline scaling properties in short and long times, respectively. Namely, the interesting feature is observed in both CD and HN relaxations. Let us consider the process $U_\gamma(S_\gamma(t))$ as a subordinator indexed by $\gamma$,
i.\,e. the process is obtained from a $L\gamma S$ random process. Following the temporal decay (\ref{eqv1}) of a given mode $k$, the relaxation function takes the form
\begin{equation}
\phi_{\rm CD}(t)=\int^\infty_1e^{-z\,t/\tau_p}\,\frac{z^{-1}\,
(z-1)^{-\gamma}}{\Gamma(\gamma)\Gamma(1-\gamma)}\,dz\,.
\label{eqv19}
\end{equation}
Here the subscript CD is not by chance. It shows a direct
connection of the relaxation function with the CD law \cite{jonscher83}.
In fact, the one-sided Fourier transform (\ref{eqii8}) gives
\begin{displaymath}
\phi^*_{\rm CD}(\omega)= \frac{1}{[1+i\omega/\omega_p]^\gamma}\,,
\quad 0<\gamma\leq 1\,.
\end{displaymath}
It should be mentioned that the method of subordination suggests
also one more scenario leading to the CD relaxation. It is based
on the inverse tempered $L\alpha S$ process (see \cite{sw09} in more details) that will discuss below.

\subsection{Relaxation process from compound subordinators}\label{parve}

\begin{figure}
\centerline{\includegraphics[clip,width=1.1\columnwidth]{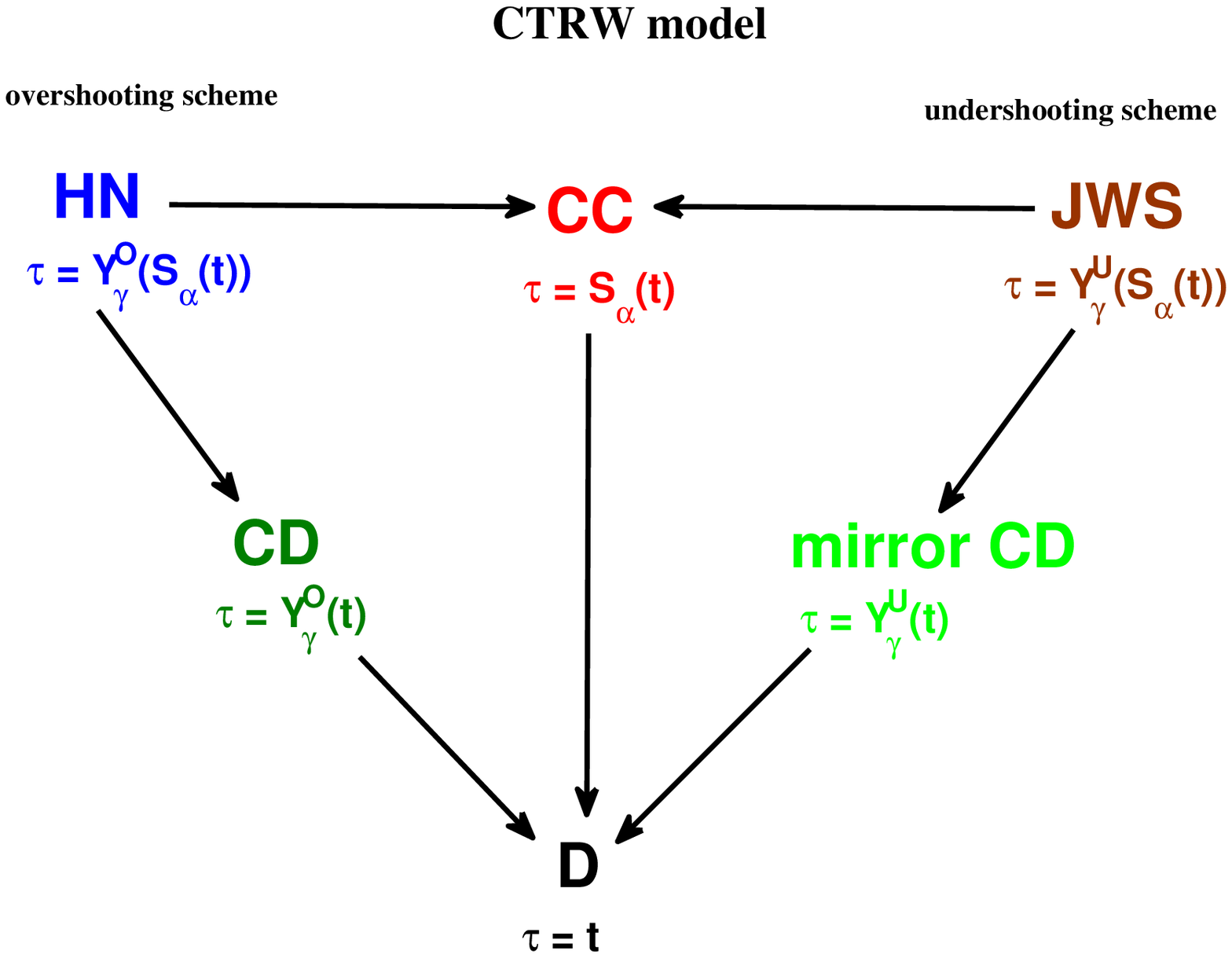}}
\caption{(Colour on-line). Operational times $\tau$ corresponding to experimental evidence. In this graph the indices
$\alpha$ and $\gamma$ correspond to parameters of subordinators in the anomalous diffusion model (see more details in the paper text).}
\label{time_scheme}
\end{figure}

As it has been already shown above, the CC and CD relaxations are only special cases of the more general HN law. To get that law, the operational time $S_\alpha(t)$ of the CC
diffusion mechanism has to be modified \cite{wjmwt10} by means
of coupling between jumps and interjump times in the underlying
 decay of a given mode, representing excitation undergoing
diffusion in the relaxing system. The relaxation response will be characterized by short-
and long-time power laws with different fractional exponents (as
in the HN case) only if the anomalous diffusion scenario is based
on a compound operational time. To construct such an operational
time, denote conveniently the discussed above processes $U^-_\gamma(S_\gamma(t))$ and $U_\gamma(S_\gamma(t))$ as
$Y^U_\gamma$ and $Y^O_\gamma$, respectively. They corresponds to
the under- and overshooting subordination scenarios
\cite{wjmwt10}. Next, we can write $Z^U_{\alpha,\gamma}(t)\leq
S_\alpha(t) \leq Z^O_{\alpha,\gamma}(t)$ for $t\geq 0$, where
$Z^U_{\alpha,\gamma}(t)=Y^U_\gamma[S_\alpha(t)]$,
$Z^O_{\alpha,\gamma}(t)=Y^O_\gamma[S_\alpha(t)]$. The overshooting
subordinator $Z^O_{\alpha,\gamma}(t)$ leads (stretching the operational time $S_\alpha(t)$)
to the HN relaxation in the form
\begin{equation}
\phi_{\rm HN}(t)=\int^\infty_1 E_\alpha\Big[-(t/\tau_p)^\alpha
z\Big]\,\frac{z^{-1}\,(z-1)^{-\gamma}}{\Gamma(\gamma)\Gamma(1-\gamma)}\,dz\,.
\label{eqv20}
\end{equation}
By direct calculations of the Fourier transformation (\ref{eqii8})
the frequency-domain shape function reads
\begin{displaymath}
\phi^*_{\rm HN}(\omega)= \frac{1}{[1+
(i\omega/\omega_p)^\alpha]^\gamma}\,,\quad 0<\alpha,\gamma\leq 1\,.
\end{displaymath}
The above approach demonstrates clearly a success in
probabilistic treatment of the observed relaxation laws (see Figure~\ref{time_scheme}).
Therefore, we continue our analysis as applied to the
undershooting (compressing $S_\alpha(t)$) subordinator
$Y^U_\gamma[S_\alpha(t)]$.

In this case we obtain the JWS relaxation function
\begin{equation}
\phi_{\rm JWS}(t)=\int^1_0 E_\alpha\Big[-(t/\tau_p)^\alpha
z\Big]\,\frac{z^{\gamma-1}\,(1-z)^{-\gamma}}{\Gamma(\gamma)\Gamma(1-\gamma)}\,dz\,,
\label{eqv21}
\end{equation}
which can be identified as a special case of the three-parameter Mittag-Leffler function
\cite{mh08,MatSaxHaub09}
\begin{displaymath}
E_{\alpha,\beta}^\gamma(x)=\sum_{j=0}^\infty
\frac{(\gamma,j)\,x^j}{\Gamma(j\alpha+\beta)j!}\,,\quad
\alpha,\beta>0\,,
\end{displaymath}
where $(\gamma,j)=\gamma(\gamma+1)(\gamma+2)\dots(\gamma+j-1)$ is
the Appell's symbol with $(\gamma,0)=1$, $\gamma\neq 0$. To avoid
any confusion, it should be mentioned that the two-parameter
Mittag-Leffler function $E_{\alpha,\beta}(x)$, more common in literature \cite{erd55},
is a special case of the function $E_{\alpha,\beta}^\gamma(x)$
with $\gamma=1$. From the series expansion of the simplest
Mittag-Lefller function $E_\alpha(x)$ it is easy to check by
direct calculations of Eq.(\ref{eqv21}) that
\begin{displaymath}
\phi_{\rm JWS}(t)=E_{\alpha,1}^\gamma\left[-(t/\tau_p)^\alpha\right]\,.
\end{displaymath}
This type of the relaxation function as an integral has been derived in the CTRW framework \cite{jw08,wjmwt10}, and the exact functional form has been established in \cite{swt10}. Using the relationship with the Mittag-Leffler function, we may now write the
kinetic equation for (\ref{eqv21}) in the pseudodifferential
equation form
\begin{displaymath}
\left(\frac{\partial^\alpha}{\partial t^\alpha}+
\tau_p^{-\alpha}\right)^\gamma\phi_{\rm JWS}(t)=\frac{t^{-\alpha\gamma}}{\Gamma(1-\alpha\gamma)}\,,
\end{displaymath}
where $\partial^\alpha/\partial t^\alpha$ is the Riemann-Liouville
fractional derivative \cite{Goren}, and $\phi_{\rm JWS}(0)=1$ the
initial condition. Taking the Fourier transform (\ref{eqii8}), we
get the shape function corresponding to (\ref{eqv21}) in the form
useful for fitting the atypical dielectric spectroscopy data
\begin{displaymath}
\phi^*_{\rm JWS}(\omega)=
1-\frac{1}{[1+(i\omega/\omega_p)^{-\alpha}]^\gamma}\,,\quad
0<\alpha,\gamma\leq 1\,. %\label{eqv22}
\end{displaymath}
If one multiplies both numerator and denominator of the fraction by $(i\omega/\omega_p)^{\alpha\gamma}$, this clearly leads to Eq.(\ref{eqvi8jws}).
This result points to the following relationship with the HN
relaxation function
\begin{displaymath}
\phi^*_{\rm JWS}(\omega)=1-(i\omega/\omega_p)^{\alpha\gamma}\phi^*_{\rm HN}(\omega)\,.
\end{displaymath}
For $\gamma\to 1$ the pdf of $Y^U_\gamma$ tends to
the Dirac $\delta$-function. It is easy to check $\lim_{\gamma\to
1}\phi_{\rm JWS}(t)=\phi_{\rm CC}(t)$, and the kinetic equation for
$\phi_{\rm JWS}(t)$ takes the mentioned above CC form (\ref{eqiii41cc}).

It is interesting to compare the power-law characteristics of the relaxation response for both overshooting and undershooting schemes of anomalous diffusion. For both scheme the exponents $n$ and $m$ fall in the range
$(0,1)$, and are defined by the subordinator parameters $0<\alpha<1$ and $0<\gamma<1$. The HN relaxation, resulting from the undershooting scheme, is characterized by the exponents $m=\alpha$ and
$m>1-n=\alpha\gamma$, and it fits the so-called typical dielectric spectroscopy data, while the JWS case, resulting from the overshooting scheme, demonstrates up-down with
$m=\alpha\gamma$ and $m<1-n=\alpha$ and it fits the atypical
two-power-law relaxation pattern which, as shown by the
experimental evidence (see Figure~\ref{diag-1}), cannot be neglected (see e.\,g.
\cite{jonscher83,jonscher96,hav94} and references therein). Such an atypical
behavior has been also observed by us in gallium (Ga)-doped
Cd$_{0.99}$Mn$_{0.01}$Te mixed crystals \cite{tjw10}, where sample frequency-domain data measured for
Cd$_{0.99}$Mn$_{0.01}$Te:Ga at 77K is fitted with the function
(\ref{eqvi8jws}). This material belongs to the semiconductor of group
II-VI possessing deep metastable recombination centers. Formation
of such centers in Cd$_{\rm 1-x}$Mn$_{\rm x}$Te:Ga results from
the bistability of Ga dopant which makes this mixed crystal as an
attractive material for holography and high-density data storage
(optical memories).

In the language of
subordinators we may conclude that the process $Y^U_\gamma(t)$ makes a
rescaling for small times, and the process $Y^O_\gamma(t)$ turns
on a similar rescaling for long times.
As for $0<\alpha,\gamma\leq 1$, in the case of HN relaxation the
declination of the imaginary susceptibility $\chi''(\omega)$ for
low frequencies will be greater than for high frequencies, whereas
the atypical relaxation shows an opposite relation (see Figure~\ref{HN-JWS}).

The original HN relaxation \cite{jonscher83,jonscher96} with exponents $0<\alpha,\gamma\leq 1$
satisfies $m\geq1-n$. Its modified version \cite{hav94}, proposed
to fit relaxation data with power-law exponents satisfying $m<1-n$,
assumes  $0<\alpha,\alpha\gamma\leq 1$. Unfortunately, the HN function
with $\gamma > 1$ cannot be derived within the framework of
diffusive relaxation mechanisms. Only for $\gamma\leq 1$ the origins
of the HN function can be found within the fractional Fokker-Planck
\cite{kcct04} and the CTRW \cite{wjmwt10} approaches.
The approach considered above includes all the data in the mathematically
unified approach.

\subsection{Clustered continuous time random walk. Compound counting process}

The CTRW process $R(t)$ determines the total distance reached by a
random walker until time $t$. It is characterized by a sequence of
iid spatio-temporal
random steps $(R_i,T_i), i\geq 1$. If we assume stochastic
independence  between jumps $R_i$ and waiting times $T_i$, we get
a decoupled random walk; otherwise we deal with a coupled CTRW.
The distance reached by the walker at time $t$ is given by the
following  sum
\begin{equation}
R(t)=\sum^{N_t}_{i=1}R_i\,,\label{eqvi14}
\end{equation}
where $N_t=\max\{n:\,\sum_{i=1}^n T_i\leq t\}$, counts the steps
performed up to $t>0$.

Theoretical studies of the relaxation phenomenon in the above
framework are based on the idea of an excitation undergoing
(anomalous, in general) diffusion in the system  under
consideration \cite{Metzler}. The relaxation function $\phi(t)$ is
then defined by the inverse Fourier or the Laplace transform (see Eqs.(\ref{eqv1}) and (\ref{eqv1a})) of
the diffusion front $\widetilde R (t)$ which is
\begin{displaymath}
\widetilde R (t)\stackrel{d}{=}\lim_{t_0\to0}\frac{R(t/t_0)}{f(t_0)}\,,
\end{displaymath}
where the dimensionless rescaling parameter $t_0\to 0$ and
$f(t_0)$ is appropriately chosen renormalization function. The
diffusion front $\widetilde R (t)$ approximates a position at time $t$
of the walker performing rescaled spatio-temporal steps
$(R_i/f(t_0),t_0 T_i)$. The characteristics of the
relaxation process are related to the properties of the diffusion
front resulting from assumptions imposed on the spatio-temporal
steps of the random walk. For example, the decoupled CTRW with
power-law waiting-time distributions (i.\,e., with the random
variables $T_i$ satisfying $\Pr(T_i\geq t)\sim(t/\tau_0)^{-\alpha}$
as $t\to\infty$ with some $0<\alpha<1$ and $\tau_0>0$) leads to the
CC relaxation (see Subsection~\ref{subsiic}). But the
frequency-domain CC relaxation with the corresponding
time-domain Mittag-Leffler pattern is only one of the cases
measured in various experiments with complex media, and derivation
of those more general patterns requires considering diffusion
scenarios based on a compound coupled CTRW representation. It
should be pointed out that the simple coupling of type $R_i\sim
T^p_i$ (with positive power exponent $p$) does not lead, however, behind the
CC relaxation \cite{wk96,kotul95}. In contrast, introducing a
dependence between the jumps and waiting times by a random
clustering procedure (being a stochastic generation of the well-known deterministic renormalization approaches) we can derive another relaxation
laws like the CD or HN ones
\cite{jw08,wjmwt10,swt10}. Below, we present the
clustered CTRW scenario which leads directly to the results discussed
in the preceding section.

Let $M_j$ be a sequence of iid positive integer-valued random
variables independent of the pairs $(R_i,T_i)$. Next, assume that
the jumps and  waiting times are assembled into clusters of random
sizes $M_1,M_2,\dots$. This assumption allows one to transform the
sequence of spatio-temporal steps $(R_i,T_i)$ into a new sequence
$(\overline{R}_j,\overline{T}_j)$ of random sums
\begin{eqnarray}
(\overline{R}_1,\overline{T}_1)&=&\sum_{i=1}^{M_1}(R_i,T_i)\,,\nonumber\\
(\overline{R}_j,\overline{T}_j)&=&\sum_{i=M_1+\dots+M_{j-1}+1}^{M_1+\dots+M_j}
(R_i,T_i)\,,\quad j\geq 2.\label{eqvi15}
\end{eqnarray}
Then the position $R^M(t)$ of the walker is determined by
$(\overline{R}_j,\overline{T}_j)$ and, in accordance with the general
formula (\ref{eqvi14}), it is given by
\begin{equation}
R^M(t)=\sum^{\bar\nu(t)}_{j=1}\overline{R}_j\,,\label{eqvi16}
\end{equation}
where $\bar\nu(t)=\max\{n:\,\sum_{j=1}^n\overline{T}_j\leq t\}$. The
dependence between the jumps $\overline{R}_j$ and the waiting times
$\overline{T}_j$ of the coupled CTRW process $R^M(t)$ is determined by
the distribution of the cluster sizes $M_j$.

As an example, let us consider the simple case of a random walk, when the waiting times are represented
by equal intervals in time, i.\,e. $T_i=\Delta t$. In this case we have
\begin{equation}
R(t)=\sum^{\lfloor t/\Delta t\rfloor}_{i=1}R_i\,,\label{eqvi17}
\end{equation}
where the step-clustering procedure (\ref{eqvi15}) yields then $\overline{T}_j=
M_j\Delta t$ for $j\geq 1$, and the coupled process $R^M(t)$
in  (\ref{eqvi16}) takes the following form
\begin{equation}
R^M(t)=\sum^{U^M({\bar\nu(t)})}_{i=1}R_i\,.\label{eqvi18}
\end{equation}
Here $U^M(\bar\nu(t))$ is a compound counting process obtained
from
\begin{displaymath}
U^M(n)=\sum_{j=1}^n \overline{T}_j/\Delta t =\sum_{j=1}^n M_j\,,
\end{displaymath}
and
%\begin{eqnarray*}
%\bar\nu(t)&=&\max\{n:\,\sum_{j=1}^n \overline{T}_j\leq
%t\}\\
%&=&\max\{n:\,U^M(n)\leq t/\Delta t\}\,.
%\end{eqnarray*}
\begin{displaymath}
\bar\nu(t)=\max\{n:\,\sum_{j=1}^n \overline{T}_j\leq
t\}=\max\{n:\,U^M(n)\leq t/\Delta t\}\,.
\end{displaymath}
Observe that formula (\ref{eqvi18}) is an analog of  (\ref{eqvi17}) with the
compound counting process $U^M(\bar\nu(t))$ substituting the
deterministic number $\lfloor t/\Delta t\rfloor$ of performed
jumps $R_i$. The counting process $U^M(\bar\nu(t))$ is always
less than $\lfloor t/\Delta t\rfloor$, and it is hence a special
case of the undershooting compound counting process \cite{wjmwt10}.
It is also a clear signature of  the spatio-temporal coupling
provided by the clustering procedure (\ref{eqvi15}).

The idea of compound counting processes in CTRW approach is not
new in physics.  The resulting CTRW processes were examined in the
context of the rareness hypothesis in the fractal-time random walk
models (see, e.\,g. \cite{Weissmann, Klafter}). In general, the
compound counting process cumulates random number of random
events. Physical situations where the relevance of this scheme
holds are numerous. For instance, they take into account the random
magnitude of claims' sequence in insurance risk theory \cite{jwz09}, the energy
release of individual earthquakes in geophysics or random
water inputs flowing into a dam in hydrology \cite{bwm00} where summing the
individual contributions yields the total amount of the studied
physical magnitude over certain time intervals.

\begin{figure}
\centerline{\includegraphics[clip,width=1.1\columnwidth]{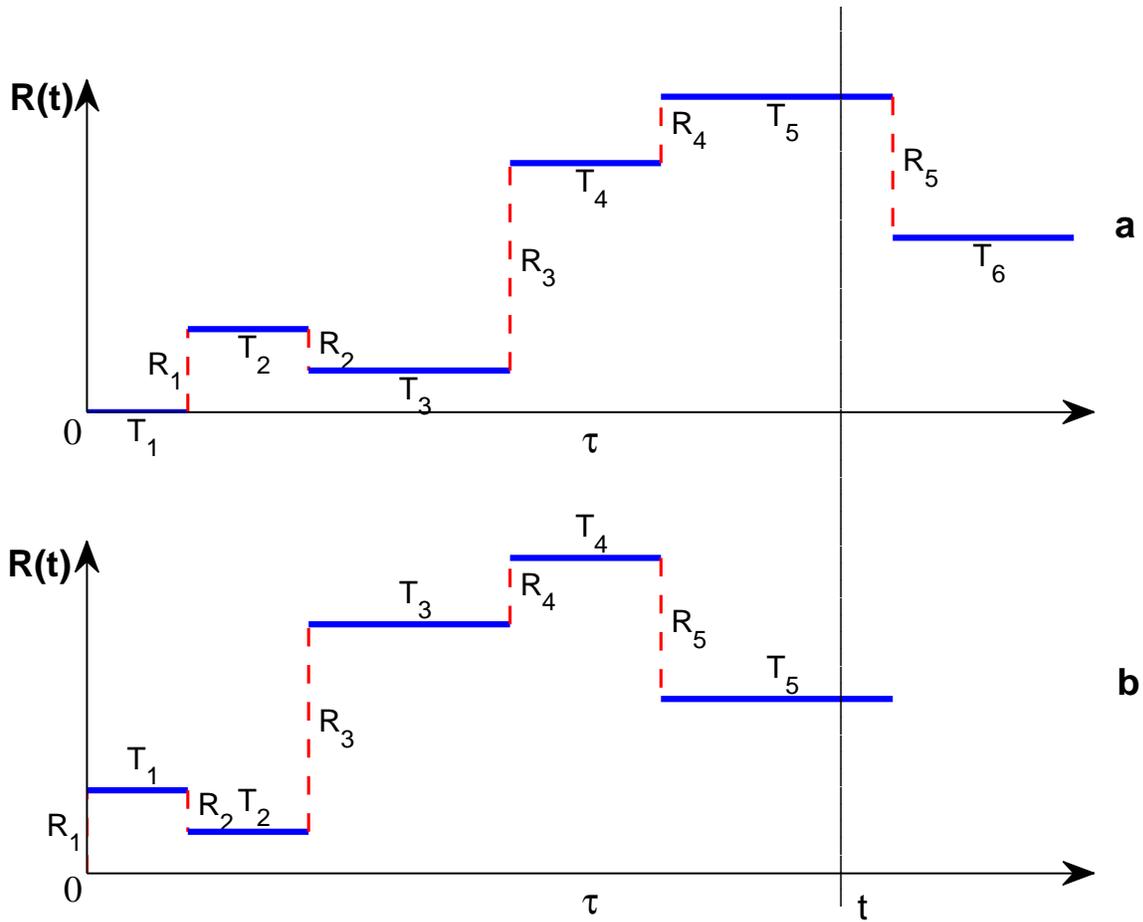}}
\caption{(Color on-line) Difference between (a) waiting-jump and (b) jump-waiting counting processes.}
\label{count-1}
\end{figure}

In the classical waiting-jump CTRW idea \cite{mw65}, in which
the jump $R_i$ occurs after the waiting-time $T_i$ (Figure~\ref{count-1}a), the random number of
the particle jumps performed by time $t>0$ is given by the renewal
counting process (\ref{eqii4c}). Then the location of a particle at
time $t$ is given by the random sum
\begin{equation}
R^-(t)=R(N_t)=\sum\limits_{i=1}^{N_t} R_{i}\,.\label{eqvi21}
\end{equation}
In the alternative jump-waiting CTRW scenario (Figure~\ref{count-1}b), the particle jump
$R_i$ precedes the waiting time $T_i$. Now the counting process $N_t+1$
gives the number of jumps by time $t$, and the particle location
at time $t$ is given by
\begin{equation}
R^+(t)=R(N_t+1)=\sum\limits_{i=1}^{N_t+1} R_{i}\,.\label{eqvi22}
\end{equation}
This is called the overshooting CTRW, or briefly OCTRW, see
\cite{jkms12}. In summary, the CTRW process $R^-(t)$ and the
OCTRW process $R^+(t)$ are obtained by subordination of the
random walk $R(n)$ to the renewal counting process $N_t$, and the
first passage process $N_t+1$, respectively. Let us mention that, in general, the subordination modifies a random process, replacing the deterministic time index by a random clock process, which usually represents a second source of uncertainty. When the jumps $R_i$ and the waiting
times $T_i$ are stochastically independent (i.\,e., uncoupled), the
CTRW and OCTRW diffusion limits and, as a consequence, the
corresponding types of relaxation, are the same.  On the other
hand, if the coupled case (i.\,e., dependent coordinates in the
random vector $(\overline{T}_i, \overline{R}_i)$) is considered, the waiting-jump and
jump-waiting schemes may lead to essentially different relaxation
patterns.

\begin{figure}
%\centerline{\includegraphics[clip,width=1.45\columnwidth]{fig_cluster_jump}}
\centerline{\includegraphics[clip,width=1.\columnwidth]{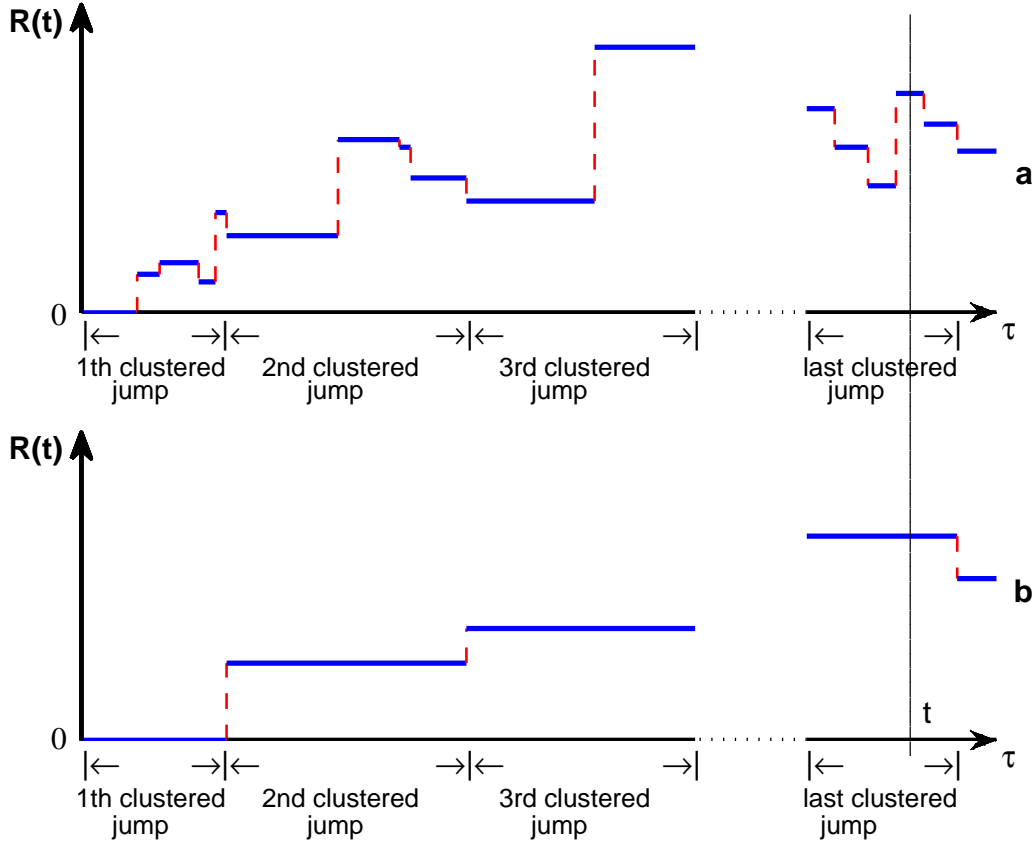}}
\caption{(Color on-line) Clustering-jump random walk: before (a) and after (b) clusterization.}
\label{cluster-1}
\end{figure}

An important and useful example of the coupled CTRW, different from the most popular L\'evy walk \cite{klafsok2011}, has been identified using the clustered CTRW concept, introduced in
\cite{wjm05} and developed further in \cite{jms11}. While in the L\'evy walk the jump size is fully determined by the waiting time $R_i\sim T^p_i$ (or equivalently, by flight duration),  in the
clustered CTRW coupling arises from random renormalization of the number
of jumps.  The clustered CTRW scheme is relevant to numerous
physical situations, including the mentioned above energy release of individual
earthquakes in geophysics, the accumulated claims in insurance
risk theory, and the random water inputs flowing into a reservoir
in hydrology \cite{Weissmann,Klafter,klafzum,huillet}. In all these cases,
summation of the individual contributions yields the total amount (in
general, random) of the studied physical magnitude over certain
time intervals.  In the clustered CTRW, the waiting time and the
subsequent jump are both random sums with the same random number
of summands, and as a consequence, this type of the CTRW is coupled,
even if the original CTRW before clustering had no dependence
between the corresponding waiting times and jumps (Figure~\ref{cluster-1}).  If the random
number of jumps in a cluster has a heavy-tailed distribution, then
the effect of clustering on the limiting distribution can be
profound.  In this case, the OCTRW jump-waiting scheme and the
traditional CTRW waiting-jump model are significantly different in
both their diffusion limits, and their governing equations
\cite{wjm05,jms11}.

Considering those cases crucial for modeling of the relaxation
phenomena, assume that the waiting times $T_i$ have a heavy-tailed
distribution with parameter $0<\alpha<1$; i.\,e., for some
$\tau_0>0$ we have
\begin{displaymath}
{\rm Pr}(T_i\geq t)\sim (t/\tau_0)^{-\alpha}
\end{displaymath}
for large $t$. Moreover, let the jump distribution be symmetric
and belong to the normal domain of attraction of a symmetric
$L\eta S$ law with the index of stability $\eta$, $0<\eta\leq 2$; i.\,e., for some $\rho_0>0$
we have
\begin{displaymath}
{\rm Pr}(|R_i|\geq x)\sim (x/\rho_0)^{-\eta}
\end{displaymath}
for large $x$ if $0<\eta < 2$ or $0<\langle {\rm D}^2
R_i\rangle=\rho_0^2<\infty$ if $\eta=2$. Then, a heavy-tailed
distribution of cluster sizes with tail exponent $0<\gamma<1$
yields different anomalous diffusion limits
\begin{equation}
\widetilde R^-(t)=C^{-1} X_\eta(Y^U_\gamma(S_{\alpha}(t/\tau_0)))\label{eqvi23}
\end{equation}
and
\begin{equation}
\widetilde R^+(t)=C^{-1}X_\eta(Y^O_\gamma(S_{\alpha}(t/\tau_0)))\label{eqvi24}
\end{equation}
of clustered CTRW and clustered OCTRW, respectively \cite{wjmwt10}
and \cite{jms11}. Here $C$ is a positive constant dependent on
the tail exponents $\alpha$, $\eta$ and proportional to the
scaling  parameter $\rho_0$. The parent process
$X_{\eta}(\tau)$ is a symmetric $L\eta S$ process
(recall, for $\eta=2$ it is just the standard Brownian
motion). The directing process $S_{\alpha}(t)$ is defined
as
\begin{displaymath}
S_{\alpha}(t)=\inf\{\tau\geq 0\,:\,
U_{\alpha}(\tau)>t\}\,,
\end{displaymath}
where $U_{\alpha}(t)$ is a strictly increasing
$L\alpha S$ process with the stability index $\alpha$.
The undershooting $Y_\gamma^U(t)$ and overshooting $Y_\gamma^O(t)$
processes have the same distributions as the processes $
U_{\alpha}^-(S_{\alpha}(t))$ and $U_{\alpha}(
S_{\alpha}(t))$, respectively, for the value of parameter $\alpha$
equal to $\gamma$.

For the clustered CTRW and OCTRW models, the anomalous diffusion
fronts $\widetilde R^\pm(t)$ in (\ref{eqvi23}) and
(\ref{eqvi24}) have frequency-domain shape
functions that can be identified respectively as the JWS and HN functions, see \cite{wjmwt10,swt10}. The characteristic material constant $\omega_p$,
appearing in both functions, takes the form
\begin{equation}
\omega_p=\frac{|Ck|^{\eta/\alpha}}{\tau_0}\,.\label{eqvi25}
\end{equation}
The diffusion fronts (\ref{eqvi23}) and (\ref{eqvi24})
allow us to develop analytical formulas (see Eq.(\ref{eqv1})) for the corresponding
time-domain relaxation functions $\phi_{\rm JWS}(t)$ and
$\phi_{\rm HN}(t)$  in terms of the three-parameter Mittag-Leffler
function \cite{mh08,MatSaxHaub09}.

Observe that
\begin{equation}
\langle e^{ik\widetilde R^{\pm}(t)}\rangle=\int_{-\infty}^{\infty}
e^{ikx}p^{\pm}(x,t)dx\,,\label{eqvi26}
\end{equation}
where $p^{\pm}(x,t)$ are pdfs of the anomalous diffusion
processes $\widetilde R^\pm(t)$. Hence, the Fourier-Laplace (FL)
images ${\cal I}_{\bf FL}(p^{\pm})(k,s)$ of the functions
$p^{\pm}(x,t)$ are just the Laplace images of the corresponding
time-domain JWS and HN relaxation functions. According to the
results in \cite{jms11,sw10,wsjms12}, we can derive
$p^{-}(x,t)$ as a mild solution (see about mild solutions of differential equations in \cite{bbm05} for more details) of the following fractional pseudo-differential equation
\begin{displaymath}
(C^{\eta}\partial_x^{\eta}+\tau_0^\alpha\partial_t^\alpha)^\gamma
p^{-}(x,t)=\delta(x)\frac{(t/\tau_0)^{-\alpha\gamma}}{\Gamma(1-\alpha\gamma)},
\end{displaymath}
where $\delta(x)$ is the Dirac delta function. Equivalently,
\begin{equation}
{\cal L}(\phi_{\rm JWS})(s)={\cal I}_{\bf
FL}(p^{-})(k,s)=\frac{s^{\alpha\gamma-1}}{(s^\alpha+|Ck|^\eta/\tau_0^\alpha)^\gamma}\,.\label{eqvi27}
\end{equation}
and hence
\begin{equation}
{\cal L}(\phi_{\rm
JWS})(s)=\frac{s^{\alpha\gamma-1}}{(s^\alpha+\omega_p^\alpha)^\gamma}\,\label{eqvi28}
\end{equation}
with $\omega_p^\alpha=|Ck|^\eta/\tau_0^\alpha$. Using the following
Laplace transformation \cite{mh08}
\begin{displaymath}
{\cal L}\left[t^{\beta-1}E^\gamma_{\alpha,\beta}(\pm\lambda
t^\alpha)\right]=\frac{s^{\alpha\gamma-\beta}}{(s^\alpha\mp\lambda)^\gamma}\,,\quad \alpha,\beta >0\,,
\end{displaymath}
we obtain
\begin{equation}
\phi_{\rm JWS}(t)=E^\gamma_{\alpha,1}\left[-
(t/\tau_p)^\alpha\right]\,.\label{eqvi29}
\end{equation}
Similarly, the pdf $p^{+}(x,t)$ is a mild solution of equation
\begin{eqnarray*}
\lefteqn{(C^{\eta}\partial_x^{\eta}+\tau_0^\alpha\partial_t^\alpha)^\gamma p^{+}(x,t)=}\\
&=\frac {\gamma}{C\Gamma(1-\gamma)}\int_0^\infty
u^{-\gamma-1}\,p^X(x/C,u)\,\int_0^{\tau_0^{\alpha} u}
p^S(t,\tau)\,d\tau\,du,&
\end{eqnarray*}
where $p^X(x,t)$ and $p^S(t,\tau)$ are the pdfs of $X_{\eta}(t)$ and $S_{\alpha}(t)$, respectively
\cite{jms11,sw10}. This implies that
\begin{equation}
{\cal I}_{\bf
FL}(p^{+})(k,s)=\frac{1}{s}\Bigg\{1-\left(\frac{|Ck|^\eta/\tau_0^\alpha}{s^\alpha+|Ck|^\eta/\tau_0^\alpha}\right)^\gamma\Bigg\}\,.\label{eqvi30}
\end{equation}
As a consequence, we have
\begin{equation}
{\cal L}(\phi_{\rm
HN})(s)=\frac{1}{s}\Bigg\{1-\left(\frac{\omega_p^\alpha}{s^\alpha+\omega_p^\alpha}\right)^\gamma\Bigg\}\,\label{eqvi31}
\end{equation}
and
\begin{equation}
\phi_{\rm HN}(t)=1-(t/\tau_p)^{\alpha\gamma}
E^\gamma_{\alpha,\alpha\gamma+1}\left[-
(t/\tau_p)^\alpha\right]\,.\label{eqvi32}
\end{equation}
The short- and long-time behaviors of these functions, which
exactly follow the high- and low-frequency power laws
(see Subsection~\ref{subsection_ib}), read
\begin{eqnarray*}
1-\phi_{\rm HN}(t)\sim\left(t/\tau_p\right)^{\alpha\gamma}/\Gamma(\alpha\gamma+1)\quad &{\rm for}&\quad t\ll\tau_p\,,\\
\phi_{\rm
HN}(t)\sim\gamma\left(t/\tau_p\right)^{-\alpha}/\Gamma(1-\alpha)\quad
&{\rm for}&\quad t\gg\tau_p\,,
\end{eqnarray*}
and
\begin{eqnarray*}
1-\phi_{\rm JWS}(t)\sim\gamma\left(t/\tau_p\right)^{\alpha}/\Gamma(\alpha+1)\quad &{\rm for}&\quad t\ll\tau_p\,,\\
\phi_{\rm
JWS}(t)\sim\left(t/\tau_p\right)^{-\alpha\gamma}/\Gamma(1-\alpha\gamma)\quad
&{\rm for}&\quad t\gg\tau_p\,.
\end{eqnarray*}
Thus, the clustered CTRWs and their diffusion limits illuminate the role of random
processes in the parametrization of relaxation phenomena. The
$L\eta S$ parent process $X_{\eta}(\tau)$, that models
particle jumps, does not change the exponents of the relaxation
power laws. It only affects the material constant $\tau_p$ by
determining the spatial features of the anomalous diffusion
$\widetilde R^{\pm}(t)$. The index $\alpha$ of the process
$S_{\alpha}(t)$ that codes the waiting times between jumps, and the
index $\gamma$ of the clustering process $Y_\gamma(t)$,
determine the power-law behavior of the relaxation function in
time and frequency. These coefficients characterize the complex
dynamics of relaxing systems.

%%%%%%%%%%%%%%
%%%%%%%%%%%%%%
\section{Extensions}
%%%%%%%%%%%%%%
%%%%%%%%%%%%%%

In real situations there are sufficiently many factors
truncating the distribution of $L\alpha S$ processes
\cite{houg86,mant94,kop95,boyar02,sww08}. This leads to the
tempered $L\alpha S$ processes which have all the moments.
As it is shown in \cite{sw09}, from the subordination by the
inverse tempered $L\alpha S$ process one can derive the
tempered diffusion equation and the relaxation function describing
the D, CC and CD types of
relaxation. The tempered diffusion has a transient character, i.\,e.
a crossover from subdiffusion at short times to normal
diffusion at long times. The transient subdiffusion has impact on
kinetics of magnetic bright points on the Sun \cite{cad99} and has
been observed in cells and cell membranes
\cite{plat02,wed09,sch11}. Physical arguments for
appearance of such effects are that subdiffusion is caused by
traps. In a finite system there is a given maximal depth of
the traps (maximal waiting time) truncating their power-law
waiting time distribution in such a way that beyond the maximal
waiting time the diffusive behavior of the complex system tends to
normal. Note also that the truncation of the waiting-time distribution
demonstrates features of weak ergodicity breaking in motion of
lipid granules \cite{jeon11}. The above mentioned relaxation
functions are only partial cases of the more universal HN law.

\subsection{Simple tempered relaxation}

The tempered $L\alpha S$ process \cite{psw05,ros07} is
characterized by the following Laplace image of its pdf
\begin{equation}
\tilde{f}_\alpha(u)=\exp\left(\delta^\alpha-(u+\delta)^\alpha\right)
\,,\label{eqvii1}
\end{equation}
where  the stability parameter $0<\alpha\leq1$ and the tempering
parameter $\delta\geq 0$ are constants. If $\delta$ equals to
zero, the tempered $L\alpha S$ process becomes simply
$L\alpha S$. In other words, the parameter $\delta$ provides
just a truncation of the ordinary, long-tailed totally skewed
$L\alpha S$ distribution. The truncation leads to the random
process having all moments finite. Formula (\ref{eqvii1}) describes
probabilistic properties of the tempered process in terms of  the
internal (operational) time. Its inverse process may be used as a
subordinator. The pdf $g_\alpha(\tau,t)$ of the subordinator depends on
the real physical time and describes the first passage over the
temporal limit $t$. Its Laplace transform reads
%\begin{eqnarray}
%&&\tilde{g}_\alpha(\tau,u)=-\frac{1}{u}\frac{\partial}
%{\partial\tau}\tilde{f}(u,\tau)=\nonumber\\
%&&\frac{(u+\delta)^\alpha-\delta^\alpha}{u}\,
%\exp\left(-\tau[(u+\delta)^\alpha-\delta^\alpha]\right)
%\,.\label{eqvii2}
%\end{eqnarray}
\begin{equation}
\tilde{g}_\alpha(\tau,u)=-\frac{1}{u}\frac{\partial}
{\partial\tau}\tilde{f}(u,\tau)=\frac{(u+\delta)^\alpha-\delta^\alpha}{u}\,
\exp\left(-\tau[(u+\delta)^\alpha-\delta^\alpha]\right)
\,.\label{eqvii2}
\end{equation}
The inverse tempered $L\alpha S$ process accounts for motion
alternating with stops so that the temporal intervals between them
are random and with heavy tails in density. The main feature of
the process is that it occurs only for short times \cite{sww08}.

Let the parent process $X(\tau)$ have the pdf $h(x,\tau)$. Then the pdf of the subordinated
process $X[S(t)]$ obeys the integral relationship between the pdfs of
the parent and directing processes, $X(\tau)$ and $S(t)$, respectively,
\begin{equation}
p(x,t)= \int^\infty_0h(x,\tau )\,g_\alpha(\tau,t)\,d\tau. \label{eqvii3}
\end{equation}
The pdf $p(x,t)$ has the most simple form in the Laplace space
\begin{equation}
\tilde{p}(x,u)= \frac{(u+\delta)^\alpha-\delta^\alpha}{u}\,
\tilde{h}(x,(u+\delta)^\alpha-\delta^\alpha). \label{eqvii4}
\end{equation}
For $\delta=0$ the above expression becomes equal to $u^{\alpha-1}\tilde{h}(x,u^\alpha)$ and hence corresponding to the pdf without tempering effects.

Let the ordinary Fokker-Planck equation
(FPE) $\partial h(x,\tau)/\partial\tau=\hat L(x)\,h(x,\tau)$ describe spatio-temporal evolution of a
particle subject to the operation time $\tau$. Acting with the operator $\hat L(x)$ on the Laplace
image $\tilde{p}(x,u)$ in (\ref{eqvii4}), we find
%\begin{eqnarray}
%\hat L(x)\,\tilde{p}(u,x)&=&[(u+\delta)^\alpha-\delta^\alpha]\,\tilde{p}(x,u
%)\nonumber\\ &-&q(x)\,\frac{[(u+\delta)^\alpha-\delta^\alpha]}{u}\,,\label{eqvii5}
%\end{eqnarray}
\begin{equation}
\hat L(x)\,\tilde{p}(u,x)=[(u+\delta)^\alpha-\delta^\alpha]\,\tilde{p}(x,u
)-q(x)\,\frac{[(u+\delta)^\alpha-\delta^\alpha]}{u}\,,\label{eqvii5}
\end{equation}
where $q(x)$ is an initial condition. Using the formal integral representation of the FPE
\begin{equation}
p(x,t)=q(x)+\int_0^td\tau\,M(t-\tau)\,\hat L(x)\,p(x,\tau)\,.\label{eqvii6}
\end{equation}
and taking the inverse Laplace transform of Eq.(\ref{eqvii5}), we obtain the explicit form
of the memory kernel $M(t)$ \cite{sww08}, namely
\begin{equation}
M(t)=e^{-\delta t}\,t^{\alpha-1}\,E_{\alpha,\,\alpha}(\delta^\alpha
t^\alpha)\,.\label{eqvii7}
\end{equation}
For $t\ll 1$ (or $\delta\to 0$) function (\ref{eqvii7}) takes the power form
$t^\alpha/\Gamma(\alpha)$. However, for  $t\gg 1$ (or
$\alpha\to 1$) $M(t)$ becomes constant and, as a result, Eq. (\ref{eqvii6}) transforms into
the integral form of the ordinary FPE.

To find the characteristics of the tempered diffusion, we consider  the Fourier transform of the diffusion process $X[S(t)]$:
\begin{equation}
\phi(t)=\left\langle e^{ikX[S(t)]}\right\rangle\,.\label{eqvii8}
\end{equation}
To expose the properties of the ``tempered relaxation'' function $\phi(t)$, we use the
frequency-domain representation (\ref{eqii8}).
Then, for the relaxation under the inverse tempered $L\alpha S$ process, the shape function
(\ref{eqii8}) takes the form
\begin{equation}
\phi^*(\omega)=\frac{1}{1-\sigma^\alpha+(i\omega/\omega_p+\sigma)^\alpha}\,,\label{eqvii10}
\end{equation}
where $0\leq\sigma\leq 1$ is a constant.

According to Eq. (\ref{eqvii10}), for $\sigma=0$ the shape function describes the
CC law. If $\alpha=1$, the function simplifies to the frequency-domain D formula. In the case of $\sigma=1$
it has the CD form. The relaxation directed by the inverse tempered
$L\alpha S$ process takes an intermediate place between the superslow relaxation and
the exponential one. Such a type of evolution is observed in
relaxation experiments (see, for example, \cite{jonscher96}).

\begin{figure}
\centerline{\includegraphics[clip,width=1.1\columnwidth]{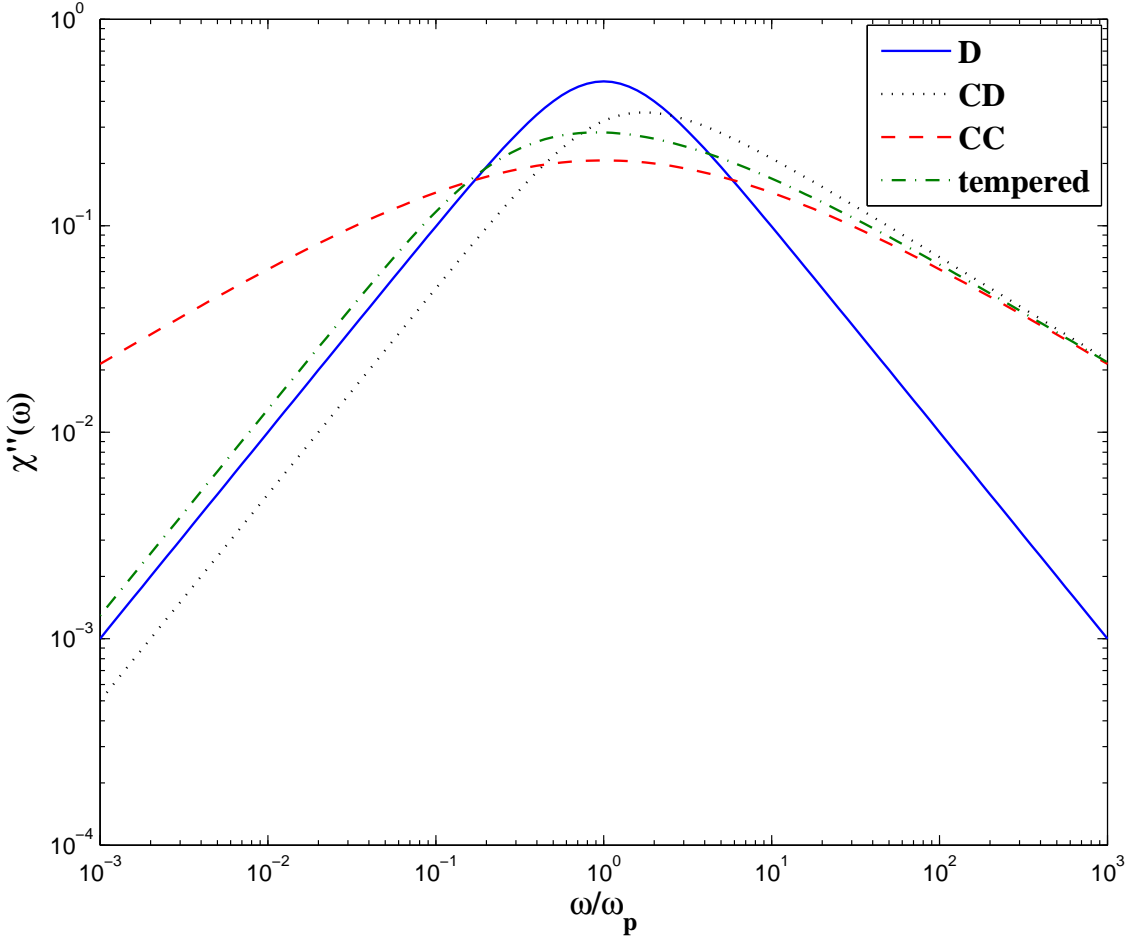}}
\caption{(Color on-line) Log-log plot of the imaginary term of the frequency-domain relaxation function
$\chi(\omega)=\chi'(\omega)-i\chi''(\omega)$ in the tempered case. Here D ($\sigma=0, \alpha=1$), CD ($\sigma=1, \alpha=0.5$), CC ($\sigma=0, \alpha=0.5$), tempered ($\sigma=0.15, \alpha=0.5$).} \label{simple_temp}
\end{figure}

It follows from experimental investigations \cite{jonscher83,jonscher96} that the complex dielectric susceptibility
$\chi(\omega)=\chi'(\omega)-i\chi''(\omega)$ of most dipolar substances demonstrates a
peak in the loss component $\chi''(\omega)$ at a characteristic frequency $\omega_p$, and it is characterized by high- ($\omega\gg\omega_p$) and low-frequency
($\omega\ll\omega_p$) dependencies (\ref{eqi1}) and (\ref{eqi2}), respectively. The tempered relaxation shows
\begin{eqnarray}
\chi_{\rm temp}'(\omega)&=&\frac{A+B\cos(C)}{A^2+2AB\cos(C)+B^2}\,,\nonumber\\ \chi_{\rm
temp}''(\omega)&=&\frac{B\sin(C)}{A^2+2AB\cos(C)+B^2}\,,\nonumber
\end{eqnarray}
where $A=1-\sigma^\alpha$, $B=(\sigma^2+\omega^2/\omega_p^2)^{\alpha/2}$ and
$C=\alpha\arctan(\omega/\delta)$. For small $\omega$ it is easy to see that
$\lim_{\omega\to 0}\chi_{\rm temp}'(\omega)\sim\omega$ and $\lim_{\omega\to 0}\chi_{\rm
temp}''(\omega)\sim 1$ whereas for large $\omega$ we get $\lim_{\omega\to\infty}\chi_{\rm
temp}'(\omega)\sim\omega^{-\alpha}$ and $\lim_{\omega\to\infty}\chi_{\rm
temp}''(\omega)\sim\omega^{-\alpha}$. This implies that
\begin{displaymath}
\lim_{\omega\to\infty}\frac{\chi_{\rm temp}''(\omega)}{\chi_{\rm
temp}'(\omega)}=\tan\Big(\frac{\alpha\pi}{2}\Big) =\cot\Big(n\frac{\pi}{2}\Big)\,,
\end{displaymath}
where $n=1-\alpha$, that is in agreement with ``energy criterion'' hypothesis formulated on the basis of experimental results \cite{jonscher83,jonscher96}.
However, for small $\omega$ we come to
\begin{displaymath}
\lim_{\omega\to 0}\frac{\chi_{\rm temp}''(\omega)}{\chi_{\rm temp}'(0)-\chi_{\rm
temp}'(\omega)}=\infty\,.
\end{displaymath}
This means that the energy lost per cycle does not have a constant relationship to the
extra energy that can be stored by a static field. Such an asymptotic
behavior suggests (Figure~\ref{simple_temp}) that the tempered relaxation takes an
intermediate place between the D, CC and CD types of relaxation.

\subsection{Tempered relaxation with clustering patterns}

In real physical systems, under influence of the surroundings, the
couplings between relaxing entities may differ from ``place to place''.
To describe this experimental result, we use two effects in the
compound subordination of random processes, namely tempering and
coupling, suggested in \cite{sw11}. Tempering of the random
subordinator influences on short-time evolution in such a
system, and coupling gives an appropriate long-term trend. It
should be also noticed that in this scheme the random processes,
applied for the subordinator construction, have finite integer
moments. Then the relaxation pattern under the tempering and the
coupling in the frequency domain reads
\begin{equation}
\phi^*(\omega)=
1-\Bigg(\frac{(i\omega/\omega_p+\sigma)^\alpha
-\sigma^\alpha}{1-\sigma^\alpha+(i\omega/\omega_p+\sigma)^\alpha}\Bigg)^\gamma\,,
\label{eqvii11}
\end{equation}
where the stability parameter $0<\alpha\leq1$, the coupling
parameter $0<\gamma\leq1$ and the tempering parameter $\sigma\geq
0$ are constants. In this case the asymptotic behavior of the
susceptibility $\chi(\omega)$ has different (independent) low- and
high-frequency power tails:
\begin{eqnarray}
&\chi_0-\chi(\infty)=\Delta\chi(\omega)\propto\left(i\omega/\omega_p
\right)^\gamma&\qquad{\rm
for}\quad \omega\ll\omega_p\,,\sigma\neq 0,\nonumber\\
&\chi(\omega)\propto\left(i\omega/\omega_p\right)^{-\alpha}&\qquad{\rm
for}\quad \omega\gg\omega_p\,. \label{eqvii12}
\end{eqnarray}
In the sense of arbitrary exponents $0<m=\gamma\leq 1$ and $0<(1-n)=\alpha\leq 1$ the
analysis of the above-mentioned relaxation pattern shows that the
same material (for example, Nylon 610 and Glycerol) can exhibit both
the $m<1-n$ and $m>1-n$ relations under different
temperature/pressure conditions, and the change between $m>1-n$
and $m<1-n$ is also observed in Polyvinylidene fluoride for
different susceptibility peaks (see Table~5.1 in \cite{jonscher83}). The form (\ref{eqvii12}) allows one to
fit the whole range of the two-power-law spectroscopy data with
independent low- and high-frequency fractional exponents being
free parameters in the model.

It is also of interest the time-domain description for the relaxation function given by (\ref{eqvii11}).
It takes the form
\begin{equation}
\phi(t)=1-\omega^\alpha_p\int_0^te^{-\sigma\tau_p^{-1}\tau}\,
x^{\alpha-1}\,G_{\alpha,\gamma}(\tau,z)\,d\tau\,, \label{eqvii13}
\end{equation}
where $G_{\alpha,\gamma}(\tau,z)$ is the Dirichlet average of the two-parameter
Mittag-Leffler function $E_{\alpha,\beta}(x)$ with $\alpha,\beta>0$ \cite{mh08},
namely
\begin{displaymath}
G_{\alpha,\gamma}(\tau,z)=
\int_0^1E_{\alpha,\alpha}[(\sigma^\alpha-
z)\tau^{-\alpha}_p\tau^\alpha]\frac{z^\gamma(1-z)^{-\gamma}\,dz}{\Gamma(\gamma)\Gamma(1-\gamma)}\,.
\end{displaymath}
Recall that many special functions of mathematical physics are
expressed in terms of an average \cite{carlson}. On the other
hand, it is useful to remark that Eq.(\ref{eqvii13}) is very similar
to the relaxation function derived in \cite{sw09}, if one accepts
$E_{\alpha,\alpha}[(\delta^\alpha-\tau^{-\alpha}_p)\tau^\alpha]$
instead of $G_{\alpha,\gamma}(\tau,z)$. This is clear because the
relaxation function of \cite{sw09} is a particular case of
Eq.(\ref{eqvii13}) with $\gamma=1$. In this connection, it should be
pointed out that for $\gamma=1$ the Dirichlet average kernel
transforms into the Dirac delta-function.

\begin{table}
\caption{Relation between the parameters $\alpha$, $\gamma$
and $\sigma$ in Eq.(\ref{eqvii11}) leading to particular forms of
relaxation fitting functions.\\ } \label{tab1} \centering
\begin{tabular}{ccccc}
\hline\noalign{\smallskip}
$0<\alpha\leq1$ & $0<\gamma\leq1$ & $\sigma\geq 0$ & Type & Ref. \\
\noalign{\smallskip}\hline\noalign{\smallskip}
any & any & any & given by (\ref{eqvii11}) & \cite{sw12} \\
1 & 1 & any & D & \cite{debye13} \\
any & 1 & 0 & CC & \cite{cole41,cole42}\\
any & 1 & 1 & CD & \cite{david50,david51}\\
any & 1 & any & pseudo CD & \cite{sww08}\\
any & any & 0 & JWS & \cite{swt10} \\
any & $\alpha$ & any & pseudo CC & -- \\
1 & any & any & mirror CD & -- \\
\noalign{\smallskip}\hline
\end{tabular}
\end{table}

Now we consider what possibilities for fitting of the experimental
data gives the shape function (\ref{eqvii11}).
Table~\ref{tab1} just serves for this purpose. Firstly, for the
corresponding values of the parameters $\alpha$, $\gamma$ and
$\sigma$, the relaxation function (\ref{eqvii11}) describes the D, CC and
CD types of relaxation. Taking $\gamma=1$, we come
to the pseudo CD relaxation derived in \cite{sww08}.
Next, if $\sigma$ equals to zero, then  Eq. (\ref{eqvii11}) takes the
form (\ref{eqvi8jws}), termed as JWS in \cite{tmk11}. The special place is accepted by two
remained cases called the pseudo CC and the mirror CD. They look
like the conventional CC and CD types of relaxation, respectively,
but there are also differences which we discuss below.

\begin{figure}
\centerline{\includegraphics[clip,width=1.15\columnwidth]{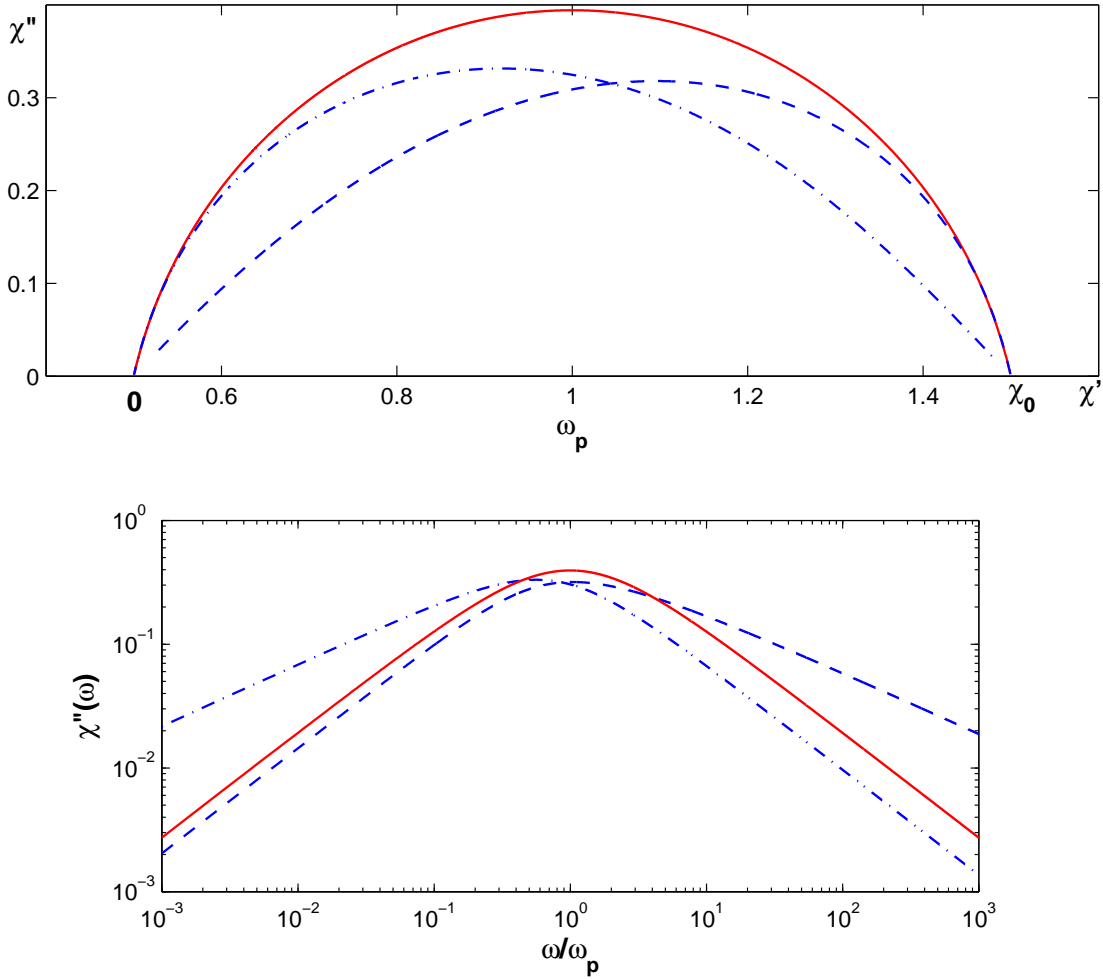}}
\caption{(Color on-line) Upper picture demonstrates the Cole-Cole
plot. Solid line shows the CC pattern with $\alpha=0.5$, dash line
shows the general tempered relaxation pattern (see Eq.~(\ref{eqvii11})
and Table~\ref{tab1}) with $\alpha=0.5$, $\gamma=0.85$ and
$\sigma=0.5$ whereas dot-dash line parameters take values of
$\alpha=0.85$, $\gamma=0.5$ and $\sigma=0.5$. Bottom picture
indicates the corresponding loss term of the susceptibility
$\chi(\omega)$ in dependence of frequency.} \label{gt-1}
\end{figure}

The most general case of the relaxation fitting function
Eq.(\ref{eqvii11}) assumes arbitrary values of its parameters in the
framework of their boundaries $0<\alpha\leq1$, $0<\gamma\leq1$ and
$\sigma\geq 0$. What features does it demonstrate? The loss curves
given by this case in a log-log representation will be asymmetric
with respect to the loss peak frequency (material constant)
$\omega_p$, but they will be more flattened than the true CD curve
and its mirror image (reflection) with a comparable asymmetry. The
same is observed for the CC plot (Figure~\ref{gt-1}). The function (\ref{eqvii11}) makes it possible to give an expression of $\chi(\omega)$ corresponding with a CC plot \cite{bb78}
making arbitrary angles $\pi\gamma/2$ and $\pi\alpha/2$ with the
$\chi'(\omega)$ axis at the low-frequency and high-frequency
ends, respectively (see more details in \cite{sw12}). Under $\gamma=1$ and $\sigma=1$ the expression (\ref{eqvii11}) reduces to the CD case, and for $\alpha=1$ we arrive at the mirror CD picture. The classical D
relaxation time appears for $\alpha=\gamma=1$. As it is shown
in \cite{bb78}, the shape of the CC plot is wholly determined by
the intersection angles with the $\chi'(\omega)$ axis, and
they often are obtained from experimental data of relaxation.

To sum up, the relaxation pattern given by Eq.(\ref{eqvii11})
satisfies the ``energy criterion'' \cite{jonscher83} indicating
origins of scaling properties in the evolution of complex
stochastic systems, namely $\lim_{\omega\to
0}\chi''(\omega)/(\chi'(0)-\chi'(\omega))=\tan(m\pi/2)$ and
$\lim_{\omega\to\infty}\chi''(\omega)/\chi'(\omega)=\cot(n\pi/2)$.
This means that the energy lost per cycle has a constant
relationship to the extra energy that can be stored by a static
field. However, the D response is behind this property.
Interestingly, the CD response and its mirror reflection
support the criterion only either on high frequencies or on low
ones, respectively, as their behavior on either short or long
times is similar to the D response.

%%%%%%%%%%%%%%
%%%%%%%%%%%%%%
\section{Subordination impact in understanding of relaxation scenarios}\label{secvii}
%%%%%%%%%%%%%%
%%%%%%%%%%%%%%

The subordination approach takes a central place in the theory of relaxation based on the stochastic point of view. The main feature of all the dynamical processes in complex systems is their random background. Subordination describes a transformation of one random process to another where the complex system under consideration is just such a converter.

Consider the kinetic equation for a general type of
Markovian processes in the form
\begin{equation}
\frac{dp\,(t)}{dt}=\hat{\bf W}p\,(t)\,,\label{eqviii1}
\end{equation}
where $\hat{\bf W}$ denotes the transition rate operator. Depending on the context \cite{chou11}, the operator
$\hat{\bf W}$ can be a differential operator with respect to space
coordinates in the case of diffusion or a matrix of transition
rates for relaxation phenomena. This equation defines the
probability $p(t)$ for the system evolution from one state into
others. Next, we determine a new process governed by an inverse
infinitely divisible process with the Laplace image (\ref{eqiv5}), namely
\begin{displaymath}
p_\Psi(t)=\int_0^\infty g(t,\tau)\,p(\tau)\,d\tau\,.
\end{displaymath}
The Laplace transform $\tilde p_\Psi(s)$ reads
\begin{displaymath}
\tilde p_\Psi(s)=\int_0^\infty e^{-st}\,p_\Psi(t)\,dt
\end{displaymath}
and leads to
\begin{eqnarray}
\hat{\bf W}\,\tilde
p_\Psi(s)&=&\frac{\Psi(s)}{s}\,\hat{\bf
W}\,\tilde p\,(\Psi(s))\nonumber\\
&=&\frac{\Psi(s)}{s}\,\left\{\Psi(s)\,\tilde p\,(\Psi(s))-p\,(0)\right\}\nonumber\\
&=&\Psi(s)\,\tilde
p_\Psi(s)-\frac{\Psi(s)}{s}\,p\,(0)\,.\label{eqviii2}
\end{eqnarray}
From this it follows
\begin{equation}
p_\Psi(t)=p\,(0)+\int^t_0d\tau\,M(t-\tau)\,\hat{\bf W}\,p_\Psi(\tau)\,,\label{eqviii3 8}
\end{equation}
where the kernel $M(t)$ is the memory function defined earlier by formula (\ref{eqiv8a}). This analysis shows resemblance between the diffusion scenarios and the scheme of two-state relaxation.

To fix the spatial structure of clusters in time, it is useful to
pass from the subordinated random processes to the CTRW limit of
random variables in one instance. The procedure results in factorization of the space and time characteristics. In what follows, the mixture of
symmetric $L\eta S$ (standard normal for $\eta=2$)
and completely asymmetric $L\alpha S$ laws has
the same distribution as the diffusion front $\tilde
R(t)=X_\eta(S_\alpha(t))$, where $X_\eta(\tau)$ belongs to the
class of symmetric $L\eta S$ processes ($0<\eta\leq2$), and it is
subordinated by the inverse $L\alpha S$ process $S_\alpha(t)$
($0<\alpha\leq1$). Really, using the 1/$\eta$-self-similarity property of
$X_\eta(\tau)$, i.\,e.
\begin{displaymath}
X_\eta(\tau)\stackrel{d}{=}\tau^{1/\eta}X_\eta(1)\,,
\end{displaymath}
and the $\alpha$-self-similarity of the
independent process $S_\alpha(t)$ as
\begin{displaymath}
S_\alpha(t)\stackrel{d}{=}[t/U_\alpha(1)]^\alpha\,,
\end{displaymath}
where
\begin{displaymath}
U_\alpha(\tau)\stackrel{d}{=}\tau^{1/\alpha}U_\alpha(1)
\end{displaymath}
is the $1/\alpha$-self-similarity of the $L\alpha S$ motion
$T_\alpha(\tau)$, we obtain
%\begin{eqnarray*}
%X_\eta(S_\alpha(t))&\stackrel{d}{=}&(S_\alpha(t))^{1/\eta}X_\eta(1)\\
%&\stackrel{d}{=}&t^{\alpha/\eta}[U_\alpha(1)]^{-\alpha/\eta}X_\eta(1)\\
%&=&C_\alpha C_\eta t^{\alpha/\eta}Q_{\alpha,\eta}\,,
%\end{eqnarray*}
\begin{displaymath}
X_\eta(S_\alpha(t))\stackrel{d}{=}(S_\alpha(t))^{1/\eta}X_\eta(1)
\stackrel{d}{=}t^{\alpha/\eta}[U_\alpha(1)]^{-\alpha/\eta}X_\eta(1)
=C_\alpha C_\eta t^{\alpha/\eta}Q_{\alpha,\eta}\,,
\end{displaymath}
what gives the desired mixture of the symmetric $L\eta S$ and completely
asymmetric $L\alpha S$ laws. Here $ C_\alpha, C_\eta$ are constants, and
$Q_{\alpha,\eta}$ is a random variable connected with the pdf $\vartheta_\alpha(x)$ ($0<\alpha\leq1$), which
describes random variables inverse to $L\alpha S$ ones. Observe that the Laplace transform of $\vartheta_\alpha(x)$ gives the well-known one-parameter Mittag-Leffler function, i.\,e.
\begin{displaymath}
E_\alpha(-z)=\int_0^\infty\exp(-zx)\,\vartheta_\alpha(x)\,dx\,.
\end{displaymath}
%Note that the symmetric process $H_\eta(\tau)$ has the following characteristic function
%\begin{displaymath}
%\langle e^{ikH_\eta(\tau)}\rangle=e^{-C_\eta k^\eta\tau}\,,\quad C_\eta>0\,.
%\end{displaymath}
According to above relations, the relaxation function takes the CC
temporal form
%\begin{eqnarray}
%\phi_{\rm CC}(t)&=&\langle\exp(ik C_\alpha C_\eta t^{\alpha/\eta}Q_{\alpha,\eta}\rangle\nonumber\\
%&=&\int_0^\infty\exp(-k^\eta C_\alpha C_\eta t^\alpha x)\,\vartheta_\alpha(x)\,dx\,\nonumber\\
%&=&E_\alpha\left[-(t/\tau_p)^\alpha\right]\,,\label{eqviii4}
%\end{eqnarray}
\begin{eqnarray}
\phi_{\rm CC}(t)&=&\langle\exp(ik C_\alpha C_\eta t^{\alpha/\eta}Q_{\alpha,\eta}\rangle\nonumber\\
&=&\int_0^\infty\exp(-k^\eta C_\alpha C_\eta t^\alpha x)\,\vartheta_\alpha(x)\,dx=E_\alpha\left[-(t/\tau_p)^\alpha\right]\,,\label{eqviii4}
\end{eqnarray}
obtained earlier in \cite{wk96} for $\eta=2$. The time constant $\tau_p=(Ck)^{-\eta/\alpha}$, where $C=(C_\alpha C_\eta)^{1/\eta}$. Note that the random variable
$Q_{\alpha,\eta}$ plays the same role (in a generalized sense) as relaxation rates of clusters
in the space representation considered above. In this case the
cluster structure dominates only on the microscopic level what
cannot be said about cooperative regions.

The diffusion front $\tilde
R(t)=X_\eta(Y_\gamma[S_\alpha(t)])$ is determined by the compound
subordination \cite{wjmwt10,sw10}, where the process
$S_\alpha(t)$ subordinates the process $Y_\gamma(y)$ independent on
$S_\alpha(t)$, and the process $Y_\gamma(y)=U_\gamma[S_\gamma(y)]$
itself is expressed in terms of $U_\gamma(s)$, being a strictly
$L\gamma S$ motion. Hence, we have
\begin{displaymath}
S_\gamma(s)=\inf\{s:\,U_\gamma(s)>y\}\,,
\end{displaymath}
being inverse to $U_\gamma(s)$.
As the process $U_\gamma(y)$ is one-self-similar that we get
\begin{displaymath}
Y^O_\gamma(y)\stackrel{d}{=}yY^O_\gamma(1)\,.
\end{displaymath}
Fixing of the space structure (clusters and super-clusters) in the system at one
instance is equivalent to
%\begin{eqnarray*}
%X_\eta(Y^O_\gamma[S_\alpha(t)])\stackrel{d}{=}(Y^O_\gamma[S_\alpha(t)])^{1/\eta}
%X_\eta(1)\\
%\stackrel{d}{=}t^{\alpha/\eta}[U_\alpha(1)]^{-\alpha/\eta}[Y^O_\gamma(1)]^{1/\eta}X_\eta(1)\,.
%\end{eqnarray*}
\begin{displaymath}
X_\eta(Y^O_\gamma[S_\alpha(t)])\stackrel{d}{=}(Y^O_\gamma[S_\alpha(t)])^{1/\eta}
X_\eta(1)
\stackrel{d}{=}t^{\alpha/\eta}[U_\alpha(1)]^{-\alpha/\eta}[Y^O_\gamma(1)]^{1/\eta}X_\eta(1)\,.
\end{displaymath}
The random variable $Y^O_\gamma(1)\stackrel{d}{=}1/{\cal B}_\gamma$
is reciprocal to the beta-distributed variable ${\cal B}_\gamma$.
The beta distribution with parameters $\gamma$ and $1-\gamma$ is
known also as the generalized arcsine distribution. In this
scenario the HN relaxation function is written as
\begin{equation}
\phi_{\rm HN}(t)=\int^\infty_0E_\alpha(-C_\alpha C_\eta k^\eta t^\alpha x)\,h_\gamma(x)\,dx\,,\label{eqviii5}
\end{equation}
where $h_\gamma(x)=[\Gamma(\gamma)\Gamma(1-\gamma)]^{-1}x^{-1}(x-1)^{-\gamma}$
is supported for $x>1$ and 0 otherwise. Here the  integrand
function $E_\alpha\left[-(t/\tau_p)^\alpha x\right]$ describes an evolution in
the distributions of microscopic quantities (like relaxation rates of
clusters) whereas the distribution $h_\gamma(x)$ relates to the
mesoscopic level of relaxation rates in super-clusters. The analogous analysis may be also provided for the JWS relaxation, but this makes the similar derivations, so we omit them.

In this context it would be useful to recall that the subordinator
$U_\alpha(t)$ remains the type of relaxation exponential {\it per se}
\cite{sokolov00}. As it has been shown in \cite{mw06a}, the stretched
exponential relaxation pattern appears from the other subordinator
undergoing the one-self-similarity
\begin{displaymath}
V_\alpha(t)\stackrel{d}{=}tU_\alpha(1)\,.
\end{displaymath}
In this case the complex system has neither cluster nor
super-cluster structures, but all relaxation rates of system
entities obey a $L\alpha S$ distribution.

Taking $\gamma=1$, the diffusion front $\widetilde
R(t)=X_\eta(Y_\gamma[S_\alpha(t)])$ is simplified to $\widetilde
R(t)=X_\eta(S_\alpha(t))$. Then Eq.(\ref{eqviii5}) transforms into
Eq.(\ref{eqviii4}), and we get a cluster structure in the disordered
system without any contribution of super-clusters. In this
case the relaxation function is the CC law. On the other hand, for
$\alpha=1$ the diffusion front reads $\widetilde R(t)=X_\eta(Y_\gamma(t))$.
In this situation the interactions among super-clusters play a dominating role. Their heavy tails of relaxation rates lead to the CD relaxation pattern.

\begin{table} \caption{Survival probability of the initial non-equilibrium state of a system in different probability approaches.\\ }
  \label{tab_outcome}
	\centering{
  \begin{tabular}{|c|c|c|}\hline
	  &  & \\
   Probability & Survival probability of & Key notations\\
 approach & the non-equilibrium state &  \\
    &  & \\  \hline
		   &  & \\
     master &  & \\
    equation &  $\phi(t)=e^{-\int^t_0 r(s)\,ds}$  & Eq.(\ref{eqiii3b}) \\
		&  & \\    \hline
		relaxation &  &   \\
    rate & $\phi(t)=\langle e^{-\tilde\beta t}\rangle$ & Eq.(\ref{eqiii5}) \\
    distribution &  & \\  \hline
    & & \\
    two-state & $\phi(t)=\langle e^{-\omega_p\tau}\rangle$ & Eq.(\ref{eqiv6a}) \\
		relaxation &  & \\
    & & \\ \hline
   &  &  \\
  diffusion  &  $\phi(t)=\begin{cases} \langle e^{ik\tilde R(t)}\rangle\\ \langle e^{-k\bar{R}(t)}\rangle  \end{cases}$ & Eqs.(\ref{eqv1}), (\ref{eqv1a}) \\
	  &  &  \\  \hline
  \end{tabular}}
\end{table}

%%%%%%%%%%%%%%
%%%%%%%%%%%%%%
\section{Outline}
%%%%%%%%%%%%%%
%%%%%%%%%%%%%%

The manifestations of many-body effects in the relaxation of dipolar systems are fundamental and universal, independent of the physical and chemical structures of their interacting entities. The interactions are usually non-trivial and anharmonic, resulting in complex (chaotic) dynamics in the phase space spanned by the coordinates and momenta of the interacting units. The presence of the many-body effects in the dynamics cannot be ignored. This means that the dynamical processes in the such systems have a stochastic background. Nevertheless, the characterization of complex systems wholly is deterministic, in the form of their universal relaxation patterns. The main feature of relaxing complex systems is that their relaxation response is non-exponential in nature. All types of the empirical functions used to fit the relaxation data exhibit the fractional-power dependence of the dielectric responses in frequency and time. It is worth noticing that this unique property is independent on any special details of  examined systems. In this framework one can expect that the macroscopic behavior of the complex systems is governed by ``averaging principles'' like the law of large numbers to be in force. However, the problem is that the macroscopic evolution of complex systems is not attributed to any particular entity taken from those forming the system. The description of the relationship between the local random characteristics of complex systems and the universal deterministic empirical laws of relaxation is performed by the limit theorems of probability theory. Applications of this approach to relaxation and transport for large classes of (physical, chemical, and other) phenomena \cite{jonscher83,hav94,jonscher96}, involving different types of self-similar random processes, has turned out to be very successful in recent years \cite{twpp08,gnpawets09,tmk11,jswbkw13,kkbnr10,rv14,otbcdfgnmnos14}. Especially, very fruitful appeared the idea of subordination description which allows one to characterize a stochastic transport of particular excitation mode in complex systems. The general probabilistic formalisms treat relaxation of the complex systems regardless of the explicit representation of local interactions, the detailed relations between relaxing entities, interaction ranges and the transport efficiencies. In a natural way, they give efficient methods for evaluating the dynamical averages of the relaxation processes. In this article, utilizing the stochastic tools, we have derived the most-known empirical relaxation laws, characterized their parameters, connected the parameters with local random characteristics of the relaxation processes, reconstructed the internal random structure of relaxing systems, justified the energy criterion, demonstrated the transition from analysis of the microscopic random dynamics in the systems to the macroscopic deterministic description by integro-differential equations. It should be noticed that the classical methods of statistical physics take into account the central limit theorem in respect to the probability distributions having finite moments. However, this concept does not help to clarify the nature of relaxation phenomena in complex systems. The above consideration has an evident advantage over the traditional models and goes behind the classical statistical physics. In our approach, to look at the problem of anomalous time evolution of complex systems from different sides, we used several alternative stochastic methods (see Table~\ref{tab_outcome}). Such an attempt has some advantages. Firstly, it sets the range of validity of the approaches. Secondly, once this range was established, it allows us to extend the results to (in principle) arbitrary relaxing systems. Thus, joining different methods may lead to a deeper understanding of the complex dynamics found in nature.

\section*{Acknowledgements}
A.S. is grateful to the Faculty of Pure and Applied Mathematics and the Hugo Steinhaus Center for pleasant hospitality during his visit in Wroc{\l}aw University of Science and Technology. He is also grateful for a partial support from the NCN Maestro grant No. 2012/06/A/ST1/00258.

\end{document}